\documentclass{mnras}

\usepackage{amsmath, amssymb, graphics, graphicx, rotating, lscape}
\usepackage{newtxtext,newtxmath}
\usepackage[T1]{fontenc}
\usepackage{ae,aecompl}

\usepackage{subfig}

\newcommand*\rad{~rad~m$^{-2}$}
\newcommand*\sigmaRM{\sigma_{\rm RM}}
\newcommand*\sigmaRMwtd{\sigma_{\rm RM, wtd}}
\newcommand*\gradRMwtd{\Delta{\rm RM_{wtd}}}
\newcommand*\DeltaRMwtd{\Delta{\rm RM_{wtd}}}
\newcommand*\DeltaRM{\Delta{\rm RM}}
\newcommand*\gradRM{\Delta{\rm RM}}

\title[Broadband polarization of radio AGN]
{Broadband, radio spectro-polarimetric study of 100 radiative-mode and jet-mode AGN}

\author[O'Sullivan et al.]{S.~P. O'Sullivan$^{1,2}$\thanks{E-mail: shane@astro.unam.mx}, C.~R.~Purcell$^{3}$, C.~S.~Anderson$^{4}$, J.~S.~Farnes$^{5,2}$, X.~H.~Sun$^{6}$, \and B.~M.~Gaensler$^{7, 2}$ \\ \\
$^{1}$Instituto de Astronom\'ia, Universidad Nacional Aut\'onoma de M\'exico (UNAM), A.P.~70-264, 04510 M\'exico, D.F., M\'exico. \\
$^{2}$ARC Centre of Excellence for All-sky Astrophysics (CAASTRO). \\
$^{3}$Research Centre for Astronomy, Astrophysics, and Astrophotonics, Macquarie University, NSW 2109, Australia. \\
$^{4}$CSIRO Astronomy and Space Science, 26 Dick Perry Avenue, Kensington WA 6151, Australia. \\
$^{5}$Department of Astrophysics/IMAPP, Radboud University, PO Box 9010, NL-6500 GL Nijmegen, The Netherlands. \\
$^{6}$School of Physics and Astronomy, Yunnan University, Kunming 650500, China. \\
$^{7}$Dunlap Institute for Astronomy and Astrophysics, The University of Toronto, 50 St.~George Street, Toronto, ON M5S 3H4, Canada. \\
}

\date{Accepted XXX. Received YYY; in original form ZZZ}
\pubyear{2015}

\begin{document}
\label{firstpage}
\pagerange{\pageref{firstpage}--\pageref{lastpage}}
\maketitle

\begin{abstract}
We present the results from a broadband (1 to 3 GHz), spectro-polarimetry study of the integrated emission from 100 extragalactic radio sources with the ATCA, selected to be highly linearly polarized at 1.4~GHz. We use a general purpose, polarization model-fitting procedure that describes the Faraday rotation measure (RM) and intrinsic polarization structure of up to three distinct polarized emission regions or `RM components' of a source. 
Overall, 37\%/52\%/11\% of sources are best fit by one/two/three RM components. However, these fractions are dependent on the signal-to-noise ratio (S/N) in polarization (more RM components more likely at higher S/N). 
In general, our analysis shows that sources with high integrated degrees of polarization at 1.4~GHz have low Faraday depolarization, are typically dominated by a single RM component, have a steep spectral index, and a high intrinsic degree of polarization. 
After classifying our sample into radiative-mode and jet-mode AGN, we find no significant difference between the Faraday rotation or Faraday depolarization properties of jet-mode and radiative-mode AGN. However, there is a statistically significant difference in the intrinsic degree of polarization between the two types, with the jet-mode sources having more intrinsically ordered magnetic field structures than the radiative-mode sources. 
We also find a preferred perpendicular orientation of the intrinsic magnetic field structure of jet-mode AGN with respect to the jet direction, while no clear preference is found for the radiative-mode sources.  
\end{abstract}

\section{Introduction}
\label{introduction}
A new window into the magnetic universe has been opened by upgraded radio-telescope facilities that allow broadband, continuum-polarization observations at high spectral resolution. The bright radio sky accessible by current facilities is dominated by active galactic nuclei (AGN) through the production of powerful jets of relativistic plasma that emit non-thermal synchrotron radiation. The observed radiation is often highly linearly polarized, which provides important information on the degree of order of the magnetic field in the emitting plasma as well as its orientation in the plane of the sky. Multi-frequency radio polarization observations allow us to measure the effect of Faraday rotation, through the rotation of the plane of linear polarization and the behaviour of the degree of polarization as a function of wavelength. The Faraday rotation measure (RM), which enables us to study the properties of magneto-ionic material along the line of sight, is defined as  
\begin{equation}
{\rm RM}_{[{\rm rad~m}^{-2}]} = 0.812 \int_{\rm source}^{\rm telescope} n_{e\,\,[{\rm cm}^{-3}]} \,\, B_{||\,\,[{\rm \mu G}]} \,\, dl_{\,\,[\rm{pc}]} 
\label{rmeqn}
\end{equation}
where $n_e$ is the free electron number density, $B_{||}$ is the line-of-sight magnetic field and $l$ is the path length through the magneto-ionic medium, in the indicated units.

Broadband radio polarization science (e.g.~Gaensler et al.~2015) is a primary driver of upcoming, pre-Square Kilometre Array (SKA), all-sky surveys (e.g.~VLA Sky Survey\footnote{https://science.nrao.edu/science/surveys/vlass/vlass}, Australian SKA Pathfinder\footnote{http://www.askap.org/possum}), as well as a key science driver for the SKA (The Origin and Evolution of Cosmic Magnetism)\footnote{http://www.skatelescope.org/magnetism}. 
Radio spectro-polarimetry observations provide unique diagnostics on magnetised structures in radio AGN and are critical for resolving degeneracies in the Faraday structure of radio sources (e.g.~Farnsworth et al.~2011, O'Sullivan et al.~2012). Indeed, accurately characterising the Faraday structure of radio AGN is also important in order to use them as reliable statistical probes of foreground magneto-ionic material, such as the cluster/group environment (Bonefede et al.~2015), the intergalactic medium (Vacca et al.~2015), intervening galaxies (Gaensler et al.~2015) and the Galactic ISM (Haverkorn et al.~2015). 

Recent broadband polarization studies of typically unresolved radio sources (e.g.~Farnes et al.~2014a, Pasetto et al.~2016) have shown the need for high precision spectro-polarimetric studies to better understand the complex behaviour of the polarization data. The development of spectro-polarimetric model-fitting techniques, such as various approaches to `QU-fitting' (e.g.~Farnsworth et al.~2011, O'Sullivan et al.~2012, Ideguchi et al.~2014, Anderson et al.~2016) have proven most successful so far in interpreting and accurately recovering the underlying Faraday structure of radio sources (Sun et al.~2015). 
However, the field of broadband radio spectro-polarimetry is still developing and several technical and scientific interpretation challenges must be overcome before this field can reach its scientific potential (c.f.~Farnsworth et al.~2011, O'Sullivan et al.~2012). 

One of our main motivations in this current work is to determine the origin of the observed difference in the integrated polarization properties of radio AGN with different host galaxy accretion states (O'Sullivan et al.~2015; OS15). The study of OS15 used a large statistical sample of 1,611 radio AGN based on narrowband data at 1.4~GHz from the NRAO VLA Sky Survey (NVSS; Condon et al.~1998), and with optical spectra classification from the Sloan Digital Sky Survey data (Best \& Heckman 2012, and references therein). In classifying the host galaxy accretion states, we follow the terminology of Heckman \& Best (2014) in which they divide AGN into `radiative-mode' and `jet-mode' AGN. The radiative-mode AGN are those with strong, high-ionisation, optical emission lines powered by accretion rates onto the central supermassive black holes in excess of $\sim$1\% of the Eddington limit, with about 10\% producing powerful radio jets (also known as `quasar-mode', `cold-mode', or `high-excitation radio galaxies'). The jet-mode AGN have weak or non-existent optical emission lines, possibly due to radiatively-inefficient accretion, with the bulk of their energetic output coming from their radio jets (also known as `radio-mode', `hot-mode', or `low-excitation radio galaxies'). Differences in the accretion states of AGN host galaxies is generally considered to be due to the large scale gaseous environment (i.e.~cold/hot gas reservoir possibly leading to radiative-mode/jet-mode AGN, Hardcastle et al.~2007). 

The radio data used in OS15 were limited in several respects. Firstly, it was at two adjacent frequency near 1.4 GHz, which severely limited the ability to investigate the effect of depolarization. Secondly, it had poor angular resolution (45"), which meant that essentially all sources were spatially unresolved and it was not possible to study the magnetic field structure of the sources. To investigate further, we have obtained broadband polarization data with the Australia Telescope Compact Array (ATCA; Wilson et al.~2011) to determine the origin of the difference in the integrated magnetic field properties of the two broad classes of radio AGN. The new observations have broad enough frequency coverage to provide high precision measurements of the amount of depolarization local to each source and 
distinguish between the intrinsic magnetic field properties of the source and the magneto-ionic material causing the depolarization. 

In Section 2, we describe the construction of our sample, the data reduction and imaging, as well as our approach to the polarization modelling. 
Section 3 contains our results and Section 4 presents our discussion. 
The conclusions are listed in Section 5. 
Throughout this paper, we assume a flat $\Lambda$CDM cosmology with 
H$_0 = 67.3$ km s$^{-1}$ Mpc$^{-1}$, 
$\Omega_M=0.315$ and $\Omega_{\Lambda}=0.685$ (Planck Collaboration et al.~2014), and define the spectral index, $\alpha$, 
such that the observed total intensity ($I$) at frequency $\nu$ follows the relation 
$I_{\nu}\propto\nu^{\rm{+}\alpha}$.

\begin{figure} 
\centering
    \includegraphics[angle=0, clip=true, trim=0cm 0cm 0cm 0cm, width=.45\textwidth]{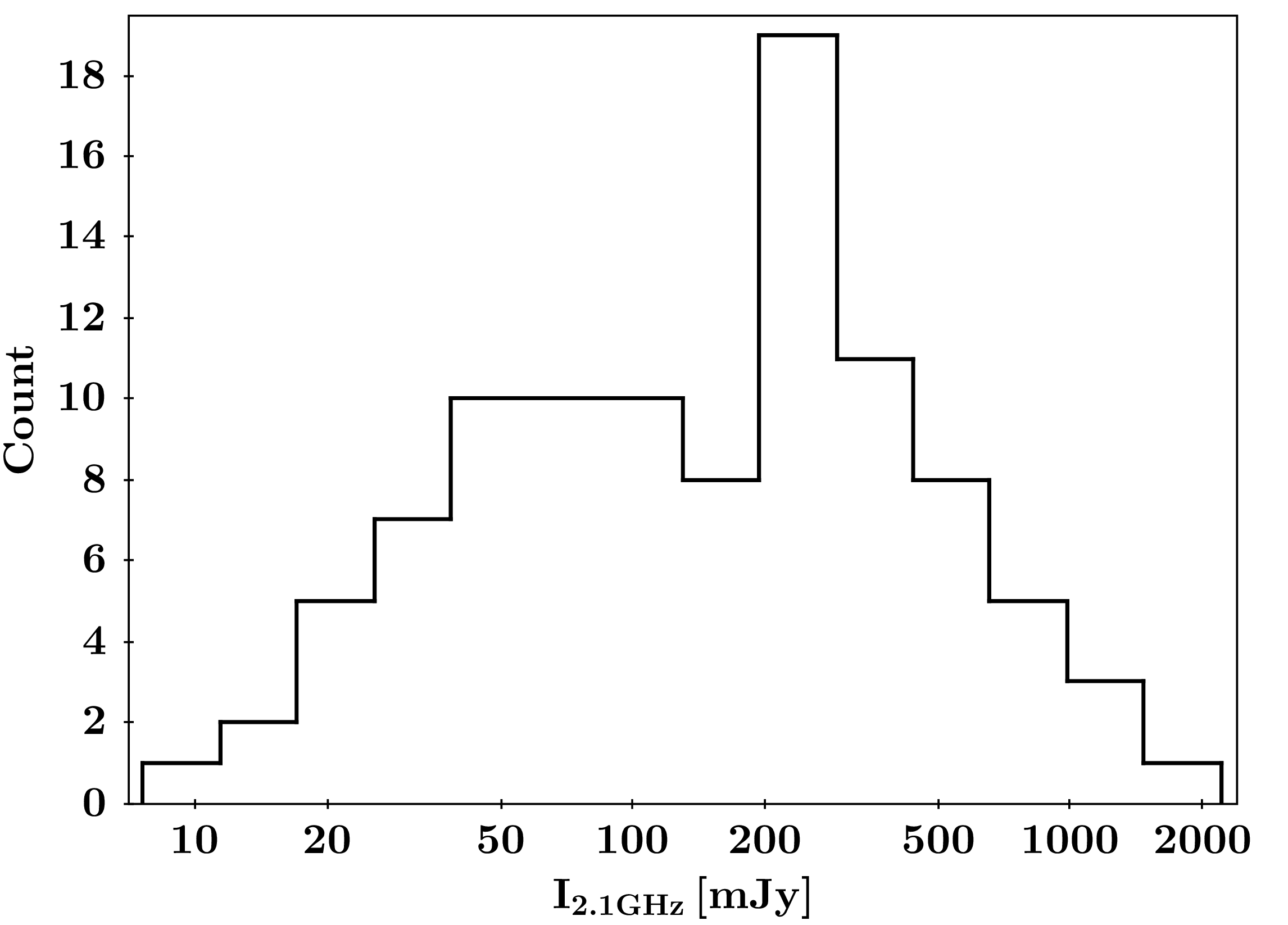} 
    \caption{ {\small Histogram of the integrated total intensity emission at 2.1~GHz, in mJy, for all sources in our sample.} }
    \label{stokesi}
\end{figure}  

\begin{figure} 
\centering
    \includegraphics[angle=0, clip=true, trim=0cm 0cm 0cm 0cm, width=.45\textwidth]{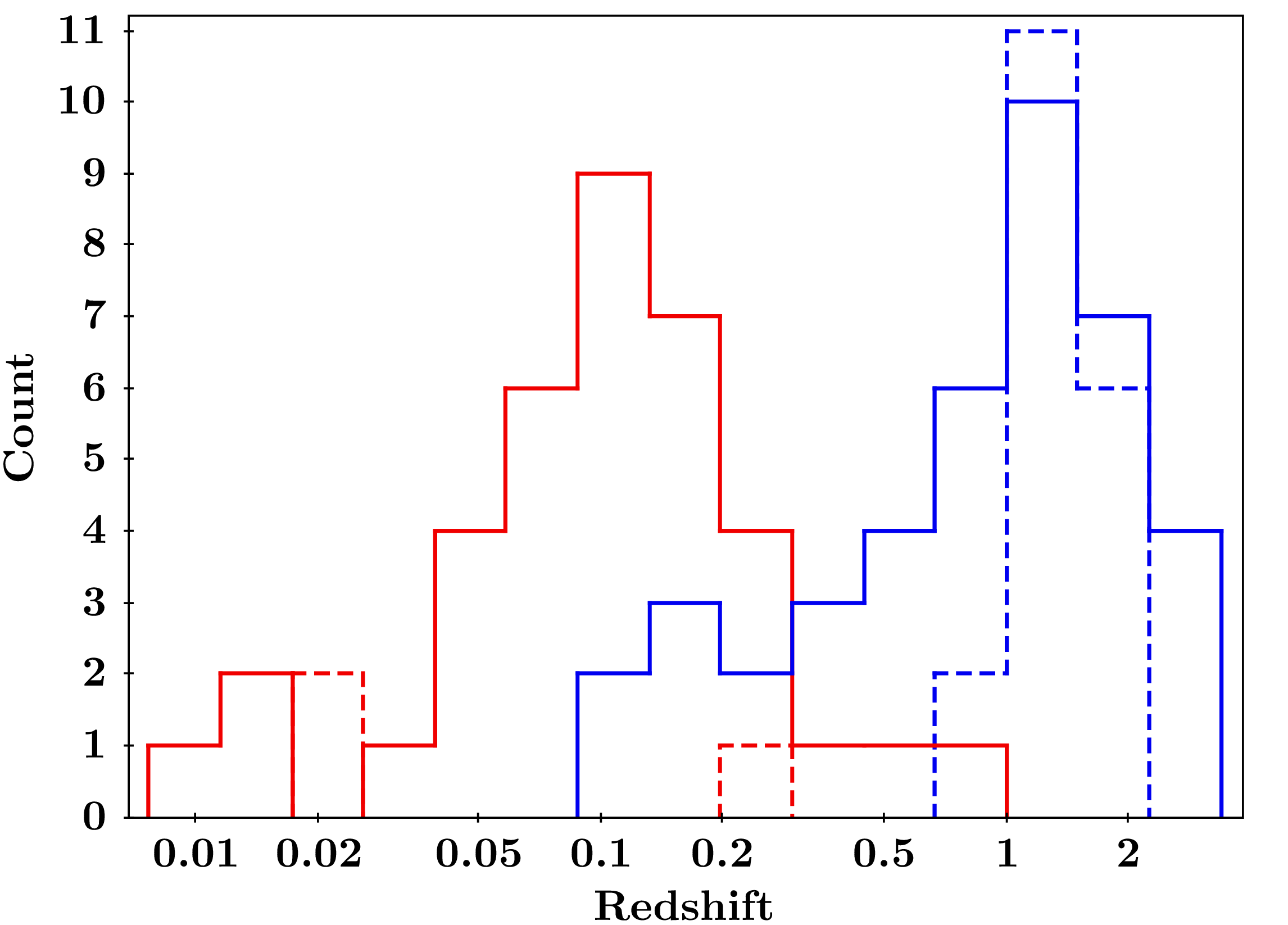}
    \caption{ {\small Histogram of the distribution of radiative-mode (blue, right) and jet-mode (red, left) sources with redshift. 
    The solid/dashed lines denote sources classified as steep/flat spectrum sources. } }
    \label{zhist}
\end{figure}      

\section{Data Analysis} \label{data}

\subsection{Sample construction and observations} \label{obs}
The target sources were selected from Hammond et al.~(2012) which provides redshifts for 4003 polarized radio sources from Taylor et al.~(2009). In this sense it is a `high-polarization' selected sample, since Taylor et al.~(2009) only included sources with peak polarized intensities greater than $\gtrsim2.4$~mJy and with degrees of polarization greater than 0.5\% (see Lamee et al.~2016 for a catalog not selected on high polarized intensity). The selected sources are limited to the declination range from $-30^\circ$ to $-40^\circ$ (the southern limit of the NVSS). The $-30^\circ$ Dec limit was chosen to maintain a high quality synthesised beam for high resolution imaging with the ATCA in the 6-km (6A) array configuration. All sources are also at least 20 degrees in latitude away from the Galactic plane. This resulted in a total of 162 sources which were observed under the C2913 program name with the ATCA in the 16~cm band (1.1 to 3.1 GHz, at 1 MHz spectral resolution) and the 4~cm band (4.5 to 6.5 GHz and 8 to 10 GHz, at 1 MHz spectral resolution), for 108 hours from 2014 April 19 to 2014 April 28. 
The required integration time for each source varied based on detecting the band-averaged integrated polarized flux of each source at $\gtrsim10$ times the noise level. In practice, many sources had detections much greater than this level due to the additional requirement of good $uv$-coverage ($\sim6$ cuts spread evenly across a 12 hour synthesis) to produce images of sufficient quality to resolve sources into core, jet and lobe components. 
Three days of observations (April 22, 23 \& 27) suffered from technical problems and the data were unreliable for the present analysis. 
This resulted in a reduction in the number of sources to a final sample of 100. The sources have integrated total intensities at 2.1~GHz ranging from 10~mJy up to 1.7~Jy. Figure~\ref{stokesi} shows a histogram of the integrated Stokes $I$ at 2.1~GHz for all sources.

The 2dF (Colless et al.~2001) and 6dF (Jones et al.~2009) optical host galaxy spectra\footnote{http://www.2dfgrs.net/, http://www.6dfgs.net/} were used to manually identify the sources as either radiative-mode (prominent emission line spectra) or jet-mode AGN (quiescent galaxy spectra lacking prominent emission lines). In all cases, the manual classification was obvious enough to not require multiple independent identifiers to avoid subjective selection of radiative-mode or jet-mode classes. This resulted in the classification of 60 radiative-mode and 40 jet-mode AGN in our sample. Figure~\ref{zhist} shows the redshift distribution of our sample, split into the radiative-mode and jet-mode classes.

\subsection{Data reduction} \label{reduction}
The data were calibrated using standard techniques in the software package {\sc Miriad}. PKS~B1934-638 was used as the bandpass and primary flux density calibrator at both 16-cm and 4~cm bands. The presence of radio-frequency interference (RFI) in the 16~cm band resulted in the removal of a large fraction of data ($\sim$30\% of the full band and effectively all data at frequencies lower than 1.3 GHz). Target sources were grouped on the sky such that each group had a nearby secondary complex gain and leakage calibration source whose time-dependent calibration solutions, at 128 MHz intervals, were applied to each target source in that group. 

\subsection{Imaging} \label{imaging}
Three broadband total intensity images were created for each source using the {\sc Miriad} task {\sc mfclean} from 1.3 to 3 GHz, 4.5 to 6.5 GHz and 8 to 10 GHz. This provided total intensity images with angular resolutions ranging from $\sim10$'' to 1'', allowing us to resolve the structure of double-lobed sources with a projected separation of $\gtrsim10$ kpc throughout our full redshift range (0.01 to 2.8). Sources that remain unresolved in the highest resolution image are most likely blazars or very compact radio galaxies (e.g.~compact symmetric objects, Gugliucci et al.~2007). 
For our spectro-polarimetric analysis, all sources were imaged with natural weighting at 10 MHz intervals from 1.3 to 3 GHz, with the data tapered to the beam-size at the lowest frequency and smoothed to a common resolution. For the present analysis to be consistent across all sources (resolved and unresolved), the integrated Stokes $I$, $Q$ and $U$ flux in each image was extracted. The source boundary was defined by convolving the 1.3 to 3 GHz total intensity image with the beam-size of the lower-resolution 10 MHz images. The $I$, $Q$ and $U$ flux densities were then integrated in each 10 MHz image down to the 10\% contour of the 1.3 to 3 GHz image. The error in the $I$, $Q$ and $U$ fluxes was determined from the rms noise in a nearby off-source area and modified for the number of independent beams sampled for the resolved sources. To prepare for the polarization model fitting, the frequency-dependent fractional $q$ and $u$ values were calculated by dividing the integrated Stokes $Q$ and $U$ by the Stokes $I$ values, as determined from a polynomial fit to the total intensity data. 
The technique of RM synthesis (Burn 1966, Brentjens \& de Bruyn 2005) was applied to generate Faraday dispersion functions for each source. These results were not used in our analysis except for obtaining an estimate of the signal-to-noise ratio of the detection of polarized emission 
and in generating band-averaged polarized intensity images. 

For the present analysis we have not conducted a spectro-polarimetric analysis of the 4~cm data. The main reason is that we cannot reliably combine the data from the 16~cm band and the 4~cm band for our polarization modelling (Section \ref{model}). Firstly, there is a significant mis-match in the short wavelength uv-coverage in the 4~cm band due to the 6~km array configuration that was used for both bands (2.7~k$\lambda$ at 1.3 GHz versus 9.4~k$\lambda$ at 4.5~GHz and 16.7~k$\lambda$ at 8~GHz). This is problematic for a large fraction of our resolved sources where we are likely resolving out emission in the 4~cm band (46 sources have angular sizes larger than the maximum angular scale at 4~cm of $\sim$22''), as well as significantly reduced signal-to-noise due to the heavily resolved polarization structure. Secondly, since we model Stokes $Q/I$ and $U/I$ without accounting for changes in the spectral index, we wish to avoid modelling the polarization data in sources where the total intensity spectral index changes in a significant manner (e.g.~from steep to flat). Thirdly, the majority of the unresolved sources have flat or inverted spectral index values which means they are subject to significant opacity effects which our polarization modelling approach does not account for, and this effect would be exacerbated by including the 4~cm band data. 
Finally, as shown in Anderson et al.~(2016), in terms of Faraday effects, the addition of 4~cm band data mainly has the effect of enhancing sensitivity to strongly depolarized components that probably arise in the complex inner regions of AGN. These components do not usually contribute significantly to the emission at 16 cm, and thus do not generally improve reconstruction of the source structure from 16~cm data.
Thus, in the proceeding analysis we only use the 4~cm data for studying the total intensity source morphologies.

\begin{figure} 
\centering
  \subfloat[ Polarization fraction ($p$) as a function of wavelength-squared ($\lambda^2$). \label{modelp}]{%
    \includegraphics[angle=0, clip=true, trim=0cm 0cm 0cm 0cm, width=.45\textwidth]{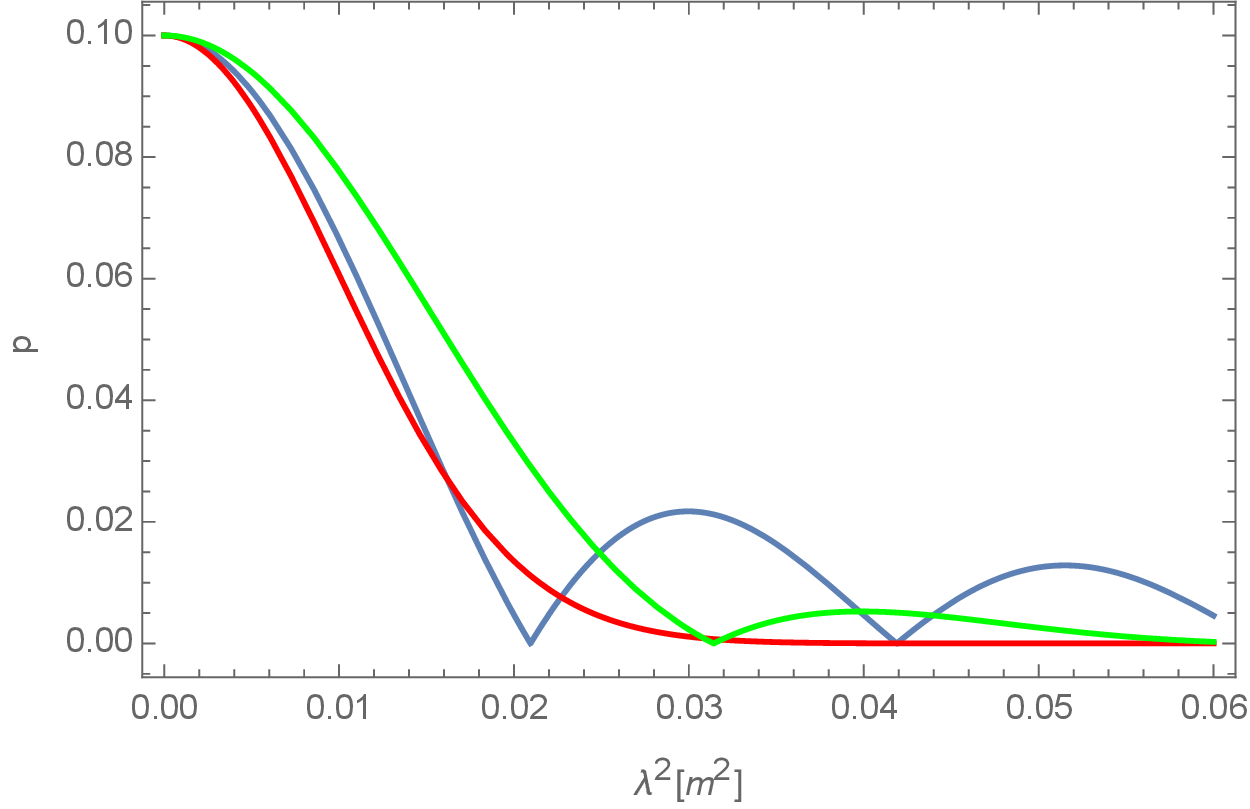}}\hfill
  \subfloat[ Faraday dispersion function (FDF): amplitude of the fourier transform of the complex polarization, $P$. \label{modelfdf}]{%
    \includegraphics[angle=0, clip=true, trim=0cm 0cm 0cm 0cm, width=.45\textwidth]{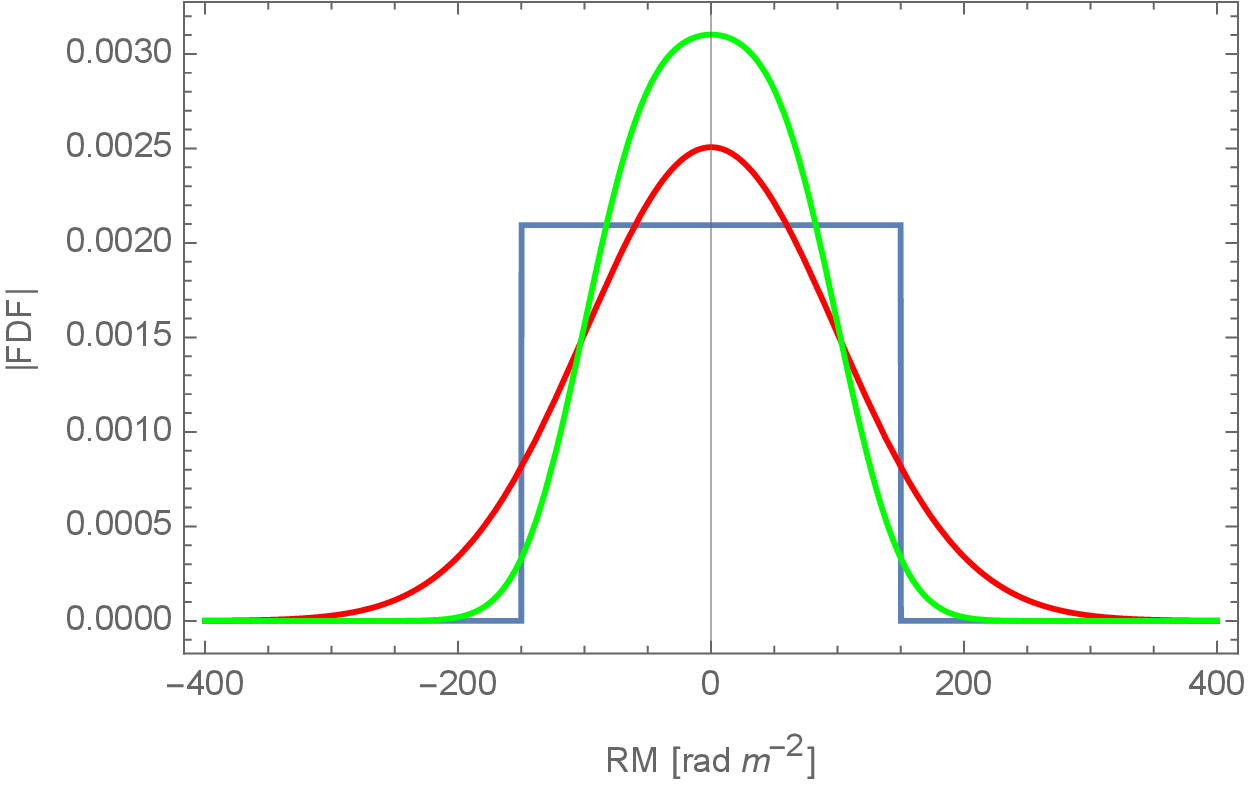}}\\
  \caption{ {\small Model behaviour for three different scenarios as described by Eqn.~\ref{modeleqn} for a source with $p=10\%$, ${\rm RM}=0$\rad, $\psi_0=0^\circ$: top-hat blue line ($\sigmaRM=0$\rad, $\DeltaRM=150$\rad), gaussian red line ($\sigmaRM=50$\rad, $\DeltaRM=0$\rad), green line ($\sigmaRM=20$\rad, $\DeltaRM=100$\rad). } } \label{modelpfdf}
\end{figure}

\subsection{Extracting the polarization and Faraday rotation parameters} \label{model}

To accurately describe the broadband polarization and Faraday rotation measure (RM) behaviour of our sources, we use the QU-fitting and model selection technique described by O'Sullivan et al.~(2012). To capture a broad range of possible Faraday rotation behaviour, we use a model for the complex polarization ($P$) which describes the effects of both random and uniform magnetic fields (e.g.~Sokoloff et al.~1998, eqn.~41). 

\begin{equation}
P_j = Q_j + iU_j = p_{0,j} \, I\, e^{2 i (\psi_{0_j}+{\rm RM}_j \lambda^2)}  \frac{\sin \Delta {\rm RM}_j \lambda^2}{\Delta {\rm RM}_j \lambda^2} e^{-2\sigma^2_{{\rm RM}_j} \lambda^4}
\label{modeleqn}
\end{equation}
where for each $j^{th}$ complex polarization component or ``RM component'', $p_{0,j}$ is the intrinsic degree of polarization, $\psi_{0_j}$ is the intrinsic polarization angle, $\Delta {\rm RM}_j$ and $\sigma_{{\rm RM}_j}$ describe RM variations from uniform and random magnetic fields, respectively, and $I$ is the integrated flux of the source. The linear polarization angle is defined as usual, $\psi=\frac{1}{2} \arctan \frac{u}{q}$. 

Figure~\ref{modelp} shows the behaviour of the polarization fraction ($p=\sqrt{q^2+u^2}$), described by Eqn.~\ref{modeleqn}, for a single RM component as a function of wavelength-squared for different combinations of $\sigma_{\rm RM}$, the RM dispersion, and $\Delta{\rm RM}$, the RM gradient (the parameters which cause depolarization). 
The parameter $\sigma_{\rm RM}$ is generally considered to describe random RM variations that are external but local to the radio source (e.g.~Laing et al.~2008), while the parameter $\Delta{\rm RM}$ can describe both internal and external Faraday depolarization effects; for example, a linear gradient in RM across the emission region, or internal Faraday rotation in a uniform field (c.f.~Sokoloff et al.~1998, Schnitzeler et al.~2015).  
In the `short-wavelength' regime ($p(\lambda)/p_0>0.5$), the $\sigmaRM$ and $\DeltaRM$ parameters produce equivalent amounts of depolarization for $\DeltaRM\sim3.22\sigmaRM$. It is also worth noting that the $\DeltaRM$ parameter in Eqn.~\ref{modeleqn} describes a linear gradient in RM across a flat beam profile, but to describe the more realistic case of a Gaussian beam profile one needs to divide $\DeltaRM$ by a factor of 1.35 (Sokoloff et al.~1998).

Figure~\ref{modelfdf} shows how the Faraday dispersion functions (i.e.~the Fourier transform of $P$) describe a range of behaviours from a Gaussian function (red) to a top-hat function (blue). 
This also describes Faraday dispersion functions similar to a ``super-Gaussian'' function, as presented by Anderson et al.~(2016). 
Delta-functions are incurred when both $\sigmaRM$ and $\DeltaRM$ are zero. 
To model more complicated behaviours than shown in Figure~\ref{modelpfdf}, we add extra RM components, up to a limit of three (e.g.~$P=P_1+P_2+P_3$). 

For each source, we considered one, two and three RM component models for fitting to the observed $q$ and $u$ data.  
We first fit `Faraday thin' models in which both the $\sigmaRM$ and $\DeltaRM$ parameters were excluded, then included $\sigmaRM$ only, then $\DeltaRM$ only, and finally including both $\sigmaRM$ and $\DeltaRM$ for each model. This resulted in a total of 12 models fitted to each source, from which we calculated both the reduced-$\chi$-squared values ($\chi^2_r$) and the Bayesian Information Criterion (BIC). The model with the lowest BIC value was selected as the best-fitting model (e.g.~Raftery 1995). 	

In order to compare the RM dispersion for each source in our sample in a consistent manner, we consider the polarization-weighted RM dispersion, defined as
\begin{equation}
\sigmaRMwtd = \sum_j  p_{0,j}  \, \sigma_{\rm RM,j}  \Bigg/ \sum_j p_{0,j} .
\label{sigmaRMwtdeqn}
\end{equation}
We also calculate the polarization-weighted RM gradient,
\begin{equation}
\DeltaRM_{\rm wtd} = \sum_j  p_{0,j}  \, \Delta {\rm RM}_j \Bigg/ \sum_j p_{0,j} .
\label{deltaRMwtdeqn}
\end{equation}

We use the Galactic foreground RM reconstruction of Oppermann et al.~(2015) to subtract the Galactic RM (GRM) contribution from our derived RM values and obtain the residual rotation measure (RRM). For sources with more than one RM component we calculate the polarization-weighted RM (${\rm RM}_{\rm wtd}$) and subtract the GRM from this value (i.e.~${\rm RRM} = {\rm RM}_{\rm wtd} - {\rm GRM}$). 
The polarization-weighted RM is calculated as 
\begin{equation}
{\rm RM}_{\rm wtd} = \sum_j p_{0,j} RM_j \Bigg/ \sum_j p_{0,j} 
\label{RMwtdeqn}
\end{equation}

Subtraction of the GRM introduces a significant additional error term to the RRM (a median error of $\sim$5~rad~m$^{-2}$ for our sample), which may mask any dependence between the RRM and another variable. Thus, when checking for correlations with RRM, we also separately check for correlations with RM and GRM. 

\begin{table*}
 \caption{Polarization model fit parameters, with errors in parentheses, for all sources, along with best-fit statistics.}
\scriptsize
 \centering
   \begin{tabular}{cccccccc}
    \hline\hline
       (1) & (2) & (3) & (4) & (5) & (6) & (7) & (8)  \\
      Name  & $p_{0(1,2,3)}$ & RM$_{(1,2,3)}$ & $\sigmaRM$$_{(1,2,3)}$ & $\DeltaRM$$_{(1,2,3)}$ & $\psi_{0(1,2,3)}$ & $\chi^2_{r}$ & BIC \\
                 &  (\%)               & (rad~m$^{-2}$) & (rad~m$^{-2}$)                 & (rad~m$^{-2}$)                &    (deg)               &                          &        \\
      \hline    
J0001-3025 & 13.5(1.1), 3.6(1.9), 0.8(0.2) & -4(1), 64(23), -149(8) & 12(1), 31(10), -- & --, --, -- & -12(3), 37(21), 38(13) & 1.1 & 1014 \\ 

J0004-4011 & 21.9(1.2), --, -- & 5(2), --, -- & 12(1), --, -- & --, --, -- & -69(2), --, -- & 1.2 & 1008 \\ 

J0009-3216 & 4.9(0.1), 0.4(0.1), -- & 7(1), -92(9), -- & --, --, -- & 17(3), --, -- & 64(1), 36(14), -- & 0.9 & 926 \\ 

J0010-3054 & 5.0(1.4), 6.7(1.1), -- & 14(15), -5(6), -- & 35(6), 12(2), -- & --, --, -- & 76(18), -17(15), -- & 1.0 & 962 \\ 

J0014-3059 & 6.7(0.1), --, -- & 5(1), --, -- & 13(1), --, -- & --, --, -- & 28(1), --, -- & 1.1 & 999 \\ 

J0015-3526 & 1.4(0.2), 1.3(0.3), -- & -2(7), -63(11), -- & --, 13(4), -- & --, --, -- & 29(14), -35(18), -- & 0.9 & 932 \\ 

J0020-3106 & 23.0(1.3), --, -- & -6(1), --, -- & 11(1), --, -- & --, --, -- & 52(2), --, -- & 1.3 & 1028 \\ 

J0021-3334 & 9.1(0.1), 0.3(0.1), -- & -15(1), -136(9), -- & --, --, -- & 31(1), --, -- & -48(1), -65(15), -- & 1.1 & 985 \\ 

J0038-3142 & 24.6(1.6), --, -- & 4(2), --, -- & 11(1), --, -- & --, --, -- & 82(3), --, -- & 1.0 & 945 \\ 

J0038-3859 & 0.5(0.04), 1.3(0.03), -- & -62(7), 29(2), -- & --, --, -- & --, --, -- & -17(10), 7(3), -- & 0.8 & 827 \\ 

J0040-3254 & 15.0(0.4), --, -- & 5(1), --, -- & 12(1), --, -- & --, --, -- & 22(1), --, -- & 1.2 & 1018 \\ 

J0044-3147 & 12.1(2.6), 4.1(2.5), -- & -1(7), -48(19), -- & --, --, -- & --, --, -- & -15(12), 71(34), -- & 1.0 & 957 \\ 

J0049-3140 & 12.1(0.3), --, -- & 1(1), --, -- & --, --, -- & --, --, -- & 40(2), --, -- & 1.0 & 933 \\ 

J0059-3152 & 8.6(0.2), --, -- & 10(1), --, -- & --, --, -- & --, --, -- & 11(1), --, -- & 1.1 & 986 \\ 

J0104-3314 & 4.6(0.1), --, -- & -4(1), --, -- & 22(1), --, -- & --, --, -- & 4(1), --, -- & 1.1 & 993 \\ 

J0108-3611 & 16.1(0.4), --, -- & 13(1), --, -- & --, --, -- & --, --, -- & -63(1), --, -- & 1.2 & 1000 \\ 

J0113-3551 & 4.7(0.1), --, -- & 5(1), --, -- & 10(1), --, -- & --, --, -- & -47(1), --, -- & 1.0 & 944 \\ 

J0115-3049 & 0.8(0.1), 1.9(0.6), -- & 21(3), 2(32), -- & 6(3), 66(10), -- & --, --, -- & 54(6), -33(17), -- & 0.8 & 869 \\ 

J0115-3307 & 14.0(0.4), --, -- & -14(1), --, -- & --, --, -- & --, --, -- & -60(2), --, -- & 1.0 & 972 \\ 

J0116-3457 & 2.5(0.6), 3.0(0.6), -- & -28(7), 9(6), -- & --, --, -- & --, --, -- & -36(16), 55(13), -- & 1.0 & 948 \\ 

J0116-3626 & 26.9(2.2), --, -- & 24(2), --, -- & 10(2), --, -- & --, --, -- & -65(4), --, -- & 1.1 & 975 \\ 

J0117-3357 & 2.2(0.7), 1.2(0.1), -- & 26(25), 4(3), -- & 48(8), --, -- & --, --, -- & -67(15), -27(9), -- & 0.7 & 838 \\ 

J0134-3843 & 2.0(0.1), 2.2(1.1), -- & 3(1), 302(54), -- & 11(1), 79(14), -- & --, --, -- & -78(1), -29(35), -- & 1.2 & 1027 \\ 

J0135-3854 & 21.1(6.5), 14.5(1.0), -- & 11(28), 3(1), -- & 68(5), 17(1), -- & --, --, -- & -59(17), 46(3), -- & 1.2 & 1029 \\ 

J0140-3642 & 3.4(0.1), --, -- & 7(1), --, -- & --, --, -- & --, --, -- & -20(2), --, -- & 0.8 & 887 \\ 

J0141-3835 & 24.9(2.1), --, -- & -2(2), --, -- & 9(2), --, -- & --, --, -- & -14(4), --, -- & 1.1 & 994 \\ 

J0144-4004 & 6.5(0.2), 1.1(0.2), -- & 6(2), 72(10), -- & --, --, -- & --, --, -- & -55(4), 67(22), -- & 0.8 & 880 \\ 

J0152-3339 & 4.9(0.2), 2.4(1.3), -- & -4(1), -135(36), -- & 11(1), 49(13), -- & --, --, -- & 21(2), 19(28), -- & 0.7 & 852 \\ 

J0153-3310 & 4.6(0.02), 7.5(0.5), -- & -2(1), -17(9), -- & --, 76(2), -- & --, --, -- & 68(0), -12(5), -- & 1.7 & 1135 \\ 

J0200-3053 & 14.7(5.1), 5.9(0.5), 4.8(2.2) & 51(33), -24(3), 1(2) & 103(9), 27(13), 13(4) & --, --, -- & -31(17), 81(13), -56(40) & 0.9 & 936 \\ 

J0203-3147 & 2.2(0.8), 6.1(0.2), -- & 294(27), 7(1), -- & 41(9), 11(1), -- & --, --, -- & -89(22), 0(2), -- & 1.0 & 956 \\ 

J0207-3857 & 6.5(0.4), 6.2(1.0), -- & 6(1), -10(16), -- & 3(3), 55(7), -- & --, --, -- & -11(2), -84(9), -- & 1.0 & 966 \\ 

J0208-3158 & 5.0(0.7), --, -- & 10(1), --, -- & 13(1), --, -- & --, --, -- & -21(2), --, -- & 1.0 & 942 \\ 

J0212-3444 & 11.4(0.9), --, -- & -1(2), --, -- & 9(2), --, -- & --, --, -- & 56(4), --, -- & 0.9 & 930 \\ 

J0216-3247 & 2.1(0.3), 2.9(0.2), -- & -67(7), 31(2), -- & --, --, -- & 96(7), 35(5), -- & -44(6), -1(3), -- & 0.6 & 824 \\ 

J0217-3113 & 9.0(0.6), --, -- & 0(2), --, -- & 17(1), --, -- & --, --, -- & -89(3), --, -- & 1.0 & 953 \\ 

J0219-3625 & 4.4(0.2), --, -- & 1(1), --, -- & 18(1), --, -- & --, --, -- & 56(1), --, -- & 1.0 & 933 \\ 

J0220-3144 & 6.5(0.3), --, -- & 0(1), --, -- & 8(1), --, -- & --, --, -- & 47(2), --, -- & 1.0 & 944 \\ 

J0222-3441 & 2.1(0.4), 1.6(0.3), -- & 10(1), 13(3), -- & --, --, -- & 100(6), 30(10), -- & 5(2), -20(8), -- & 1.3 & 1050 \\ 

J0224-3456 & 11.1(0.7), --, -- & 0(3), --, -- & --, --, -- & --, --, -- & 55(4), --, -- & 1.3 & 1046 \\ 

J0229-3643 & 2.3(0.1), --, -- & -4(1), --, -- & --, --, -- & 39(2), --, -- & -66(2), --, -- & 1.0 & 949 \\ 

J0231-3935 & 6.8(0.1), --, -- & 8(1), --, -- & 6(1), --, -- & --, --, -- & -75(1), --, -- & 0.9 & 898 \\ 

J0233-3559 & 2.8(0.1), --, -- & 20(2), --, -- & --, --, -- & --, --, -- & 1(3), --, -- & 1.0 & 943 \\ 

J0239-3331 & 22.3(0.7), --, -- & -4(1), --, -- & 13(1), --, -- & --, --, -- & 85(1), --, -- & 1.0 & 916 \\ 

J0240-3239 & 9.1(0.3), --, -- & 6(1), --, -- & 8(1), --, -- & --, --, -- & -62(2), --, -- & 1.0 & 962 \\ 

J0256-3328 & 27.5(2.8), 48.6(22.5), -- & 8(3), -37(53), -- & 14(2), 73(16), -- & --, --, -- & 3(6), 82(31), -- & 1.0 & 950 \\ 

J0258-3146 & 8.9(0.5), 3.8(1.0), -- & 2(1), -31(20), -- & 7(1), 41(10), -- & --, --, -- & -50(3), -44(16), -- & 1.0 & 949 \\ 

J0259-3940 & 1.1(0.5), 3.6(0.6), -- & -1(7), 17(7), -- & --, 30(4), -- & --, --, -- & -50(18), 77(5), -- & 0.8 & 861 \\ 

J0300-3414 & 0.8(0.2), 3.6(0.3), 0.3(0.2) & -81(11), 32(2), 115(18) & 23(5), 11(1), -- & --, --, -- & -55(14), -67(3), -2(31) & 0.8 & 934 \\ 

J0307-3037 & 4.8(0.1), --, -- & -1(1), --, -- & 10(1), --, -- & --, --, -- & -69(1), --, -- & 1.0 & 935 \\ 

J0314-3407 & 10.2(0.7), 0.8(0.4), -- & 34(1), -26(18), -- & 16(2), --, -- & --, --, -- & -26(2), -63(34), -- & 0.7 & 827 \\ 

J0326-3243 & 5.5(0.2), 0.7(0.1), -- & 19(2), -100(10), -- & --, --, -- & 35(4), --, -- & 37(2), 14(16), -- & 0.9 & 934 \\ 

J0331-3032 & 27.3(1.2), --, -- & 22(2), --, -- & 16(1), --, -- & --, --, -- & -50(2), --, -- & 1.0 & 947 \\ 

J0336-3616 & 1.9(0.2), 1.6(0.1), -- & 7(4), 16(3), -- & --, --, -- & 140(7), 39(4), -- & 42(5), 50(5), -- & 0.6 & 773 \\ 

J0338-3522 & 12.3(0.2), 0.9(0.1), -- & 26(1), 159(6), -- & 26(0), 20(3), -- & --, --, -- & -57(1), 71(7), -- & 0.7 & 821 \\ 

J0342-3147 & 8.2(0.2), 1.3(0.2), -- & 47(1), -44(2), -- & 15(1), 8(1), -- & --, --, -- & 57(1), 17(4), -- & 0.8 & 903 \\ 

J0342-3703 & 5.5(0.1), 3.9(0.3), -- & -3(1), 4(1), -- & --, --, -- & --, 78(2), -- & 47(2), 13(1), -- & 0.5 & 764 \\ 

J0344-3950 & 9.3(2.9), 3.0(1.2), -- & -18(6), 2(16), -- & 19(9), --, -- & --, --, -- & -39(17), 34(69), -- & 0.9 & 927 \\ 

J0350-3856 & 21.0(1.3), --, -- & -4(2), --, -- & 10(1), --, -- & --, --, -- & 12(3), --, -- & 1.2 & 993 \\ 

J1008-3011 & 4.2(0.1), 3.8(0.4), -- & -74(3), -59(15), -- & --, 40(6), -- & --, --, -- & -28(7), 56(7), -- & 1.2 & 1018 \\ 

J1024-3234 & 0.4(0.1), 3.3(0.6), 2.1(0.6) & 120(6), -93(3), -117(5) & --, --, -- & --, --, -- & -86(9), 87(8), -29(12) & 1.0 & 960 \\ 

J1051-3138 & 2.4(0.1), 0.5(0.1), -- & 59(1), -61(5), -- & 8(1), 20(3), -- & --, --, -- & 77(1), 17(7), -- & 1.1 & 986 \\ 

J1103-3251 & 3.3(0.5), 2.4(0.8), -- & 38(2), -3(19), -- & 6(3), 35(9), -- & --, --, -- & 72(6), -24(16), -- & 1.3 & 1050 \\ 

J1107-3043 & 4.6(0.1), 0.8(0.1), -- & -9(1), 73(6), -- & --, --, -- & --, --, -- & -62(2), -33(11), -- & 1.1 & 1003 \\ 

J1109-3732 & 28.1(0.3), 5.0(0.2), 2.2(0.2) & -41(2), 35(3), -134(4) & --, --, -- & --, --, -- & 79(1), -6(5), 21(7) & 0.6 & 813 \\ 

J1115-3051 & 2.0(0.4), 12.0(0.9), -- & 61(10), -20(2), -- & --, 8(2), -- & --, --, -- & -18(17), -33(3), -- & 1.2 & 1032 \\ 

J1117-3033 & 6.9(0.3), --, -- & 7(2), --, -- & --, --, -- & --, --, -- & 84(3), --, -- & 1.1 & 995 \\ 

J1122-3605 & 27.3(1.3), 2.6(1.4), 2.4(1.1) & -60(1), 22(13), 128(13) & 6(2), --, 11(10) & --, --, -- & -17(2), 3(27), -41(20) & 0.8 & 917 \\ 

J1125-3523 & 14.0(3.1), 4.2(0.2), -- & -137(12), -38(2), -- & 50(6), --, -- & --, --, -- & -90(9), 7(5), -- & 0.5 & 774 \\ 

J1141-3504 & 2.2(0.2), 1.1(0.3), -- & -25(2), 60(9), -- & 8(2), 21(5), -- & --, --, -- & -29(4), -36(10), -- & 0.9 & 937 \\ 

J1147-3812 & 2.0(0.1), 1.5(0.2), -- & 28(1), -67(10), -- & 9(1), 39(4), -- & --, --, -- & 24(2), -49(8), -- & 1.5 & 1073 \\ 

J1153-3009 & 6.7(0.1), 0.7(0.1), -- & -55(1), 61(6), -- & --, --, -- & --, --, -- & 59(1), -8(10), -- & 1.2 & 998 \\ 

J1154-3721 & 25.5(1.5), --, -- & -10(2), --, -- & 16(1), --, -- & --, --, -- & -74(3), --, -- & 1.0 & 959 \\

J1217-3313 & 21.3(0.9), --, -- & -88(1), --, -- & 9(1), --, -- & --, --, -- & -53(2), --, -- & 1.1 & 992 \\ 

J1228-3003 & 12.2(5.5), 1.2(0.8), 15.8(7.5) & -36(8), 88(19), -41(13) & --, 8(16), 29(9) & --, --, -- & 33(25), 78(31), 86(9) & 0.8 & 907 \\ 

J1230-3121 & 7.0(0.2), 0.6(0.1), -- & -68(1), -196(6), -- & 7(1), --, -- & --, --, -- & -83(1), 76(11), -- & 0.9 & 915 \\ 

J1233-3534 & 4.6(1.9), 10.3(0.3), -- & 199(35), -114(1), -- & 53(10), 13(1), -- & --, --, -- & -37(27), 55(1), -- & 1.2 & 1015 \\ 
      \hline
   \end{tabular}
\label{modelfits}
\end{table*}

\begin{table*}
 \contcaption{ }
 \scriptsize
 \centering
   \begin{tabular}{cccccccc}
    \hline\hline
       (1) & (2) & (3) & (4) & (5) & (6) & (7) & (8)  \\
      Name  & $p_{0(1,2,3)}$ & RM$_{(1,2,3)}$ & $\sigmaRM$$_{(1,2,3)}$ & $\DeltaRM$$_{(1,2,3)}$ & $\psi_{0(1,2,3)}$ & $\chi^2_{r}$ & BIC \\
                 &  (\%)               & (rad~m$^{-2}$) & (rad~m$^{-2}$)                 & (rad~m$^{-2}$)                &    (deg)               &                          &        \\
      \hline 
J1244-4012 & 3.1(0.1), 0.6(0.1), -- & -79(1), -224(9), -- & 3(2), 14(6), -- & --, --, -- & 9(2), 49(12), -- & 1.0 & 970 \\ 

J1246-3406 & 8.7(0.5), --, -- & -53(3), --, -- & --, --, -- & --, --, -- & 10(6), --, -- & 0.8 & 874 \\ 

J1257-3334 & 3.6(0.3), 2.2(0.3), -- & -33(3), -76(4), -- & --, --, -- & --, --, -- & 2(4), 52(6), -- & 1.1 & 983 \\ 

J1300-3253 & 1.1(0.1), 4.0(0.1), -- & 6(5), -58(1), -- & --, --, -- & --, --, -- & -28(7), 67(2), -- & 1.0 & 972 \\ 

J1301-3226 & 7.2(0.7), 0.8(0.2), 0.5(0.1) & -46(2), 7(10), 124(9) & --, --, -- & 27(9), --, 47(18) & -46(2), 32(20), -14(12) & 0.5 & 741 \\ 

J1304-3406 & 19.0(0.5), --, -- & -28(1), --, -- & 6(1), --, -- & --, --, -- & 51(1), --, -- & 1.0 & 954 \\ 

J1307-3207 & 3.5(0.2), 1.1(0.2), 0.4(0.1) & -16(1), -56(5), -163(8) & --, --, 12(5) & --, --, -- & 87(2), -43(7), -31(12) & 1.0 & 988 \\ 

J1307-3737 & 19.3(1.2), --, -- & -40(2), --, -- & 11(1), --, -- & --, --, -- & -79(3), --, -- & 1.0 & 949 \\ 

J1316-3338 & 4.6(0.4), 1.6(0.6), -- & -37(2), -59(3), -- & 14(2), --, -- & --, --, -- & -37(4), 47(8), -- & 1.4 & 1077 \\ 

J1324-3623 & 3.4(0.1), --, -- & -15(2), --, -- & --, --, -- & --, --, -- & -23(2), --, -- & 1.1 & 947 \\ 

J1342-3240 & 4.4(1.0), 3.1(1.2), -- & -32(5), -66(12), -- & --, 15(5), -- & --, --, -- & -61(12), 13(19), -- & 1.3 & 1006 \\ 

J1342-3601 & 21.8(0.6), --, -- & -40(1), --, -- & 11(1), --, -- & --, --, -- & 16(1), --, -- & 1.1 & 971 \\ 

J1345-3140 & 14.3(0.5), 3.1(0.5), 0.5(0.1) & -36(1), -6(4), -197(8) & --, --, -- & --, --, -- & 77(2), 83(10), -14(13) & 0.6 & 803 \\ 

J1403-3358 & 8.3(0.4), 9.4(0.8), -- & 11(1), -5(11), -- & 11(1), 53(4), -- & --, --, -- & 21(2), -71(6), -- & 1.2 & 1020 \\ 

J1411-3205 & 12.9(0.8), 4.7(1.8), -- & 6(2), 25(11), -- & 19(7), --, -- & --, --, -- & 59(16), -34(47), -- & 1.2 & 1028 \\ 

J1413-3002 & 8.8(4.8), 2.9(1.4), 2.1(0.4) & 161(40), -88(23), -12(4) & 71(12), 36(8), 7(4) & --, --, -- & -63(29), -30(24), 5(8) & 0.8 & 894 \\ 

J1433-3141 & 31.1(0.9), --, -- & -1(1), --, -- & 11(1), --, -- & --, --, -- & -81(1), --, -- & 1.0 & 939 \\ 

J1438-3122 & 2.6(0.3), 5.1(0.5), -- & 1(2), 2(12), -- & 6(3), 46(5), -- & --, --, -- & 19(6), 84(6), -- & 1.0 & 955 \\ 

J1457-3539 & 6.8(4.6), 3.0(0.1), -- & 31(61), 55(1), -- & 102(15), 4(1), -- & --, --, -- & -82(36), -3(1), -- & 1.0 & 968 \\ 

J1511-3242 & 5.1(0.5), 1.8(0.7), -- & 39(2), -8(14), -- & 9(1), 32(10), -- & --, --, -- & -12(3), -84(13), -- & 1.0 & 978 \\ 

J1522-2730 & 2.2(0.1), 1.5(0.1), -- & 61(1), 11(1), -- & --, --, -- & --, --, -- & -42(1), 77(2), -- & 1.4 & 1061 \\ 

J1524-3012 & 2.4(0.7), 1.3(0.9), -- & 43(4), 32(9), -- & --, --, -- & --, --, -- & 85(17), -28(28), -- & 1.0 & 962 \\ 

J1527-3014 & 2.6(0.1), 0.5(0.1), -- & 19(1), -242(8), -- & --, --, -- & --, --, -- & 22(3), 40(15), -- & 0.9 & 925 \\
      \hline
   \end{tabular}
\end{table*}

\begin{figure} 
\centering
    \includegraphics[angle=0, clip=true, trim=0cm 0cm 0cm 0cm, width=.45\textwidth]{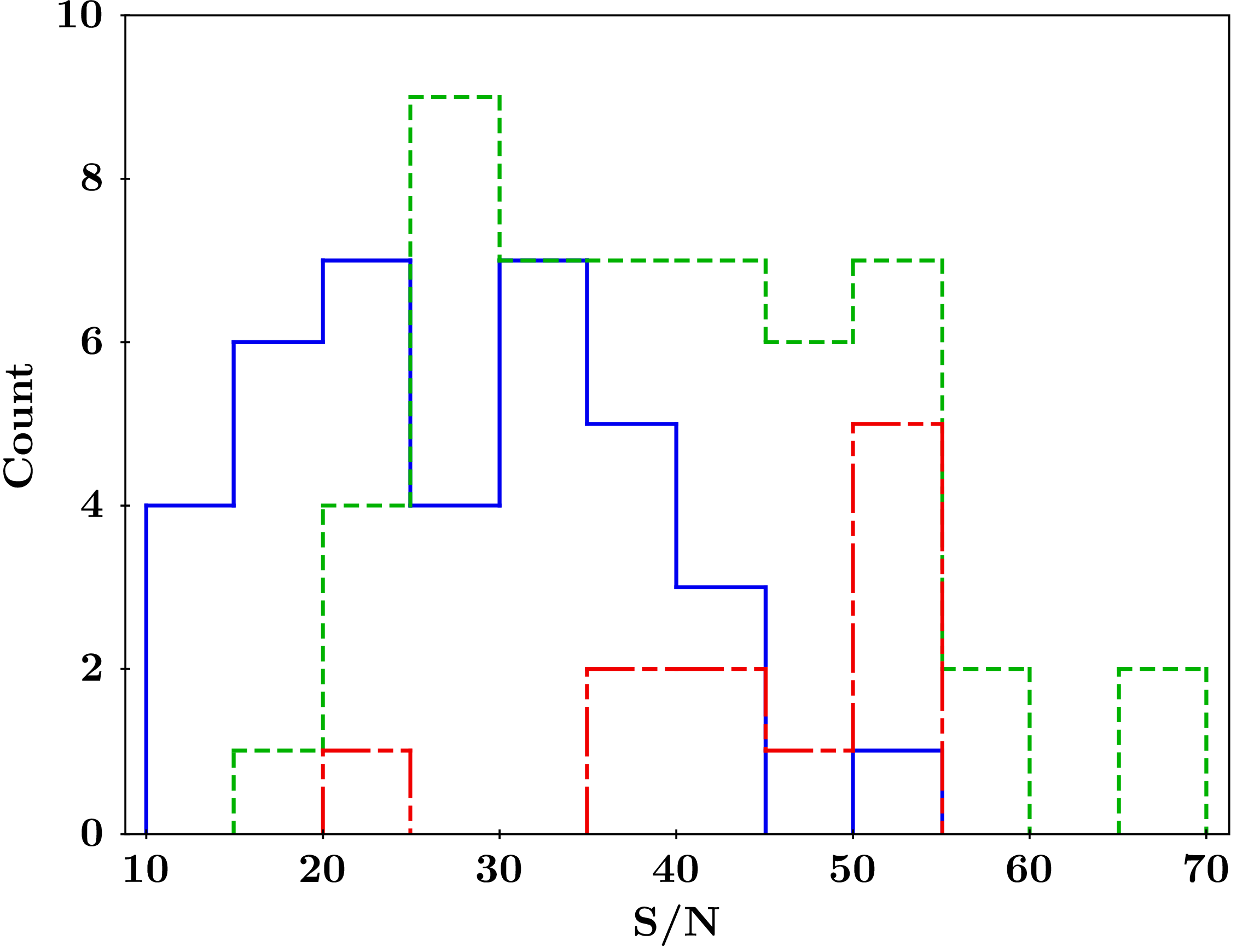} 
    \caption{ {\small Histogram of the signal-to-noise ratio in polarization (S/N) for all sources, split by the number of RM components. 
    Solid blue: one RM component, dashed green: two RM components, dash-dot red: three RM components). } }
    \label{S/N}
\end{figure}    

\begin{figure} 
\centering
    \includegraphics[angle=0, clip=true, trim=0cm 0.2cm 0cm 1cm, width=.5\textwidth]{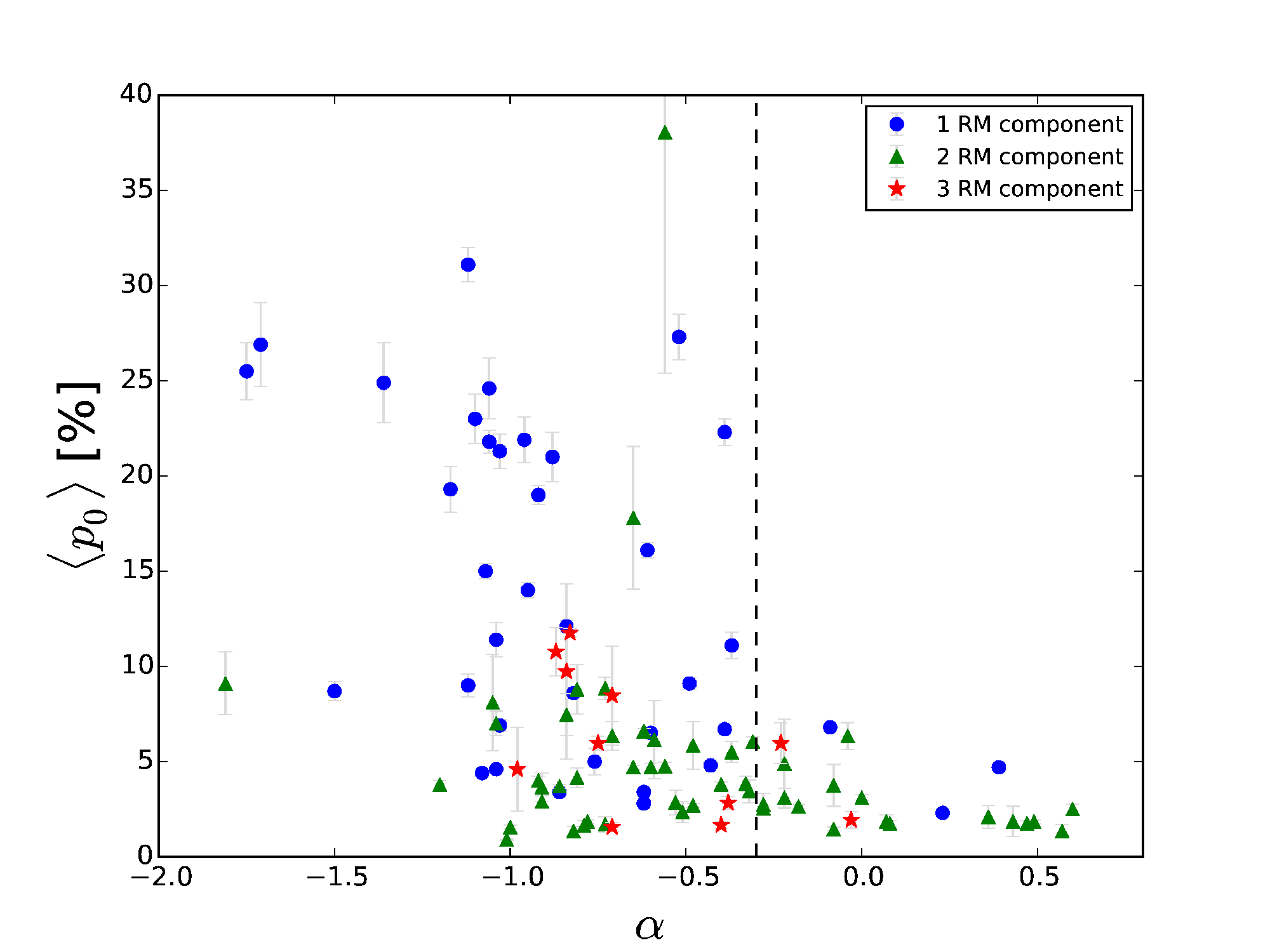}
    \caption{ {\small Plot of the mean intrinsic degree of polarization, $\langle p_0 \rangle$, versus the total intensity spectral index ($\alpha$) for each source. 
    Blue circles: one RM component, green triangles: two RM components, red stars: three RM components. } }
    \label{spixp0}
\end{figure}

\section{Results}
We first present a general overview of the results for our full sample of 100 sources, before looking at particular source properties in more detail and comparing the polarization and Faraday rotation properties of the radiative-mode and jet-mode classes. 

Table~\ref{modelfits} presents the best-fit parameters for all sources ($p_{0,j}$, RM$_j$, $\sigma_{{\rm RM},j}$, $\DeltaRM_j$, $\psi_{0,j}$), including the $\chi^2_{r}$ and BIC best-fit statistics. Figure~\ref{quplots1} shows the best-fit models overlaid on the $q(\lambda^2)$ and $u(\lambda^2)$ data for each source. 
All best-fitting models provide excellent descriptions of the data with $\chi^2_{r}$ ranging from 0.5 to 1.7, with a median value of 1.0.  
A summary of the key properties of the sources in our sample is shown in Table~\ref{modelfitsextra}. 
Out of 100 sources, 37\% are best fit by one RM component, 52\% with two RM components and 11\% with three RM components. 
Within the limit of our measurements, 24\% of sources are classified as `Faraday thin', meaning that they have no measurable depolarization (i.e.~$\sigmaRM$ and $\gradRM$ are zero for all RM components). Only 9\% of all sources require one or more $\gradRM$ parameters, while the majority (67\%) of fits require one or more RM components with non-zero values of $\sigmaRM$. No source had a best-fit model that required a combination of both $\sigmaRM$ and $\gradRM$ parameters. 

Figure~\ref{S/N} shows the distribution of signal-to-noise ratio (S/N) in polarization for all sources. The S/N was estimated from the peak of the RM synthesis spectrum after correction for polarization bias (George et al.~2012), with the rms noise level calculated from the Faraday dispersion function of $q$ and $u$ in regions far from the peak. 
By splitting the sources into histograms for one, two and three RM component sources, we can see the effect of S/N on the number of fitted RM components. The one RM component sources dominate at low S/N ($< 25$), while the two and three RM component sources dominate at higher S/N.  The median S/N for all sources is 35.5. The median S/N ratios for 1, 2 and 3 RM components are 27.4, 39.7 and 49.7, respectively. 

To compare the intrinsic degree of polarization between sources, we also calculate the mean intrinsic degree of polarization, $\langle p_0 \rangle$, for each source (Table~\ref{modelfitsextra}). 
In Figure~\ref{spixp0}, we show the value of $\langle p_0 \rangle$ for each source versus the spectral index ($\alpha$). This shows the clear difference in $p_0$ of flat-spectrum versus steep-spectrum sources, and that no sources dominated by synchrotron self-absorption ($\alpha\gtrsim 0.0$) violate the theoretical limit of $p_0\sim10\%$. It also shows that $\alpha=-0.3$ (dashed vertical line) is a reasonable division for polarized emission coming from optically thin emission versus synchrotron self-absorbed regions. We adopt $\alpha=-0.3$ as the division between steep-spectrum and flat-spectrum sources throughout. We did not find that varying this division value, for example to $\alpha=-0.5$, significantly affected our results or conclusions. 
The median value of $\langle p_0 \rangle$ for all sources is 5\% with a median error from the model-fits of 0.7\%. The median value of $\langle p_0 \rangle$ for flat and steep-spectrum sources is 6.6\% and 2.6\%, respectively. 

For our full sample, the median values of RRM, $\sigmaRMwtd$, and $\gradRMwtd$ are 9.7~rad~m$^{-2}$, 14.1~rad~m$^{-2}$, and 60.6~rad~m$^{-2}$, with median errors of 7.5~rad~m$^{-2}$, 1.8~rad~m$^{-2}$, and 3.3~rad~m$^{-2}$, respectively  (Table~\ref{modelfitsextra}). One can immediately see that the typical RRM error is similar to the value of the RRM, making any potential physical relationships with RRM very difficult to detect. For the $\sigmaRMwtd$ and $\gradRMwtd$ medians, we have excluded the Faraday thin sources.

Sources that are spatially resolved (75\%) have a median angular size of 28'' and a median spectral index of $\alpha_{med} = -0.82$, ranging from $-1.81$ to $-0.18$. The median linear size of resolved sources is 102.2~kpc. This is consistent with the expectation that the majority of polarized sources selected at 1.4~GHz are steep-spectrum sources and have linear sizes $\gg10$~kpc. Of the 25\% of sources that are still unresolved at 9 GHz (at $\sim3"$ angular resolution), 80\% have $\alpha>-0.3$ (i.e.~most unresolved sources in our sample are flat-spectrum blazars).

\begin{table*}
 \caption{Basic properties of all sources in the sample.}
 \scriptsize
 \centering
   \begin{tabular}{cccccccccccccccc}
    \hline\hline
       (1) & (2) & (3) & (4) & (5) & (6) & (7) & (8) & (9) & (10) & (11) & (12) & (13) & (14) & (15) & (16)  \\
Name  & RA & DEC & $\alpha$ &   $I_{2.1\,{\rm GHz}}$ &  $z$ &  $|$RRM$|$ &  $\sigmaRMwtd$ &  $\gradRMwtd$ &  $\langle p_{0} \rangle$ &  S/N &  \# &  $\beta$ &  Morph &  $l$ &  Rad/Jet \\
      \hline 
J0001-3025 & 0.472 & -30.419 & -0.23 & 105 & 1.303 & 8.0 & 15.1 & 0.0 & 6.0 & 42 & 3 & -0.27 & ext & 25 & 0 \\ 

J0004-4011 & 1.191 & -40.196 & -0.96 & 25 & 0.098 & 3.0 & 12.1 & 0.0 & 21.9 & 23 & 1 & -0.92 & cmplx & 19 & 1 \\ 

J0009-3216 & 2.399 & -32.277 & -0.18 & 286 & 0.025 & 8.4 & 0.0 & 15.8 & 2.7 & 50 & 2 & -0.09 & ext & 399 & 1 \\ 

J0010-3054 & 2.646 & -30.904 & -0.48 & 141 & 0.999 & 3.2 & 21.8 & 0.0 & 5.8 & 35 & 2 & 0.06 & ext & 181 & 0 \\ 

J0014-3059 & 3.658 & -30.989 & -0.39 & 150 & 2.785 & 0.1 & 13.0 & 0.0 & 6.7 & 43 & 1 & -0.88 & unres & $<24$ & 0 \\ 

J0015-3526 & 3.808 & -35.447 & -0.82 & 102 & 1.296 & 38.8 & 6.2 & 0.0 & 1.3 & 16 & 2 & 0.07 & dbl & 126 & 0 \\ 

J0020-3106 & 5.077 & -31.115 & -1.10 & 28 & 0.242 & 8.6 & 10.8 & 0.0 & 23.0 & 21 & 1 & -0.43 & dblc & 153 & 1 \\ 

J0021-3334 & 5.400 & -33.576 & -0.65 & 233 & 0.129 & 22.5 & 0.0 & 29.9 & 4.7 & 50 & 2 & -0.38 & ext & 15 & 0 \\ 

J0038-3142 & 9.733 & -31.711 & -1.06 & 15 & 0.234 & 2.4 & 11.1 & 0.0 & 24.6 & 18 & 1 & -0.80 & dblc & 58 & 1 \\ 

J0038-3859 & 9.612 & -38.996 & -1.01 & 752 & 0.593 & 1.6 & 0.0 & 0.0 & 0.9 & 27 & 2 & 0.95 & dbl & 53 & 0 \\ 

J0040-3254 & 10.073 & -32.911 & -1.07 & 74 & 2.100 & 3.2 & 12.4 & 0.0 & 15.0 & 36 & 1 & -0.80 & dblc & 339 & 0 \\ 

J0044-3147 & 11.021 & -31.785 & -1.05 & 24 & 2.788 & 13.2 & 0.0 & 0.0 & 8.1 & 33 & 2 & 0.35 & dbl & 41 & 0 \\ 

J0049-3140 & 12.302 & -31.677 & -0.84 & 27 & 2.640 & 0.0 & 0.0 & 0.0 & 12.1 & 28 & 1 & -0.10 & ext & 163 & 0 \\ 

J0059-3152 & 14.966 & -31.872 & -0.82 & 64 & 1.231 & 4.8 & 0.0 & 0.0 & 8.6 & 32 & 1 & 0.01 & ext & 140 & 0 \\ 

J0104-3314 & 16.165 & -33.248 & -1.04 & 224 & 0.380 & 5.9 & 22.0 & 0.0 & 4.6 & 31 & 1 & -2.26 & dbl & 29 & 0 \\ 

J0108-3611 & 17.021 & -36.199 & -0.61 & 21 & 0.122 & 5.8 & 0.0 & 0.0 & 16.1 & 36 & 1 & -0.22 & dbl & 14 & 1 \\ 

J0113-3551 & 18.316 & -35.864 & 0.39 & 112 & 1.220 & 2.5 & 10.2 & 0.0 & 4.7 & 41 & 1 & -0.58 & unres & $<25$ & 0 \\ 

J0115-3049 & 18.944 & -30.822 & 0.57 & 254 & 1.414 & 2.6 & 47.9 & 0.0 & 1.4 & 24 & 2 & 0.83 & unres & $<26$ & 0 \\ 

J0115-3307 & 18.948 & -33.123 & -0.95 & 27 & 1.910 & 15.5 & 0.0 & 0.0 & 14.0 & 28 & 1 & -0.28 & dbl & 224 & 0 \\ 

J0116-3457 & 19.153 & -34.950 & -0.28 & 44 & 1.960 & 7.7 & 0.0 & 0.0 & 2.8 & 26 & 2 & 0.96 & unres & $<25$ & 0 \\ 

J0116-3626 & 19.101 & -36.445 & -1.71 & 10 & 0.173 & 20.3 & 9.6 & 0.0 & 26.9 & 16 & 1 & -1.48 & cmplx & 102 & 1 \\ 

J0117-3357 & 19.436 & -33.952 & -0.73 & 123 & 0.647 & 16.7 & 30.9 & 0.0 & 1.7 & 28 & 2 & -1.07 & ext & 70 & 0 \\ 

J0134-3843 & 23.633 & -38.726 & 0.36 & 400 & 2.140 & 154.6 & 46.9 & 0.0 & 2.1 & 48 & 2 & -0.60 & unres & 25 & 0 \\ 

J0135-3854 & 23.899 & -38.907 & -0.65 & 58 & 2.080 & 3.5 & 46.8 & 0.0 & 17.8 & 30 & 2 & -0.61 & ext & 76 & 0 \\ 

J0140-3642 & 25.180 & -36.709 & -0.86 & 65 & 0.964 & 1.9 & 0.0 & 0.0 & 3.4 & 23 & 1 & -0.09 & ext & 226 & 0 \\ 

J0141-3835 & 25.432 & -38.586 & -1.36 & 21 & 0.178 & 6.8 & 8.8 & 0.0 & 24.9 & 13 & 1 & -0.69 & dblc & 218 & 1 \\ 

J0144-4004 & 26.205 & -40.077 & -1.20 & 79 & 0.782 & 12.6 & 0.0 & 0.0 & 3.8 & 27 & 2 & -0.37 & dblc & 335 & 0 \\ 

J0152-3339 & 28.135 & -33.664 & -0.91 & 89 & 0.618 & 53.7 & 23.2 & 0.0 & 3.7 & 23 & 2 & -0.26 & dbl & 120 & 1 \\ 

J0153-3310 & 28.292 & -33.174 & -0.31 & 653 & 0.612 & 17.5 & 47.1 & 0.0 & 6.1 & 66 & 2 & 0.68 & unres & $<20$ & 0 \\ 

J0200-3053 & 30.051 & -30.891 & -0.71 & 1725 & 0.677 & 20.3 & 68.1 & 0.0 & 8.5 & 52 & 3 & -0.90 & dblc & 157 & 0 \\ 

J0203-3147 & 30.802 & -31.797 & -0.81 & 55 & 1.442 & 77.8 & 18.7 & 0.0 & 4.2 & 33 & 2 & -0.36 & ext & 124 & 0 \\ 

J0207-3857 & 31.815 & -38.951 & -0.04 & 168 & 0.254 & 9.1 & 28.0 & 0.0 & 6.3 & 60 & 2 & 0.72 & unres & $<12$ & 1 \\ 

J0208-3158 & 32.062 & -31.968 & -0.76 & 49 & 0.124 & 0.7 & 13.3 & 0.0 & 5.0 & 31 & 1 & -1.08 & ext & 403 & 1 \\ 

J0212-3444 & 33.191 & -34.748 & -1.04 & 40 & 0.297 & 9.5 & 8.8 & 0.0 & 11.4 & 13 & 1 & -0.78 & dbl & 270 & 1 \\ 

J0216-3247 & 34.201 & -32.795 & 0.60 & 130 & 1.331 & 19.2 & 0.0 & 60.6 & 2.5 & 47 & 2 & 0.16 & unres & $<25$ & 0 \\ 

J0217-3113 & 34.396 & -31.217 & -1.12 & 32 & 1.140 & 7.3 & 16.9 & 0.0 & 9.0 & 25 & 1 & -1.15 & dblc & 207 & 0 \\ 

J0219-3625 & 34.760 & -36.428 & -1.08 & 550 & 0.489 & 3.8 & 17.9 & 0.0 & 4.4 & 30 & 1 & -1.73 & dblc & 465 & 0 \\ 

J0220-3144 & 35.142 & -31.738 & -0.60 & 90 & 1.871 & 6.7 & 8.1 & 0.0 & 6.5 & 35 & 1 & -0.42 & dbl & 68 & 0 \\ 

J0222-3441 & 35.735 & -34.691 & 0.07 & 1087 & 1.490 & 4.2 & 0.0 & 69.7 & 1.8 & 42 & 2 & -1.81 & unres & $<25$ & 0 \\ 

J0224-3456 & 36.125 & -34.943 & -0.37 & 39 & 0.150 & 6.4 & 0.0 & 0.0 & 11.1 & 10 & 1 & -1.20 & ext & 340 & 1 \\ 

J0229-3643 & 37.368 & -36.733 & 0.23 & 270 & 2.115 & 9.7 & 0.0 & 39.4 & 2.3 & 32 & 1 & -0.73 & unres & $<25$ & 0 \\ 

J0231-3935 & 37.966 & -39.596 & -0.09 & 378 & 1.646 & 3.1 & 6.0 & 0.0 & 6.8 & 51 & 1 & -0.29 & unres & $<25$ & 0 \\ 

J0233-3559 & 38.324 & -35.988 & -0.62 & 207 & 0.252 & 13.9 & 0.0 & 0.0 & 2.8 & 19 & 1 & -0.80 & cmplx & 281 & 1 \\ 

J0239-3331 & 39.911 & -33.531 & -0.39 & 47 & 0.203 & 10.6 & 13.1 & 0.0 & 22.3 & 25 & 1 & -0.90 & dblc & 96 & 1 \\ 

J0240-3239 & 40.163 & -32.666 & -0.49 & 58 & 0.288 & 1.7 & 8.4 & 0.0 & 9.1 & 38 & 1 & -0.64 & dbl & 201 & 0 \\ 

J0256-3328 & 44.198 & -33.474 & -0.56 & 53 & 0.108 & 29.7 & 52.0 & 0.0 & 38.0 & 21 & 2 & -1.20 & dblc & 109 & 1 \\ 

J0258-3146 & 44.525 & -31.774 & -0.71 & 180 & 1.830 & 16.0 & 16.8 & 0.0 & 6.3 & 45 & 2 & -0.59 & unres & $<25$ & 0 \\ 

J0259-3940 & 44.861 & -39.677 & -0.51 & 596 & 0.066 & 10.0 & 23.1 & 0.0 & 2.3 & 38 & 2 & -1.36 & cmplx & 34 & 1 \\ 

J0300-3414 & 45.151 & -34.235 & -0.71 & 397 & 1.704 & 3.0 & 11.9 & 0.0 & 1.6 & 38 & 3 & -0.42 & dblc & 76 & 0 \\ 

J0307-3037 & 46.785 & -30.625 & -0.43 & 259 & 1.106 & 12.3 & 9.9 & 0.0 & 4.8 & 45 & 1 & -0.65 & ext & 149 & 0 \\ 

J0314-3407 & 48.637 & -34.128 & -0.37 & 246 & 0.067 & 10.6 & 15.2 & 0.0 & 5.5 & 37 & 2 & -1.08 & ext & 61 & 1 \\ 

J0326-3243 & 51.563 & -32.723 & 0.00 & 73 & 0.699 & 18.8 & 0.0 & 31.0 & 3.1 & 29 & 2 & -0.46 & unres & $<21$ & 0 \\ 

J0331-3032 & 52.924 & -30.548 & -0.52 & 31 & 0.128 & 8.4 & 15.6 & 0.0 & 27.3 & 25 & 1 & -1.28 & dblc & 50 & 1 \\ 

J0336-3616 & 54.225 & -36.268 & 0.08 & 617 & 1.541 & 0.8 & 0.0 & 93.6 & 1.8 & 30 & 2 & -1.70 & unres & $<25$ & 0 \\ 

J0338-3522 & 54.701 & -35.372 & -0.62 & 497 & 0.113 & 15.2 & 25.1 & 0.0 & 6.6 & 41 & 2 & -2.11 & dbl & 238 & 0 \\ 

J0342-3147 & 55.742 & -31.785 & -0.56 & 224 & 1.824 & 5.8 & 14.1 & 0.0 & 4.8 & 40 & 2 & -0.97 & dblc & 357 & 0 \\ 

J0342-3703 & 55.523 & -37.056 & -0.60 & 1306 & 0.284 & 8.0 & 0.0 & 32.2 & 4.7 & 53 & 2 & -0.56 & dbl & 26 & 0 \\ 

J0344-3950 & 56.071 & -39.843 & -0.59 & 284 & 0.178 & 11.5 & 14.3 & 0.0 & 6.2 & 28 & 2 & -1.23 & dblc & 142 & 0 \\ 

J0350-3856 & 57.538 & -38.937 & -0.88 & 30 & 0.147 & 5.5 & 9.8 & 0.0 & 21.0 & 21 & 1 & -1.38 & dblc & 149 & 1 \\ 

J1008-3011 & 152.230 & -30.187 & -0.92 & 261 & 1.067 & 10.7 & 19.0 & 0.0 & 4.0 & 38 & 2 & 1.22 & dblc & 366 & 0 \\ 

J1024-3234 & 156.002 & -32.571 & -0.03 & 363 & 1.568 & 6.6 & 0.0 & 0.0 & 1.9 & 51 & 3 & 0.51 & unres & $<25$ & 0 \\ 

J1051-3138 & 162.770 & -31.637 & -0.08 & 813 & 1.429 & 19.2 & 9.8 & 0.0 & 1.5 & 54 & 2 & -0.41 & unres & $<25$ & 0 \\ 

J1103-3251 & 165.881 & -32.855 & -0.53 & 551 & 0.356 & 4.9 & 17.8 & 0.0 & 2.8 & 49 & 2 & -0.06 & dblc & 399 & 0 \\ 

J1107-3043 & 166.934 & -30.727 & -0.48 & 244 & 0.740 & 8.2 & 0.0 & 0.0 & 2.7 & 50 & 2 & 0.26 & ext & 18 & 0 \\ 

J1109-3732 & 167.490 & -37.539 & -0.83 & 252 & 0.010 & 4.4 & 0.0 & 0.0 & 11.8 & 54 & 3 & -0.19 & dblc & 21 & 1 \\ 

J1115-3051 & 168.990 & -30.860 & -1.04 & 74 & 0.092 & 12.7 & 6.9 & 0.0 & 7.0 & 27 & 2 & -0.08 & dblc & 79 & 1 \\ 

J1117-3033 & 169.306 & -30.562 & -1.03 & 47 & 0.087 & 22.4 & 0.0 & 0.0 & 6.9 & 16 & 1 & -0.79 & cmplx & 50 & 1 \\ 

J1122-3605 & 170.617 & -36.087 & -0.87 & 82 & 0.091 & 15.0 & 6.1 & 0.0 & 10.8 & 39 & 3 & -0.01 & dblc & 529 & 1 \\ 

J1125-3523 & 171.482 & -35.389 & -1.81 & 302 & 0.033 & 46.4 & 38.6 & 0.0 & 9.1 & 28 & 2 & -1.49 & cmplx & 91 & 1 \\ 

J1141-3504 & 175.460 & -35.069 & -0.79 & 229 & 0.184 & 19.9 & 12.5 & 0.0 & 1.7 & 41 & 2 & -0.25 & dbl & 50 & 1 \\ 

J1147-3812 & 176.756 & -38.203 & 0.47 & 1354 & 1.048 & 19.5 & 21.6 & 0.0 & 1.8 & 45 & 2 & -0.71 & unres & $<24$ & 0 \\ 

J1153-3009 & 178.455 & -30.164 & -0.86 & 352 & 1.376 & 12.5 & 0.0 & 0.0 & 3.7 & 39 & 2 & -0.16 & dbl & 35 & 0 \\ 

J1154-3721 & 178.586 & -37.354 & -1.75 & 14 & 0.101 & 6.8 & 16.2 & 0.0 & 25.5 & 22 & 1 & -1.55 & ext & 84 & 1 \\

J1217-3313 & 184.311 & -33.232 & -1.03 & 45 & 0.056 & 20.9 & 9.1 & 0.0 & 21.3 & 35 & 1 & -0.41 & dblc & 90 & 1 \\ 

J1228-3003 & 187.179 & -30.051 & -0.84 & 268 & 0.749 & 18.5 & 16.0 & 0.0 & 9.7 & 51 & 3 & -0.35 & dblc & 484 & 0 \\ 

J1230-3121 & 187.687 & -31.357 & -0.40 & 242 & 2.276 & 17.1 & 6.4 & 0.0 & 3.8 & 50 & 2 & 0.01 & unres & $<25$ & 0 \\ 

J1233-3534 & 188.488 & -35.579 & -0.84 & 113 & 0.154 & 87.0 & 25.6 & 0.0 & 7.4 & 48 & 2 & -0.66 & dbl & 40 & 1 \\ 
      \hline
   \end{tabular}
\label{modelfitsextra}
\end{table*}

{\tiny
\begin{table*}
 \contcaption{ }
 \scriptsize
 \centering
   \begin{tabular}{cccccccccccccccc}
    \hline\hline
       (1) & (2) & (3) & (4) & (5) & (6) & (7) & (8) & (9) & (10) & (11) & (12) & (13) & (14) & (15) & (16)  \\
Name  & RA & DEC & $\alpha$ &   $I_{2.1\,{\rm GHz}}$ &  $z$ &  $|$RRM$|$ &  $\sigmaRMwtd$ &  $\gradRMwtd$ &  $\langle p_{0} \rangle$ &  S/N &  \# &  $\beta$ &  Morph &  $l$ &  Rad/Jet \\
      \hline 
J1244-4012 & 191.122 & -40.213 & -0.78 & 257 & 0.191 & 35.9 & 4.8 & 0.0 & 1.8 & 36 & 2 & 0.08 & dblc & 62 & 0 \\ 

J1246-3406 & 191.556 & -34.110 & -1.50 & 33 & 0.088 & 10.2 & 0.0 & 0.0 & 8.7 & 10 & 1 & -0.94 & dbl & 125 & 1 \\ 

J1257-3334 & 194.336 & -33.578 & -0.91 & 702 & 0.190 & 5.5 & 0.0 & 0.0 & 2.9 & 32 & 2 & -0.92 & dbl & 84 & 0 \\ 

J1300-3253 & 195.176 & -32.886 & -0.28 & 199 & 1.256 & 3.2 & 0.0 & 0.0 & 2.5 & 52 & 2 & -0.14 & unres & $<25$ & 0 \\ 

J1301-3226 & 195.252 & -32.441 & -0.38 & 466 & 0.017 & 12.4 & 0.0 & 25.9 & 2.8 & 50 & 3 & -0.21 & dblc & 35 & 1 \\ 

J1304-3406 & 196.140 & -34.110 & -0.92 & 39 & 0.051 & 9.9 & 6.3 & 0.0 & 19.0 & 27 & 1 & -0.29 & dbl & 49 & 1 \\ 

J1307-3207 & 196.813 & -32.133 & -0.40 & 410 & 1.211 & 0.9 & 1.0 & 0.0 & 1.7 & 54 & 3 & -0.30 & ext & 140 & 0 \\ 

J1307-3737 & 196.849 & -37.619 & -1.17 & 20 & 0.124 & 9.3 & 11.4 & 0.0 & 19.3 & 20 & 1 & -0.63 & dblc & 160 & 1 \\ 

J1316-3338 & 199.033 & -33.650 & -0.22 & 941 & 1.210 & 0.7 & 10.4 & 0.0 & 3.1 & 53 & 2 & -0.02 & unres & $<25$ & 0 \\ 

J1324-3623 & 201.015 & -36.393 & -0.62 & 147 & 0.318 & 10.0 & 0.0 & 0.0 & 3.4 & 16 & 1 & 0.35 & dblc & 120 & 0 \\ 

J1342-3240 & 205.526 & -32.680 & -0.08 & 199 & 0.791 & 7.1 & 6.3 & 0.0 & 3.8 & 40 & 2 & 0.18 & unres & $<22$ & 0 \\ 

J1342-3601 & 205.703 & -36.029 & -1.06 & 106 & 0.073 & 8.6 & 11.2 & 0.0 & 21.8 & 39 & 1 & -0.54 & dblc & 86 & 1 \\ 

J1345-3140 & 206.266 & -31.671 & -0.75 & 340 & 0.143 & 5.4 & 0.0 & 0.0 & 6.0 & 41 & 3 & -0.52 & ext & 183 & 1 \\ 

J1403-3358 & 210.911 & -33.980 & -0.73 & 161 & 0.014 & 23.0 & 33.2 & 0.0 & 8.9 & 42 & 2 & 0.29 & ext & 13 & 1 \\ 

J1411-3205 & 212.948 & -32.087 & -0.81 & 91 & 0.041 & 25.8 & 13.8 & 0.0 & 8.8 & 33 & 2 & -0.57 & cmplx & 19 & 1 \\ 

J1413-3002 & 213.382 & -30.045 & -0.98 & 347 & 0.065 & 102.3 & 54.0 & 0.0 & 4.6 & 23 & 3 & 0.05 & dbl & 44 & 1 \\ 

J1433-3141 & 218.364 & -31.695 & -1.12 & 109 & 0.058 & 4.8 & 11.1 & 0.0 & 31.1 & 37 & 1 & -0.68 & dblc & 59 & 1 \\ 

J1438-3122 & 219.545 & -31.374 & -0.33 & 329 & 1.287 & 6.6 & 32.4 & 0.0 & 3.8 & 31 & 2 & 0.02 & ext & 48 & 0 \\ 

J1457-3539 & 224.361 & -35.653 & -0.22 & 631 & 1.424 & 14.3 & 72.3 & 0.0 & 4.9 & 70 & 2 & 0.16 & unres & $<25$ & 0 \\ 

J1511-3242 & 227.960 & -32.715 & -0.32 & 354 & 1.105 & 14.0 & 15.2 & 0.0 & 3.4 & 56 & 2 & -0.00 & unres & $<25$ & 0 \\ 

J1522-2730 & 230.657 & -27.503 & 0.49 & 777 & 1.294 & 12.1 & 0.0 & 0.0 & 1.9 & 54 & 2 & 1.81 & unres & $<26$ & 0 \\ 

J1524-3012 & 231.139 & -30.206 & 0.43 & 148 & 0.019 & 11.3 & 0.0 & 0.0 & 1.9 & 44 & 2 & 0.52 & unres & $<1$ & 1 \\ 

J1527-3014 & 231.908 & -30.250 & -1.00 & 84 & 0.969 & 46.3 & 0.0 & 0.0 & 1.5 & 21 & 2 & -0.50 & dbl & 105 & 1 \\ 
      \hline
   \end{tabular} \\ 
\scriptsize{Note: 1 - Source name; 2 - Right Ascension in J2000 coordinates; 3 - Declination in J2000 coordinates; 4 - Spectral index; 5 - Integrated total intensity at 2.1~GHz, in mJy; 
                    6 - Redshift; 7 - Absolute value of the residual rotation measure, in rad~m$^{-2}$; 8 - polarization-weighted RM dispersion, in rad~m$^{-2}$; 
                    9 - polarization-weighted RM gradient, in rad~m$^{-2}$; 10 - mean intrinsic degree of polarization, in per cent; 11 - polarization signal-to-noise ratio; 12 - Number of RM components; 
                    13 - polarization spectral index; 14 - Total intensity morphology; 15 - Linear size, in kpc. 16 - Radiative-mode: 0, Jet-mode: 1.}
\end{table*}
}

\subsection{Total intensity morphology and the number of RM components}
We use the 2.1~GHz, 5.5~GHz and 9~GHz images (Figure~\ref{images1}) to classify the sources into broad morphological types. We label sources with distinct double-lobed structure as `double' (e.g.~Fig.~\ref{images1}f) and `double$+$core' when they show a bright inner jet/core component (e.g.~Fig.~\ref{images1}g). Sources that display a complex radio structure we denote as `complex' (e.g.~Fig.~\ref{images1}b), while sources that have a bright core with weaker extended emission are labelled as `extended' (e.g.~Fig.~\ref{images1}n). Unresolved sources are labelled `unresolved'. We consider these morphology classifications as good descriptions of all the source structures observed in our sample. 
Overall, 20\% of sources are classified as `double', 28\% as `double$+$core', 22\% as `extended', 7\% as `complex', and 25\% as `unresolved' (Table~\ref{cmpntsmorph}). 

One interesting comparison to make is between the total intensity morphology classification and the number of RM components, in order to see if it might be possible to infer the physical structure of the sources from integrated polarization observations. Table~\ref{cmpntsmorph} shows the different percentages of the morphological classes for one, two and three RM components. The values in parentheses are the percentages when excluding sources with signal-to-noise ratios lower than the median value (S/N $< 35.5$). 
Overall, there is not a one-to-one correspondence between the number of RM components and the morphological class. 
However, we see that the majority of `double' sources have two RM components, with an increasing fraction at high S/N (71\% have two RM components). 
The `double$+$core' class have the largest number of three RM components sources, but appear roughly equally likely to be described by one, two or three RM components. Both the `extended' and `complex' morphological classes have a preference for two RM component models. 
The majority of `unresolved' sources are also described by two RM components (80\% of cases). It is worth noting that most of the `unresolved' class are flat-spectrum sources, so in some cases the two RM components may be parameterising the frequency-dependent polarized emission from synchrotron self-absorbed regions (e.g.~Porth et al.~2011). 

\begin{table}
 \caption{Percentages of the number of RM components for sources with different total intensity morphologies.}
 \centering
   \begin{tabular}{ccccc}
    \hline\hline
       (1) & (2) & (3) & (4) & (5)  \\
      Morphology   & 1 RM & 2 RM &  3 RM    & All        \\
                           & (\%)   &  (\%)  &    (\%) &  (\%)                \\
      \hline            
unresolved          & 16(15)  & 80(80) & 4(5)       &  25(40) \\
double                & 40(29) & 55(71) & 5(0)       &  20(14) \\
double$+$core   & 50(23) & 29(31) & 21(46)   &  28(26) \\
extended            & 35(11) & 50(56) & 15(33)   &  20(18) \\
complex             & 57(0)  & 43(100) & 0(0)       &  7(2) \\
      \hline
   \end{tabular}\\
\scriptsize{Note: Values in parentheses are for only those sources with S/N greater than the median value for the sample of 35.5. 
In columns (2), (3) and (4), `1/2/3 RM' is shorthand for one/two/three RM components. In column (5), `All' refers to all sources in the sample. }
\label{cmpntsmorph}
\end{table}

\begin{table}
 \caption{Median values of the polarization, Faraday rotation properties and linear size for 
sources with different total intensity morphologies.}
 \centering
   \begin{tabular}{ccccc}
    \hline\hline
       (1) & (2) & (3) & (4) & (5)  \\
      Morphology   & ${\langle p_0 \rangle}_{\rm med}$         & ${\sigmaRM}_{\rm wtd, med}$       &     RRM$_{\rm med}$       & $l_{\rm med}$    \\
                            &                (\%)                                         &         (rad~m$^{-2}$)                        &         (rad~m$^{-2}$)       &         (kpc) \\
      \hline            
unresolved          & 2.8    & 15.2  & 9.7      & --      \\
double                 & 6.5    & 12.5  & 12.5   & 68    \\
double$+$core    & 10.8  & 12.4  & 9.3      &  149  \\
extended             & 5.5    & 18.7  & 6.8      & 140      \\
complex              & 8.8    & 13.8   & 20.3   & 50       \\
\hline
   \end{tabular}\\
\label{morphmeds}
\end{table}

In Table~\ref{morphmeds}, we list the median values of $\langle p_0 \rangle$, RRM and $\sigmaRMwtd$ for the different morphological classes. 
The `unresolved' class has the lowest median $\langle p_0\rangle$ of 2.8\%, which is unsurprising given that these are mainly flat spectrum sources. 
The `double$+$core' class has the highest median $\langle p_0\rangle$ of 10.8\%, which is significantly different than the `double' and `extended' classes which have similar median $\langle p_0\rangle$ values of $\sim$6\% (considering the median error in $\langle p_0\rangle$ is 0.7\%). 
There is no significant difference in RRM or $\sigmaRMwtd$ between the morphological classes, considering the median error in RRM and $\sigmaRMwtd$ of 7.5~rad~m$^{-2}$ and 1.8~rad~m$^{-2}$, respectively.  

\begin{figure} 
\centering
    \includegraphics[angle=0, clip=true, trim=0cm 0cm 0cm 1cm, width=.48\textwidth]{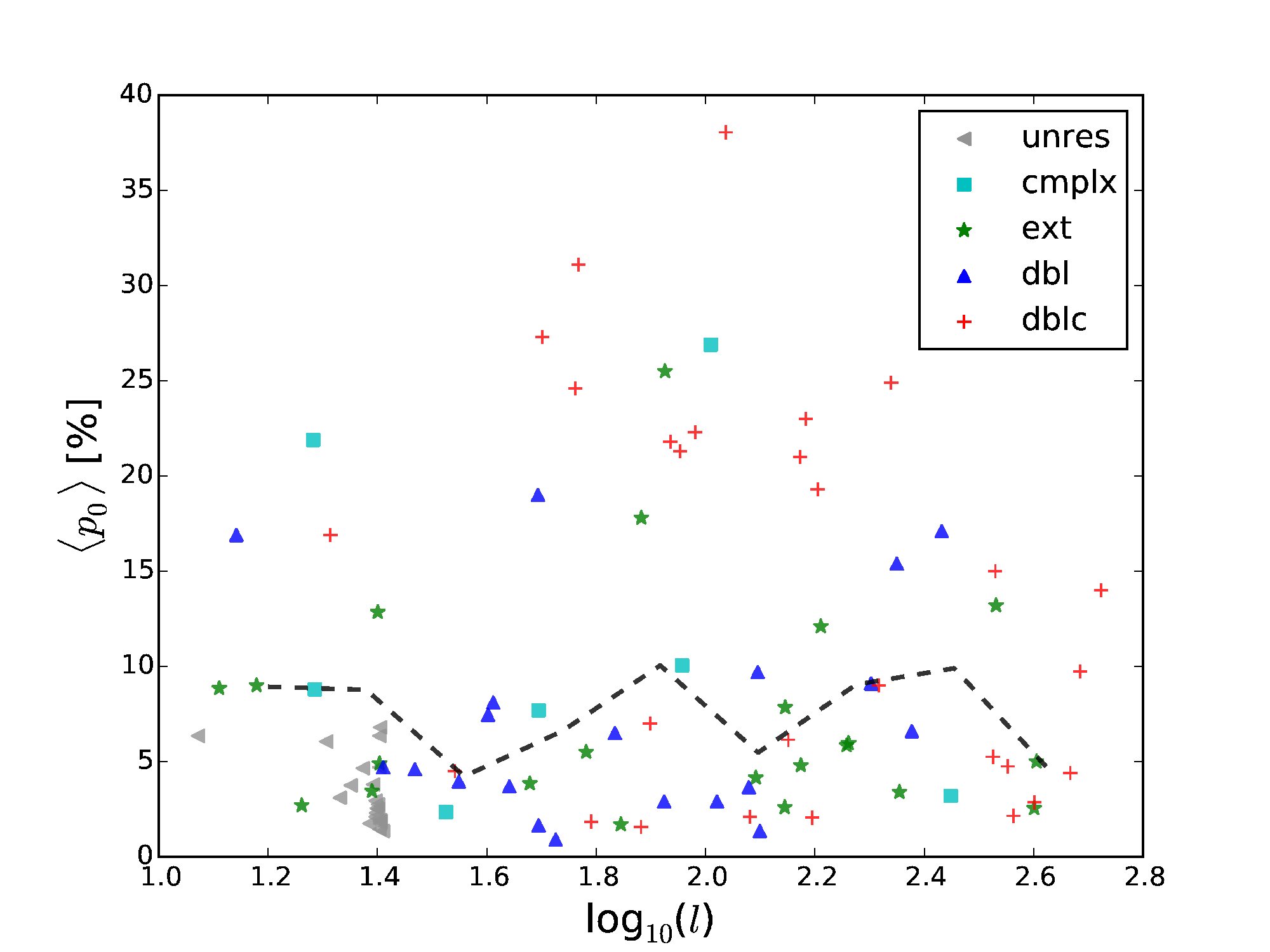} 
    \caption{ {\small Plot of the mean intrinsic degree of polarization, $\langle p_0 \rangle$, versus the linear size ($l$) of each source. 
    The coloured symbols represent the different morphological classes defined in the text. } }
    \label{lp0morph}
\end{figure}

\begin{figure} 
\centering
    \includegraphics[angle=0, clip=true, trim=0cm 0cm 0cm 1cm, width=.48\textwidth]{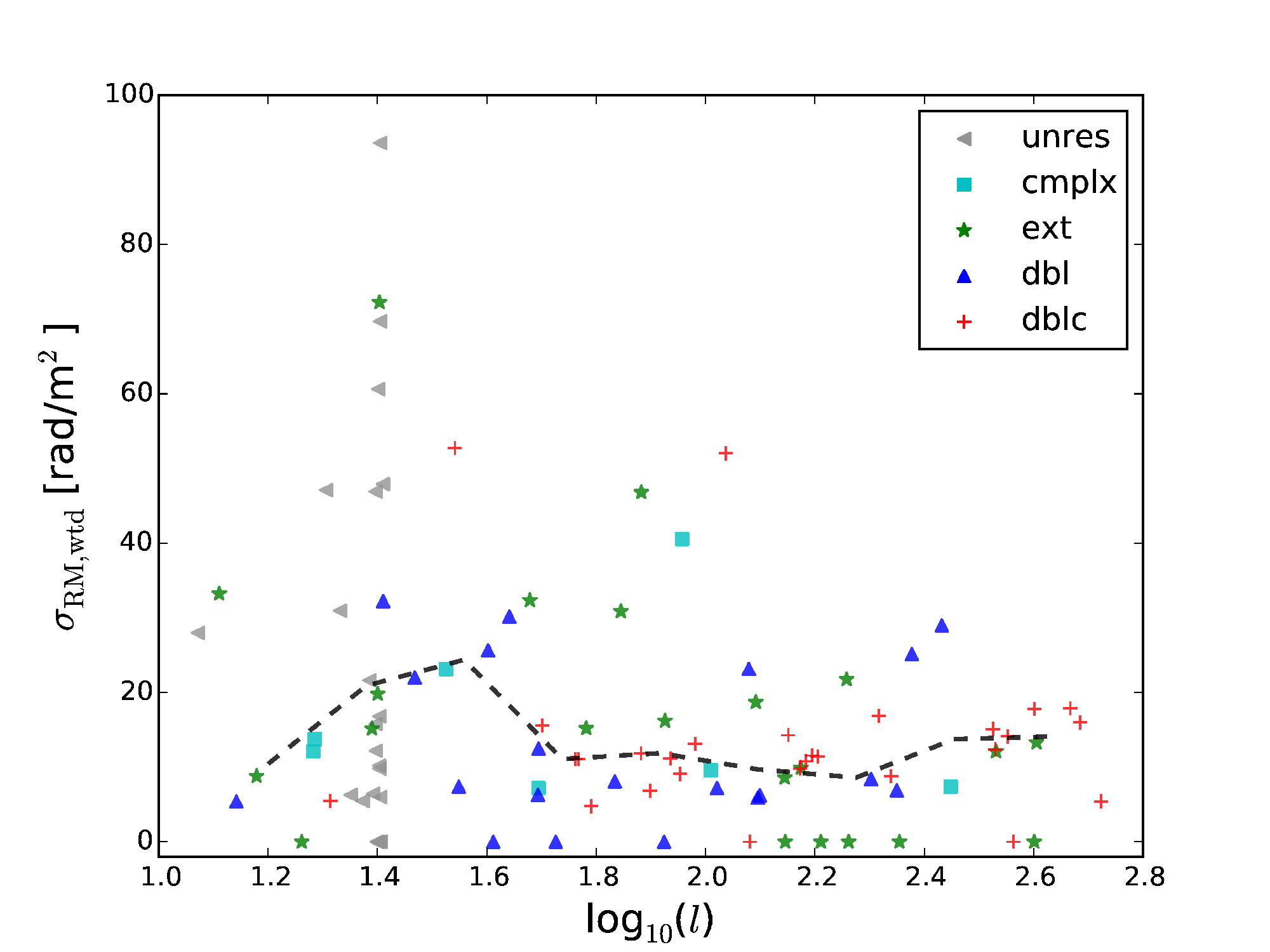} 
    \caption{ {\small Plot of the polarization-weighted RM dispersion, $\sigmaRMwtd$, versus the linear size ($l$) of each source. 
    The coloured symbols represent the different morphological classes defined in the text. } }
    \label{lsigmaRMmorph}
\end{figure}

\begin{figure} 
\centering
    \includegraphics[angle=0, clip=true, trim=0cm 0cm 0cm 1cm, width=.48\textwidth]{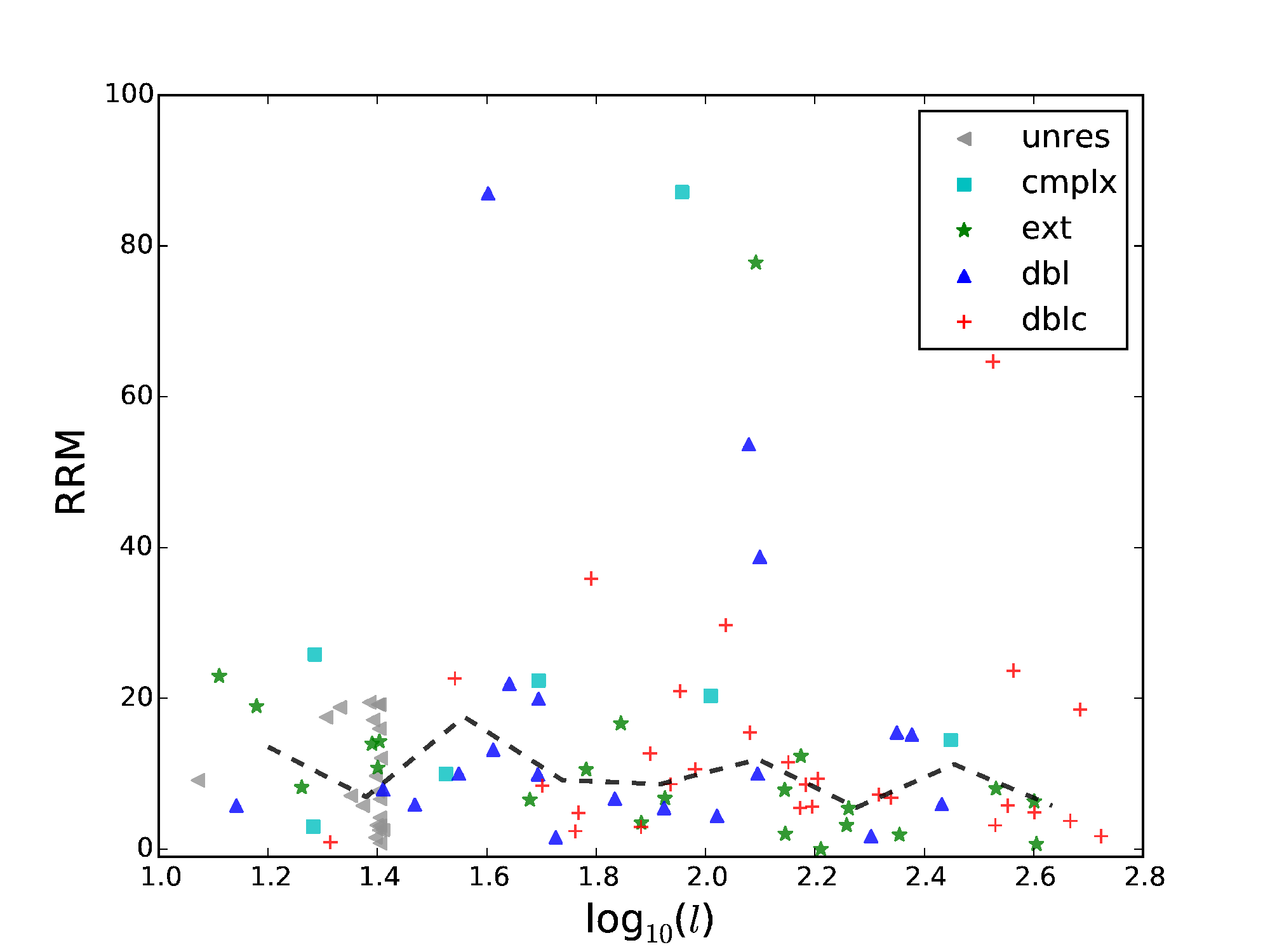} 
    \caption{ {\small Plot of the residual rotation measure, RRM, versus the linear size ($l$) of each source. 
    The coloured symbols represent the different morphological classes defined in the text. } }
    \label{lRRMmorph}
\end{figure}

In terms of linear size ($l$), the `double$+$core' and `extended' classes are the largest, with median linear sizes of 149~kpc and 140~kpc, respectively. 
The `double' and `complex' classes are typically more compact with median linear sizes of 68~kpc and 50~kpc, respectively. 
In Figures~\ref{lp0morph}, \ref{lsigmaRMmorph}, and \ref{lRRMmorph}, we plot $\langle p_0\rangle$, $\sigmaRMwtd$, and RRM versus the linear size. We indicate the different morphological classes with different symbols. No correlations are found between these parameters and the linear size, as can be seen from the running medians (dashed lines). We expect that much larger sample sizes are needed to robustly determine any dependence of $\langle p_0\rangle$, $\sigmaRMwtd$, or RRM with linear size.

\subsection{Radiative-mode AGN versus Jet-mode AGN} \label{radjetresults}
Our broadband polarization modelling allows us to determine both the intrinsic magnetic field properties ($p_0, \psi_0$) and the Faraday rotation properties (RM, $\sigmaRM$, $\DeltaRM$) in a robust manner. This enables us to directly investigate the physical origin of the observed difference in the integrated degree of polarization at 1.4~GHz ($p_{\rm 1.4 GHz}$) between radiative-mode and jet-mode AGN, as found in OS15. Namely that the jet-mode AGN can achieve values of $p_{\rm 1.4 GHz}$ of up to 30\%, while the radiative-mode AGN are limited to $p_{\rm 1.4 GHz}\lesssim15\%$. 
The same behaviour is observed in our current, smaller sample (see Fig.~\ref{sigmaRMp1400herglerg}), while we also find a significant difference in $p_{\rm 1.4 GHz}$ between the steep-spectrum radiative-mode and jet-mode sources (KS test p-value: 0.0001). 
The key question we wish to answer is whether the difference in $p_{\rm 1.4 GHz}$ is influenced more by the intrinsic magnetic field properties or the Faraday depolarization of the different source types.

There are significantly more flat-spectrum radiative-mode sources (19/60) than flat-spectrum jet-mode sources (3/40).
The intrinsic limit of $p_{0{\rm ,max}}\sim10\%$ for flat-spectrum sources (as opposed to $p_{0{\rm ,max}}\sim70\%$ for steep-spectrum sources) is one obvious reason why radiative-mode sources would have lower degrees of polarization, on average, than jet-mode sources. However, it would not explain why some radiative-mode sources do not achieve such high degrees of polarization as the jet-mode sources. 
Therefore, in comparing the intrinsic polarization and Faraday rotation properties of the two types of sources, we only consider those sources with $\alpha<-0.3$ (i.e.~where the polarized emission is likely coming from an optically thin region of the source). This leaves us with a small but reasonable statistical sample of 41 radiative-mode sources and 37 jet-mode sources. 

The redshift distribution of the radiative-mode and jet-mode sources are very different (Fig.~\ref{zhist}) and should be kept in mind when interpreting the results presented below. However, Section~\ref{sec_z} shows why the different redshift distributions of the radiative-mode and jet-mode sources are unlikely to significantly alter our conclusions. 
The median linear size of the resolved radiative-mode and jet-mode sources is 140~kpc and 86~kpc, respectively. This is consistent with the results from the much larger sample of OS15, and with the expectation that the radiative-mode sources produce more powerful (and larger) radio jets (Heckman \& Best 2014). 

\begin{table}
 \caption{Morphologies of steep-spectrum radiative-mode and jet-mode sources.}
 \centering
   \begin{tabular}{ccc}
    \hline\hline
       (1) & (2) & (3)   \\
      Morphology   & Radiative-mode       & Jet-mode         \\
      \hline            
unresolved          & 12.2\%        & 0\%    \\ 
double                 & 26.8\%        & 24.3\%  \\
double$+$core    & 31.7\%       & 40.5\%  \\
extended             & 29.3\%       & 16.2\%  \\
complex               & 0\%            & 19.0\%  \\
\hline
   \end{tabular}\\
\label{radjetmorph}
\end{table}

The morphological classes of the steep-spectrum radiative-mode and jet-mode sources are quite similar (see Table~\ref{radjetmorph}). The main differences are that there are no radiative-mode sources classified as `complex' compared to 19\% of jet-mode sources, and 10\% of the radiative-mode sources remain unresolved. However, this difference could be explained in terms of angular resolution since all the `complex' sources are at $z < 0.25$, and the unresolved sources have a median $z=2.3$. Therefore, some of the high redshift `unresolved', and also `extended', radiative-mode sources may display more complex morphologies with better angular resolution. There is also slightly more `double$+$core' class jet-mode sources, and the jet-mode `double$+$core' sources have a median $\langle p_0 \rangle$ of 21.8\% while the radiative-mode `double$+$core' sources have a median $\langle p_0 \rangle$ of only 4.4\%. This emphasises that the difference in polarization properties between jet-mode and radiative-mode sources is not due to radio morphology alone. 

\subsubsection{Magneto-ionic properties}
In order to investigate the turbulent magnetised environments of sources with different numbers of RM components in a consistent manner, we use the polarization-weighted RM dispersion, $\sigmaRMwtd$ (Eqn.~\ref{sigmaRMwtdeqn}). This provides a characterisation of the amount of Faraday depolarization for each source, and allows us to assess its impact on the observed degree of polarization at 1.4~GHz for both the radiative-mode and jet-mode sources. 

\begin{figure} 
\centering
    \includegraphics[angle=0, clip=true, trim=0cm 0.5cm 0cm 0cm, width=.45\textwidth]{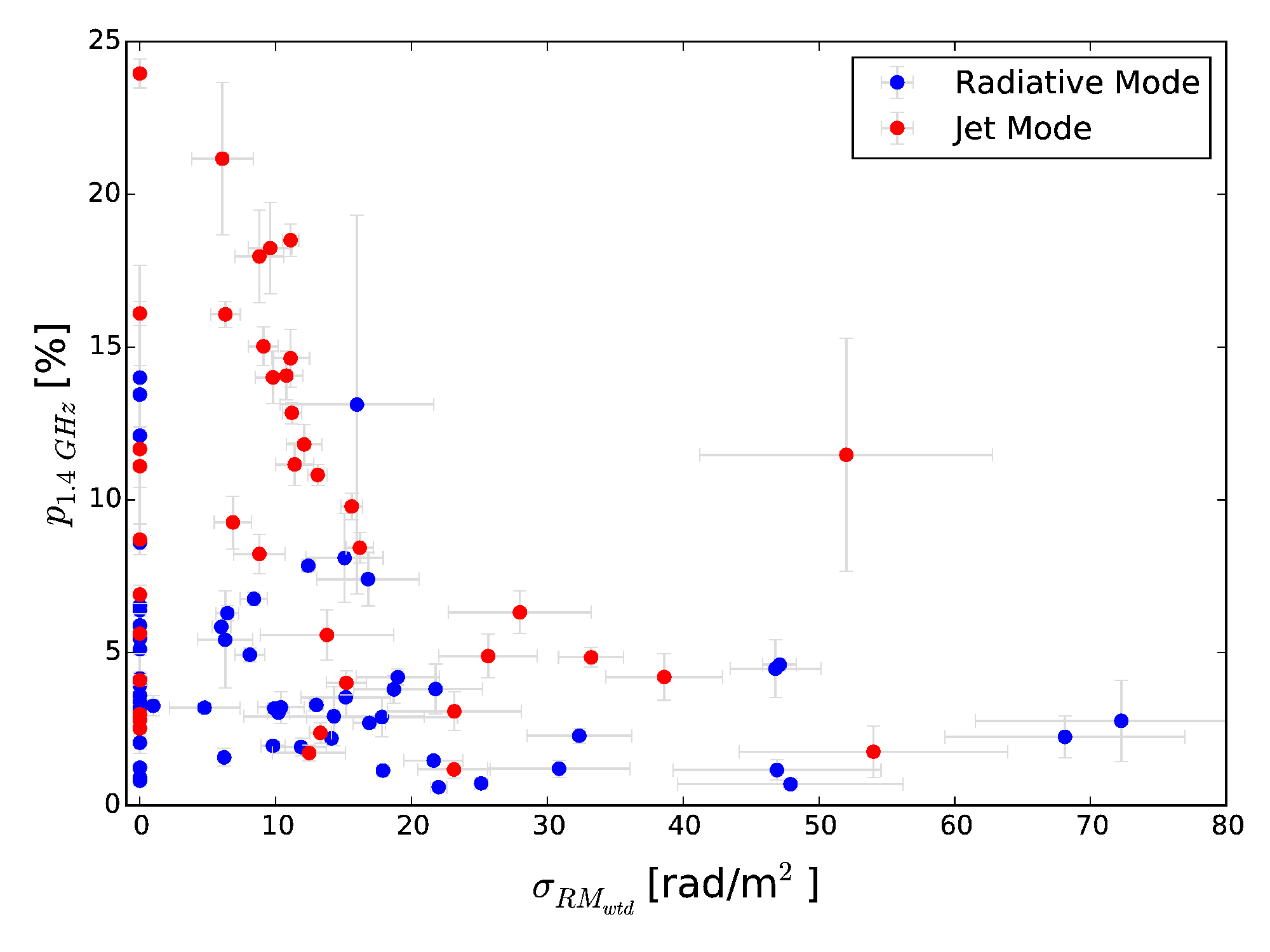} 
    \caption{ {\small Plot of the integrated degree of polarization at 1.4~GHz ($p_{\rm 1.4\,GHz}$) vs.~the polarization-weighted 
    RM dispersion ($\sigmaRMwtd$). Radiative-mode (blue) and jet-mode sources (red). } }
    \label{sigmaRMp1400herglerg}
\end{figure}

In Figure~\ref{sigmaRMp1400herglerg} we plot the degree of polarization at 1.4 GHz ($p_{\rm 1.4\,GHz}$) versus the polarization-weighted RM dispersion ($\sigmaRMwtd$) for radiative-mode (blue) and jet-mode (red) AGN. There is a clear absence of sources with a high degree of polarization at 1.4 GHz and high RM dispersions, showing that Faraday depolarization is an important factor at 1.4~GHz, as would be expected. A Spearman correlation test shows there is an anti-correlation, with a correlation coefficient of $-0.44$ with a p-value of 0.0002. 
However, there are also several sources with low RM dispersions that also have low degrees of polarization at 1.4 GHz, showing the Faraday depolarization is not the main cause of the low $p_{\rm 1.4\,GHz}$ for these sources. 
There are 19 sources for which $\sigmaRMwtd=0$ (11 radiative-mode and 8 jet-mode). These are sources which are either truly Faraday thin, or the signal-to-noise was too low for the model-fitting to recover the true $\sigmaRM$ value(s). 

\begin{figure} 
\centering
    \includegraphics[angle=0, clip=true, trim=0cm 0.0cm 0cm 1cm, width=.5\textwidth]{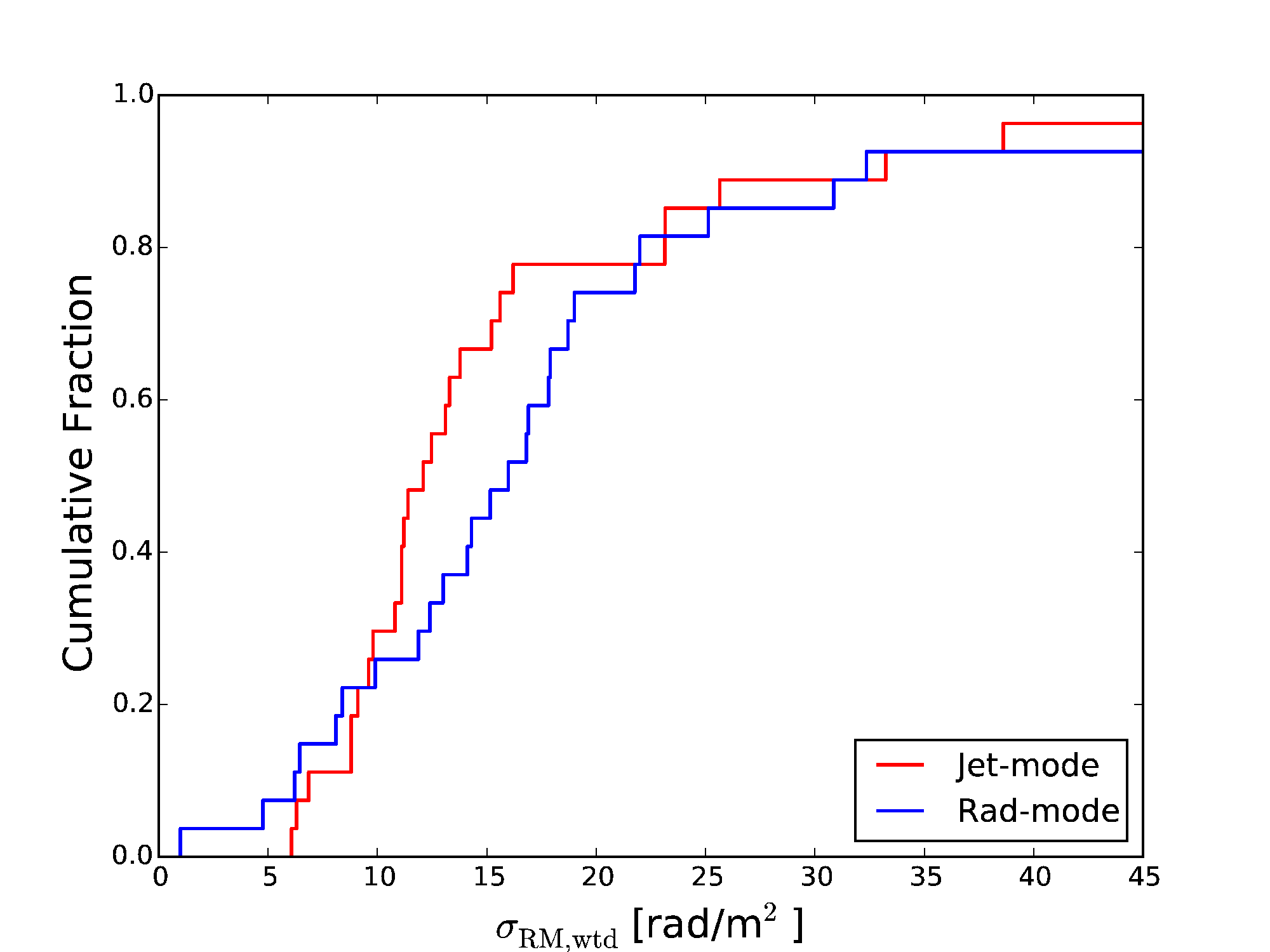} 
    \caption{ {\small Empirical cumulative distribution function (ECDF) of the polarization-weighted RM dispersion ($\sigmaRMwtd$) for 
    radiative-mode (blue) and jet-mode sources (red). } }
    \label{ecdfsigmaRM}
\end{figure} 

\begin{figure} 
\centering
    \includegraphics[angle=0, clip=true, trim=0cm 0.1cm 0cm 1.1cm, width=.5\textwidth]{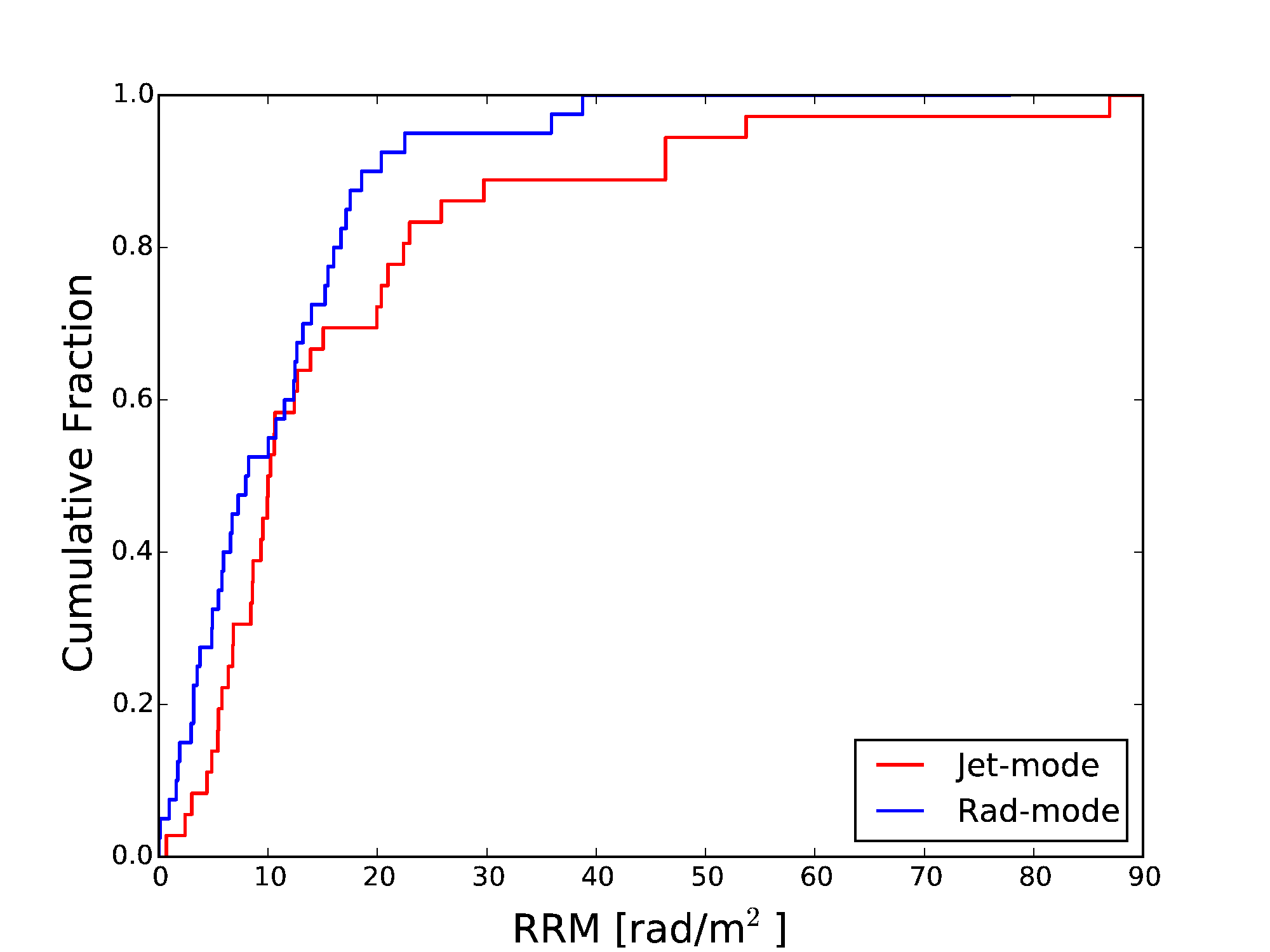} 
    \caption{ {\small ECDF of the residual rotation measure (RRM) for radiative-mode (blue) and jet-mode sources (red). } }
    \label{ecdfRRM}
\end{figure}

In Figure~\ref{ecdfsigmaRM}, we directly compare the values of $\sigmaRMwtd$ for radiative-mode and the jet-mode sources in an empirical cumulative distribution function (ECDF), excluding sources where $\sigmaRMwtd=0$. 
A two-sided KS test gives no indication that the underlying distributions are different (p-value: 0.17), even when sources with $\sigmaRMwtd=0$ are included (p-value: 0.60).
Another means by which to check if the magneto-ionic properties of radiative-mode and jet-mode sources are different is to use the RRM. Figure~\ref{ecdfRRM} shows the ECDF for the two types, and it is clear from the figure, and a KS test confirms, that there is no difference in RRM between the two samples (p-value: 0.29). Similarly, we find no difference in RM between the radiative-mode and jet-mode sources (KS test p-value: 0.78). 

We can investigate the RM local to the source in another way for the 24 radiative-mode and 11 jet-mode sources that have two RM components. In several cases it is likely that the two RM components correspond to the two lobes of the radio source. Thus, we define ${\rm dRM}=| {\rm RM}_1 - {\rm RM}_2 |$ in order to get an estimate of the variation in the RM between the two lobes, which is most likely caused by magneto-ionic material in the local source environment. However, no significant difference between the radiative-mode and jet-mode sources is seen (KS p-value: 0.86). 
One may need to be somewhat cautious here in interpreting dRM because, as was found from simulations in Sun et al.~(2015), even if the input RM difference of two components is small, the fitted RM difference ($| {\rm RM}_1 - {\rm RM}_2 |$) could be significantly larger in some cases. 

\subsubsection{The effect of the number of RM components}
\label{RMcmpnts}
\begin{figure} 
\centering
    \includegraphics[angle=0, clip=true, trim=0cm 0cm 0cm 0cm, width=.45\textwidth]{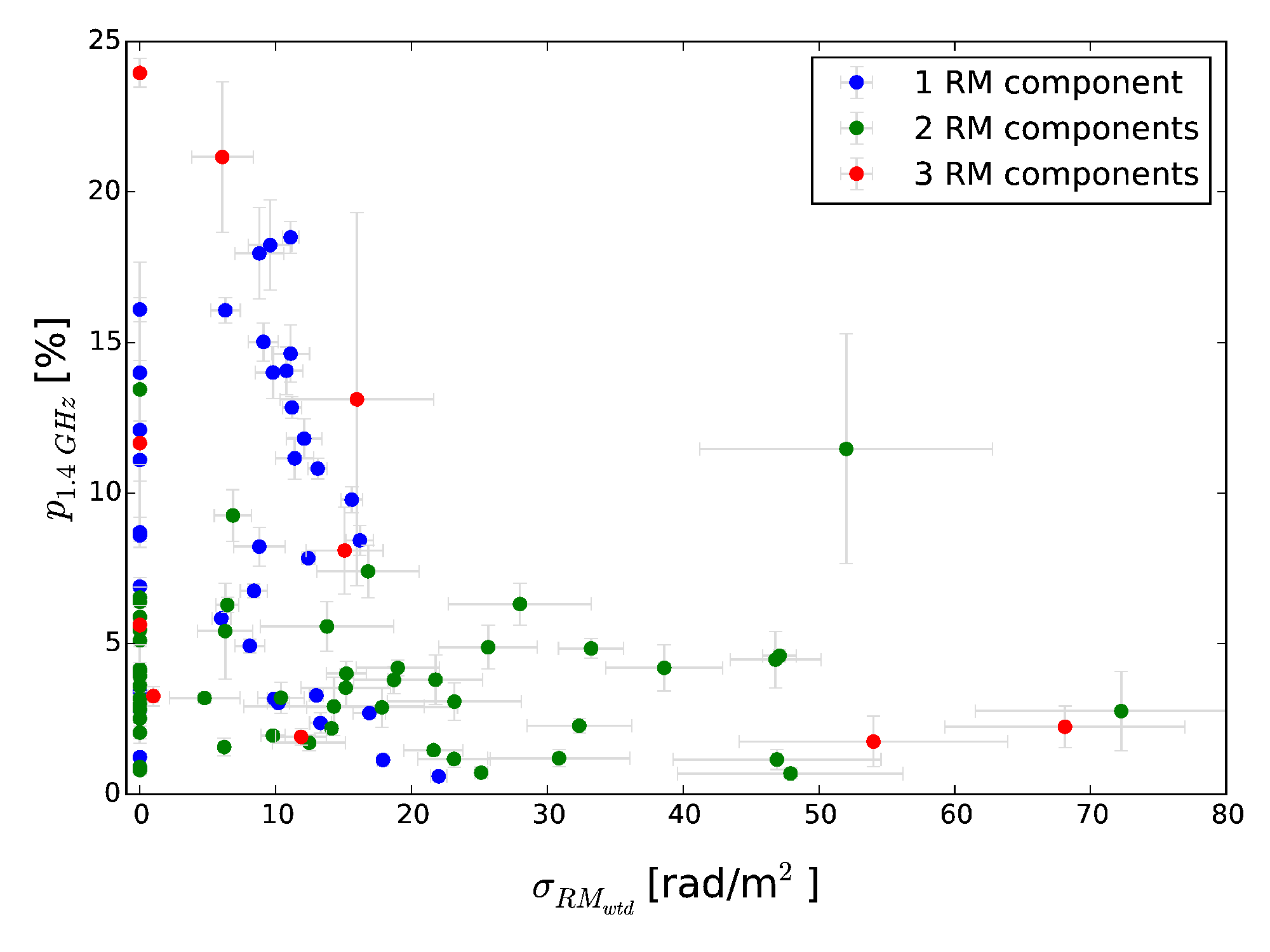} 
    \caption{ {\small Plot of the integrated degree of polarization at 1.4~GHz ($p_{\rm 1.4\,GHz}$) versus the polarization-weighted 
    RM dispersion ($\sigmaRMwtd$), split into one RM component (blue), two RM component (green) and three RM component (red) sources. }  }
    \label{sigmaRMp1400cmpnts}
\end{figure}

One might expect that the number of RM components in each source could effect the observed integrated degree of polarization at 1.4~GHz in light of the vector nature of the linearly polarized emission. For example, low values of $p_{1.4\, {\rm GHz}}$ for sources with two or three RM components could be caused by destructive interference between the components. 
Therefore, we again show $p_{1.4\, {\rm GHz}}$ vs.~$\sigmaRMwtd$ in Figure~\ref{sigmaRMp1400cmpnts}, but now display the number of RM components of each source with a different colour (Blue: one RM component, Green: Two RM components, Red: Three RM components). Here we see that the majority of the highly polarized sources at 1.4 GHz are dominated by a single RM component; although the two highest values have three RM components. 

\begin{table}
 \caption{Number of RM components for steep-spectrum radiative-mode and jet-mode sources.}
 \centering
   \begin{tabular}{ccc}
    \hline\hline
       (1) & (2) & (3)   \\
     \# RM components   & Radiative-mode       & Jet-mode         \\
      \hline            
1 RM component         & 56.8\%        & 31.7\%    \\ 
2 RM components        & 29.7\%        & 58.5\%  \\
3 RM components       & 13.5\%         & 9.8\%  \\
\hline
   \end{tabular}\\
\label{radjetRMcmpnts}
\end{table}

In Table~\ref{radjetRMcmpnts}, we show the percentages of one, two, and three RM components for steep-spectrum radiative-mode and jet-mode sources. 
Approximately 57\%/30\% of steep-spectrum jet-mode sources have one/two RM components, whereas $\sim$32\%/58.5\% of steep-spectrum radiative-mode sources are best fit by one/two RM components. The excess of one RM component sources over two RM components for the jet-mode sources, in comparison with the radiative-mode sources, appears to be one of the key reasons why jet-mode sources can achieve higher $p_{1.4\, {\rm GHz}}$ values. 
The fraction of three RM components is similar for both radiative-mode and jet-mode sources. 

There are several one RM component sources with $p_{1.4\, {\rm GHz}} < 10\%$. In total, a small number of these are flat-spectrum sources (3/21), indicating that the polarized emission is likely emanating from a synchrotron-self-absorbed region, and as expected, limits the the intrinsic degree of polarization to $<10$\%. Thus, we expect that the remaining optically-thin, low-polarization, one RM component sources have intrinsically disordered magnetic field structures, which we investigate in more detail next. 

\begin{figure} 
\centering
    \includegraphics[angle=0, clip=true, trim=0cm 0cm 0cm 1cm, width=.5\textwidth]{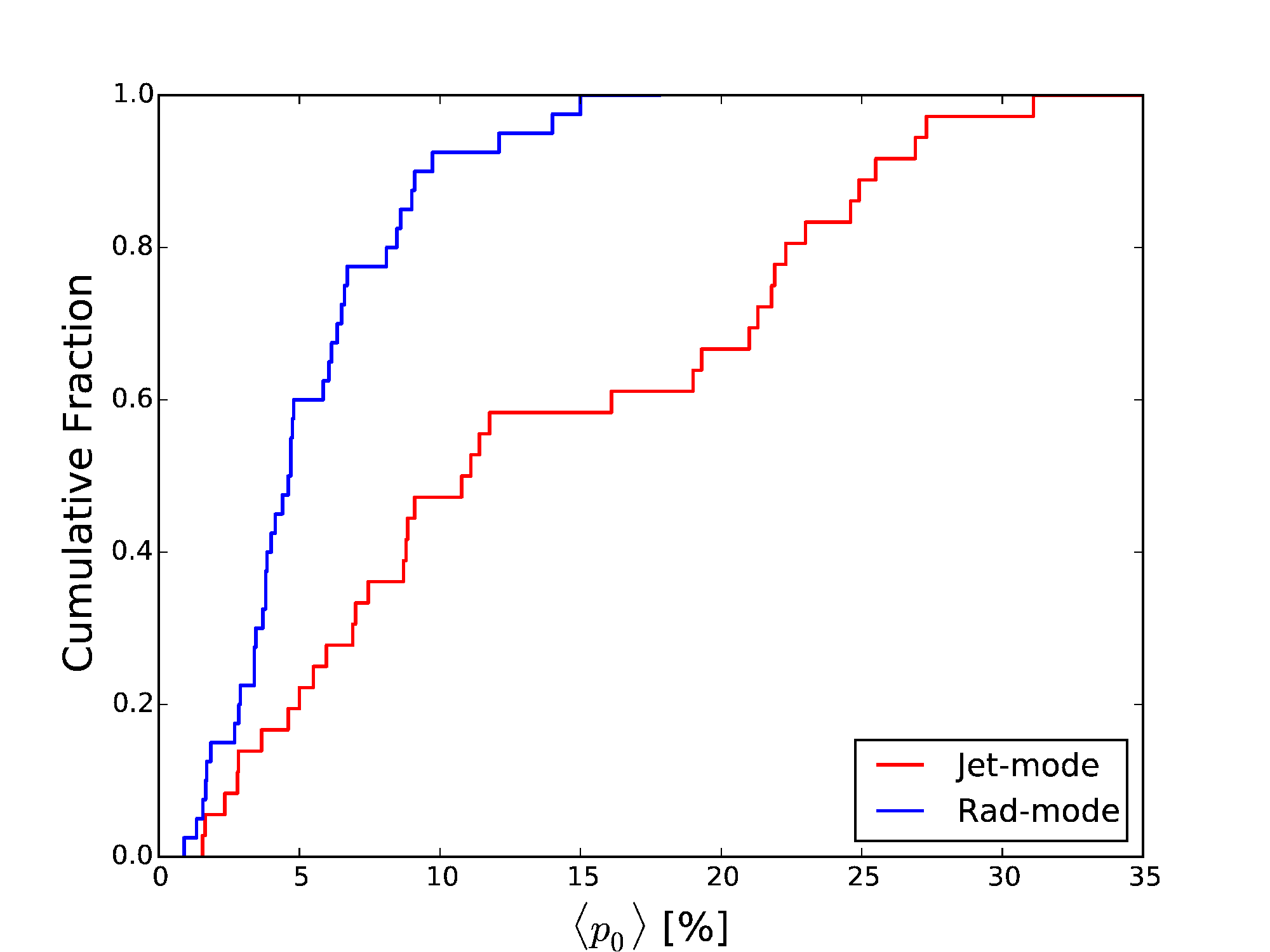} 
    \caption{ {\small ECDF of the mean intrinsic degree of polarization, $\langle p_0 \rangle$, for radiative-mode (blue) and jet-mode sources (red). } }
    \label{ecdfp0}
\end{figure} 

\subsubsection{Intrinsic magnetic field order} \label{bfieldorderherglerg}
The median value of $\langle p_0 \rangle$ for all steep spectrum sources in our sample is 6.6\%. The median value of the Faraday depolarization ($\sigmaRMwtd$) for all steep spectrum sources is $\sim$11~rad~m$^{-2}$, which is only sufficient to reduce the intrinsic degree of polarization by a factor of $\sim$1.7 at 1.4~GHz. 
This means that most radio sources in our sample are dominated by a disordered magnetic field structure. Although possible, this does not mean that the sources are dominated by a random magnetic field structure (with a uniform and symmetric jet/lobe structure) but more likely that there are significant uniform magnetic field structures but with a non-uniform and/or non-symmetric jet/lobe structure, as often seen in high-fidelity, high-resolution, polarization images of radio galaxies (e.g.~Ishwara-Chandra et al.~1998). 

Figure~\ref{ecdfp0} shows the ECDF for $\langle p_0 \rangle$ of each source, split into radiative-mode and jet-mode sources (again excluding flat spectrum sources). As is clear from the plot, the jet-mode sources have significantly larger values of $\langle p_0 \rangle$ than the radiative-mode sources. This is confirmed with a KS test that strongly suggests the two samples are not drawn from the same underlying distribution, with a p-value of 0.0001. This clearly shows that the intrinsic magnetic field order is the dominant variable causing the observed difference in $p_{1.4\, {\rm GHz}}$ between radiative-mode and jet-mode sources. 

As a note, one might naively expect that $\langle p_0 \rangle$ increases for two and three RM component sources because we are isolating individual magnetic patches in the sources. However, two and three RM component models will always have lower values of $p_{01,2,3}$ than the individual source components (if they were resolved) because each $p_{01,2,3}$ from the model-fitting is as a fraction of the total Stokes $I$, not the Stokes $I$ of each component. 

\subsubsection{Intrinsic magnetic field orientation} \label{bfieldgeometry}
Our model fitting approach also recovers the intrinsic polarization angle ($\psi_0$)\footnote{Also known as the intrinsic electric vector position angle.} of each RM component for each source. These $\psi_0$ values are of interest to compare with the jet direction, in order to determine if there is a preferential orientation of the jet/lobe magnetic field (which is orthogonal to $\psi_0$). 
We estimated the jet orientation by visually inspecting each image. For straight sources, our estimate of the jet orientation is highly reliable but for bent or complex source structures the estimate was poor or impossible in some cases. We were able to provide reliable jet angles to 61 out of the 75 resolved sources in our sample (31 radiative-mode and 30 jet-mode). In the worst cases, we still consider the large-scale jet angle to be accurate to within 20$^\circ$. 

\begin{figure} 
\centering
    \includegraphics[angle=0, clip=true, trim=0cm 0cm 0cm 0cm, width=.5\textwidth]{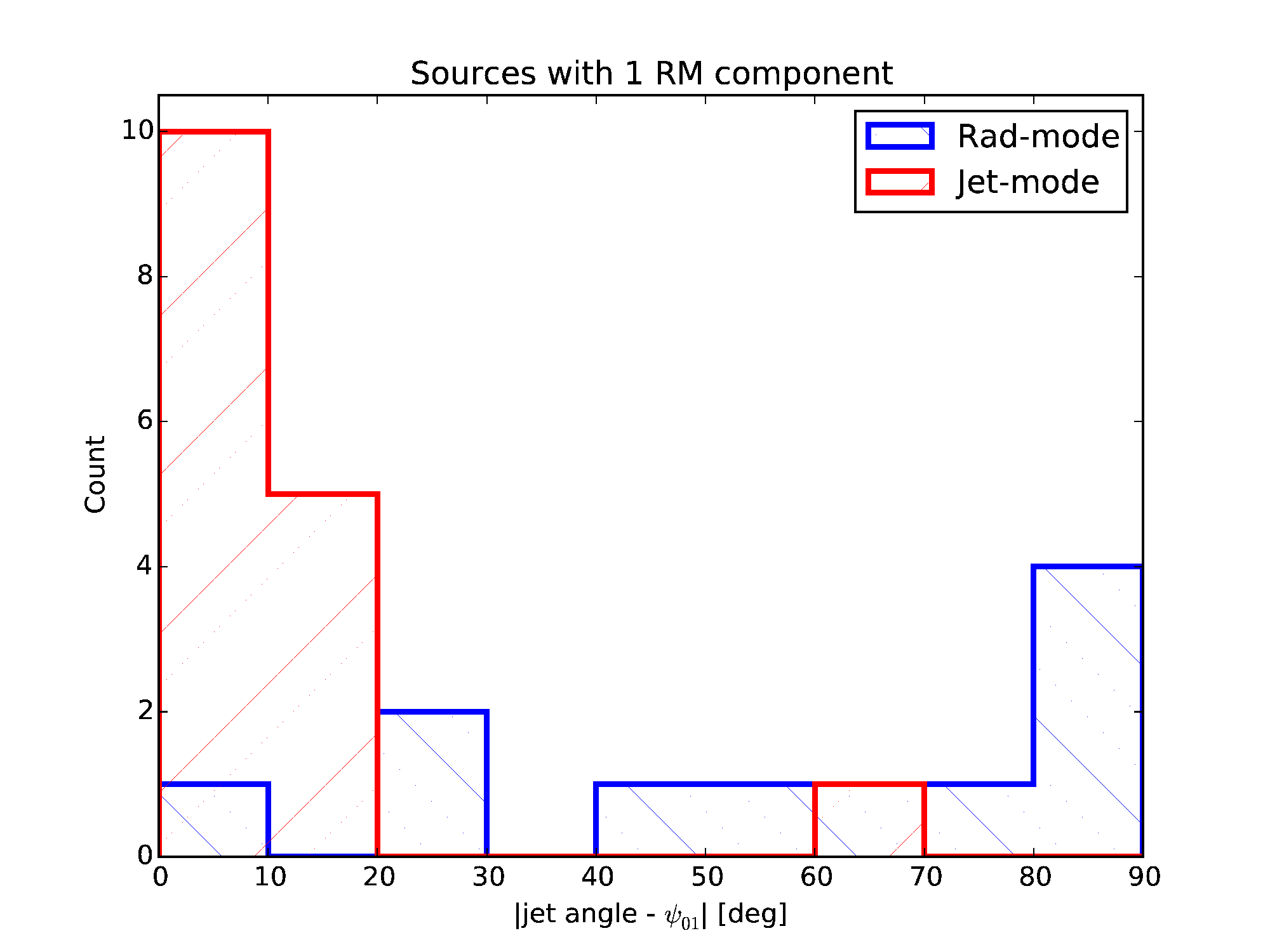} 
    \caption{ {\small Histogram of the absolute value of the difference in angle between the jet direction and the intrinsic polarization angle ($\psi_0$), 
    for one RM component sources only, and split into radiative-mode (blue) and jet-mode sources (red). } }
    \label{histjetpa1}
\end{figure} 

\begin{figure} 
\centering
    \includegraphics[angle=0, clip=true, trim=0cm 0cm 0cm 0cm, width=.5\textwidth]{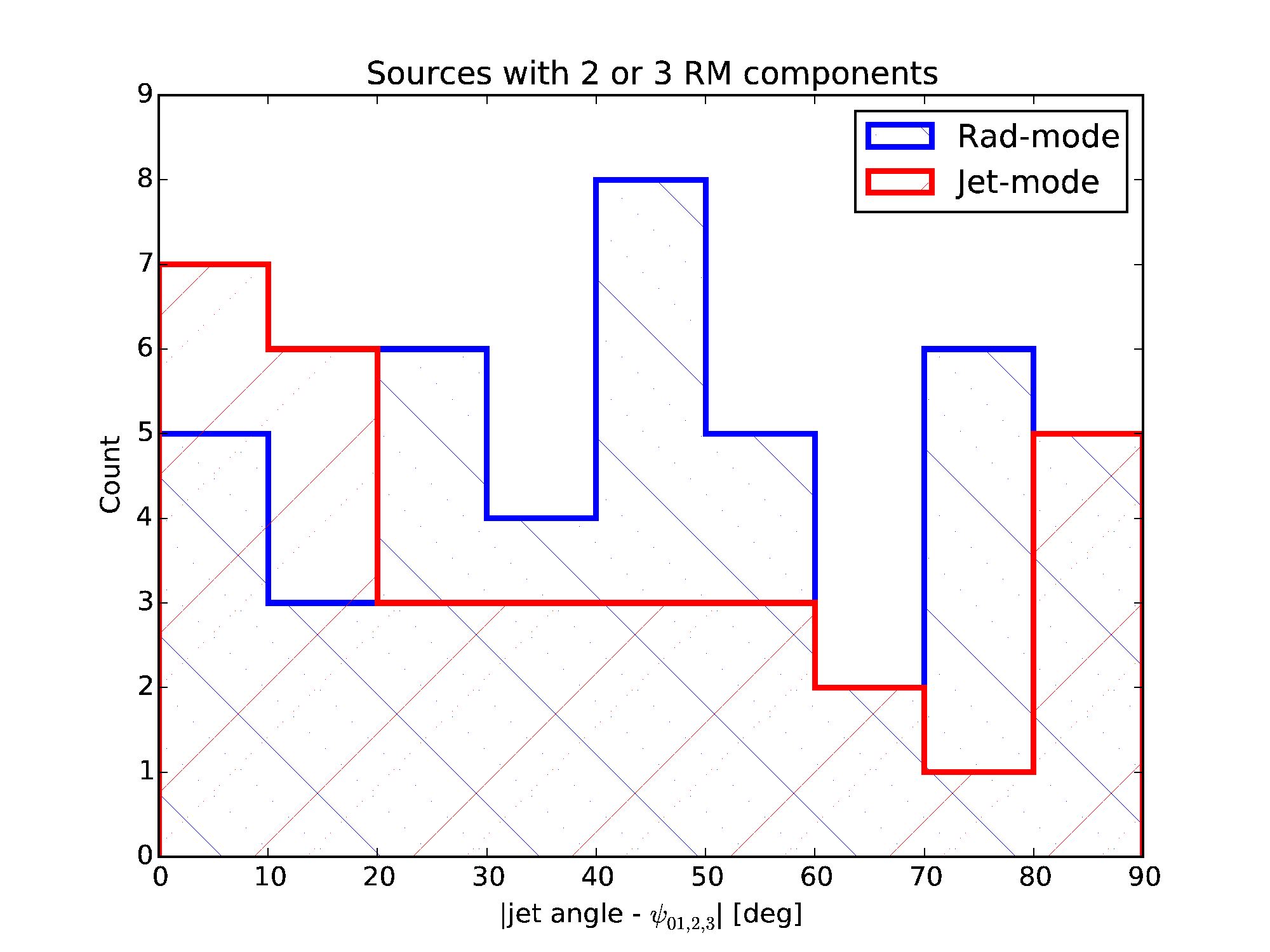} 
    \caption{ {\small Histogram of the absolute value of the difference in angle between the jet direction and the intrinsic polarization angle ($\psi_0$), 
    for two and three RM component sources only, and split into radiative-mode (blue) and jet-mode sources (red). } }
    \label{histjetpa23}
\end{figure} 

Figure~\ref{histjetpa1} shows a histogram of the absolute value of the difference between the jet angle and the intrinsic polarization angle, for one RM component sources. 
The histogram shows a clear peak near 0$^\circ$, but only for the jet-mode sources. A small excess of radiative-mode sources is seen at 90$^\circ$. The median error in $\psi_0$ for one RM component sources is 2$^\circ$.
Figure~\ref{histjetpa23} shows the same histogram but for two and three RM component sources. The peak near zero remains for the jet-mode sources but it is diluted, and a second peak appears at 90$^\circ$. There is no obvious trend for the radiative-mode sources, although there is a maximum in the histogram at $\sim$45$^\circ$. The median errors in $\psi_0$ for two and three RM component sources is 16$^\circ$ and 36$^\circ$, respectively.

For the jet-mode sources, $\sim$94\% of one RM component sources have intrinsic polarization angles that are within 20$^\circ$ of $|{\rm jet\,\,angle}-\psi_{0,1}|=0^\circ$, and $\sim$93\% of two and three RM component sources have at least one RM component with its polarization angle ($\psi_{0,1}$, $\psi_{0,2}$, or $\psi_{0,3}$) aligned within 20$^\circ$ of the jet direction. 
This means that the jet-mode sources preferentially have intrinsic magnetic field orientations ($\psi_{0}+90^\circ$) perpendicular to the jet direction. This result is compelling given that the intrinsic polarization angle is generally considered the most poorly constrained parameter in the model fitting. 
All the sources considered here are dominated by optically thin emission (since all the flat spectrum sources are unresolved and thus do not have a measured jet direction). 

\begin{figure} 
\centering
    \includegraphics[angle=0, clip=true, trim=0cm 0cm 0cm 0cm, width=.45\textwidth]{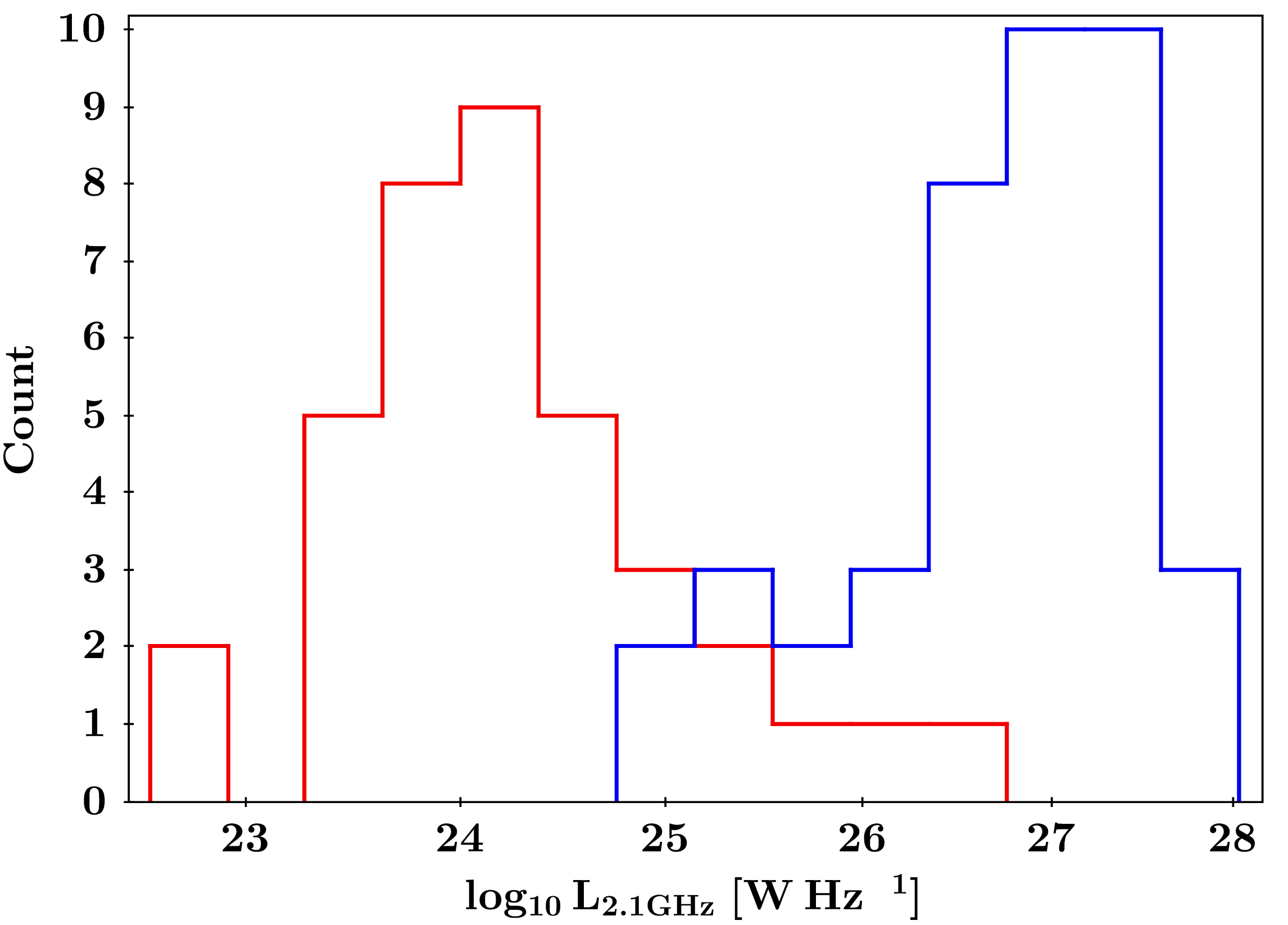} 
    \caption{ {\small Histogram of the radio luminosity at 2.1 GHz, split into radiative-mode (blue, right) and jet-mode (red, left) sources. 
    Only steep spectrum sources are included. } }
    \label{lumherglerg}
\end{figure} 

\subsubsection{Luminosity dependence}
Figure~\ref{lumherglerg} shows the luminosity distribution of the radiative-mode and jet-mode sources, with the low/high luminosity sources effectively corresponding to the jet-mode/radiative-mode sources with some overlap from $\sim10^{25} - 10^{27}$~W~Hz$^{-1}$, consistent with expectations from the 1.4~GHz radio luminosity function (e.g.~Pracy et al.~2016). 
There is a well studied anti-correlation between total intensity and $p_{1.4\, {\rm GHz}}$ for steep-spectrum sources (e.g.~Stil \& Keller 2015, Lamee et al.~2016), and since in flux-limited surveys the majority of faint sources have low luminosities, this anti-correlation can also be seen in luminosity (Banfield et al.~2011, Banfield et al.~2014, OS15). Additionally, OS15 found that this anti-correlation was driven primarily by the jet-mode sources. 
Due to the fact that the majority of jet-mode sources have low radio luminosity and the radiative-mode sources have high radio luminosity, our results in Section~\ref{radjetresults} can also be thought of in terms of the differences between low luminosity and high luminosity radio sources. However, since the accretion mode is related to the intrinsic jet power and hence the resultant radio luminosity (modulo the environment), we regard the luminosity as an observed effect/symptom, whereas the accretion mode and environment are the underlying physical causes that are responsible for the luminosity differences.
Therefore, we consider the radiative-mode/jet-mode division as the most fundamental division from which to understand the origin of the differences in the polarization properties of radio-loud AGN.

\subsection{Intrinsic polarization angle difference in two RM component models}
\begin{figure} 
\centering
    \includegraphics[angle=0, clip=true, trim=0cm 0cm 0cm 0cm, width=.5\textwidth]{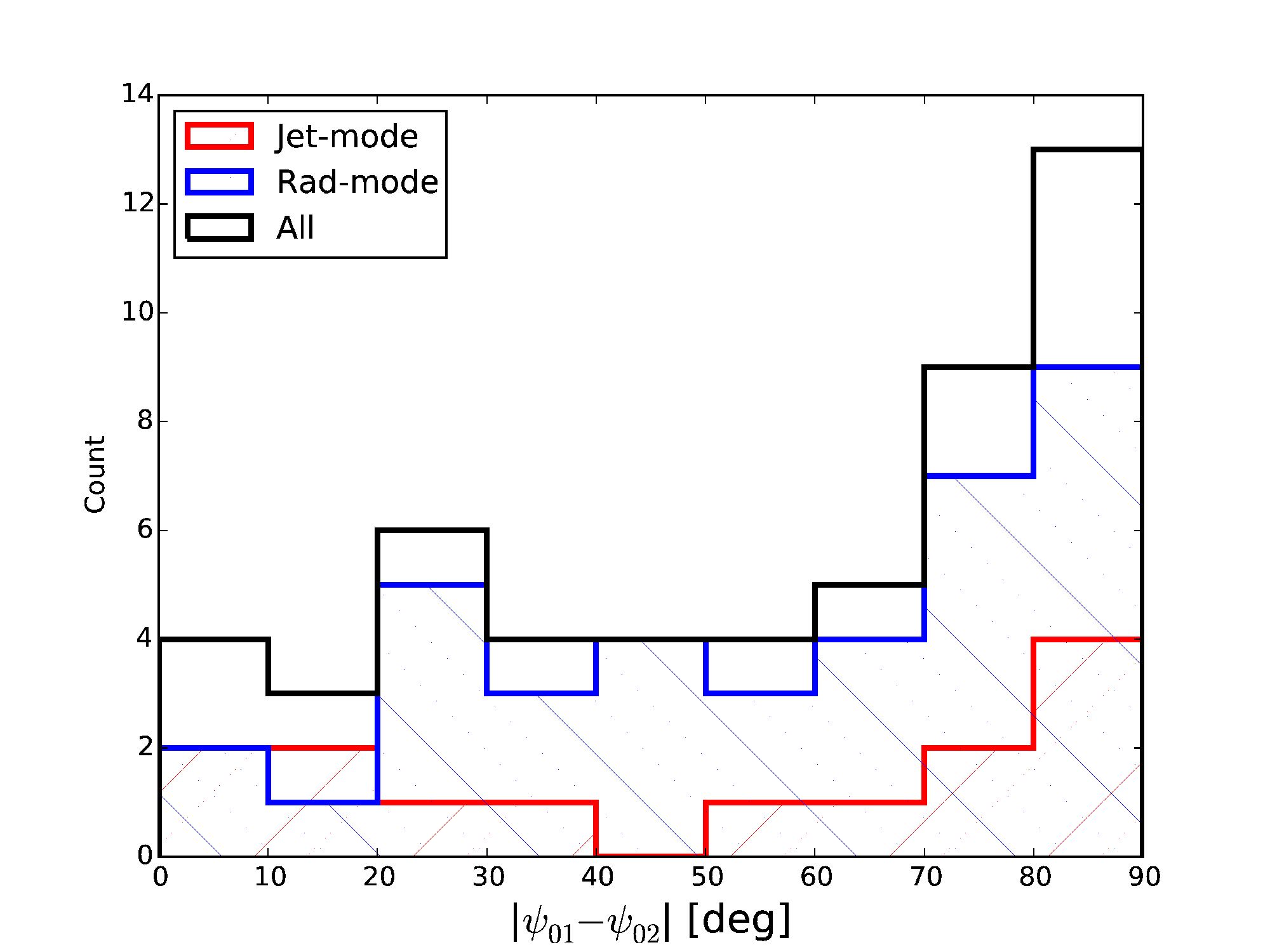} 
    \caption{ {\small Histogram of the difference between the intrinsic polarization angles ($\psi_{0,1}$, $\psi_{0,2}$) for two RM component sources 
    (Red: Jet-mode, Blue: Radiative-mode, Black: All). } }
    \label{histanglediff}
\end{figure}

A total of 52 sources have best-fit models with two RM components. In these cases it is interesting to see if there is any preference for intrinsic polarization angle differences in the best fit models. Figure~\ref{histanglediff} shows that two RM component models are most often found with a polarization angle difference close to 90$^\circ$. This effect may be important to consider with regard to the probability of fitting two RM components versus one RM component, and more generally for the statistics of polarization model-fitting results (e.g.~Farnsworth et al.~2011, Kumazaki et al.~2014, Sun et al.~2015). 
Figure~\ref{histanglediff} also shows that there is no difference in the distribution of $|\psi_{0,1}-\psi_{0,2}|$ for radiative-mode and jet-mode sources, other than radiative-mode sources having the majority of two RM component sources (Section~\ref{RMcmpnts}). 

\subsection{`Faraday thin' sources}
We find a total of 24 sources that are Faraday thin, in the sense that the RM components of these sources have sufficiently small Faraday depolarization ($\sigmaRM$, $\gradRM$) across our observing band to make it unmeasurable. For example, from Figure~\ref{sigmaRMp1400cmpnts} it appears this cutoff is approximately 5~rad~m$^{-2}$ in $\sigmaRMwtd$ for most sources. Of these Faraday thin sources, 10 have one RM component, 11 have two RM components and 3 have three RM components. There does not appear to be any strong preference for a particular source morphology for Faraday thin sources, with 5/8/3/6/2 sources with unresolved/double/double$+$core/extended/complex morphologies. There are also roughly equal percentages of Faraday thin sources with a flat spectrum (23\%) and a steep spectrum (24\%). There is a slight preference for more Faraday thin sources in the radiative-mode class (62.5\%) than the jet-mode class (37.5\%). This preference might be explained by a higher fraction of FRII sources expected in the radiative-mode class (Best et al.~2009); where the extended, compact hotspots may be expected to have small enough RM dispersions to be classified as Faraday thin in our observations.  

\subsection{RM gradients: the $\DeltaRM$ parameter}
For nine sources (7 radiative-mode, 2 jet-mode), the best-fit model requires the $\DeltaRM$ parameter in Eqn.~\ref{modeleqn}, and not $\sigmaRM$. This means that the Faraday depolarization is best described by a uniform magnetic field component, and broadly speaking, could physically mean either the presence of internal Faraday rotation or an external gradient in Faraday rotation across the emission region. Interestingly, six of these sources (J0009-3216, J0216-3247, J0222-3441, J0229-3643, J0326-3243, J0336-3616) are unresolved, flat spectrum sources (i.e.~blazars). The median value of $\gradRMwtd$ for these sources is 50~rad~m$^{-2}$.
Two of the three steep-spectrum sources (J0342-3703: Two RM components, J1301-3226: Three RM components) have well-resolved double-lobed structures in the 4~cm band images (Fig.~\ref{images3}q, \ref{images5}b), while the third steep-spectrum source (J0021-3334: Two RM components) is only marginally resolved (Fig.~\ref{images1}h). The median value of $\gradRMwtd$ for these sources is 30~rad~m$^{-2}$, and in all cases only one of the RM components has a non-zero $\DeltaRM$ value (i.e.~meaning a large difference in Faraday depolarization between the RM components). 

\subsection{Broadband polarization SEDs}

The broadband, sparsely-sampled, integrated fractional polarization behaviour versus wavelength, $p(\lambda)$, of radio sources has been studied over many years (e.g.,~Conway et al. 1974,~Farnes et al.~2014a (FGC14);~Pasetto et al.~2016). Interpreting the behaviour of $p(\lambda)$ has been difficult due to the lack of quasi-continuous and simultaneous wavelength coverage as well as spatially resolved information for large samples of radio sources. Recently, FGC14 complied one of the largest catalogs of broadband polarization data for 951 sources from major radio surveys and other polarization data published over the last 50 years (with from 3 to 56 measurements for individual sources ranging from 400 MHz up to 100 GHz). 

The common expectation for $p(\lambda)$ behaviour is a monotonic decrease with increasing wavelength (depolarization). However, 21\% of sources in the FGC14 catalog show re-polarization (increasing $p(\lambda)$ with increasing wavelength). FGC14 parameterised this behaviour by fitting a power-law to the data, $p(\lambda)\propto\lambda^\beta$, where $\beta < 0$ corresponds to depolarization and $\beta > 0$ to re-polarization. 
FGC14 were also able to divide their sample into flat and steep total-intensity spectrum sources, with $p(\lambda)$ displaying different behaviours for the two types of sources. Most of the steep spectrum sources had $\beta < 0$ while $\sim$50\% of the flat spectrum sources had $\beta >0$. 

\begin{figure} 
\centering
    \includegraphics[angle=0, clip=true, trim=0cm 0cm 0cm 0cm, width=.5\textwidth]{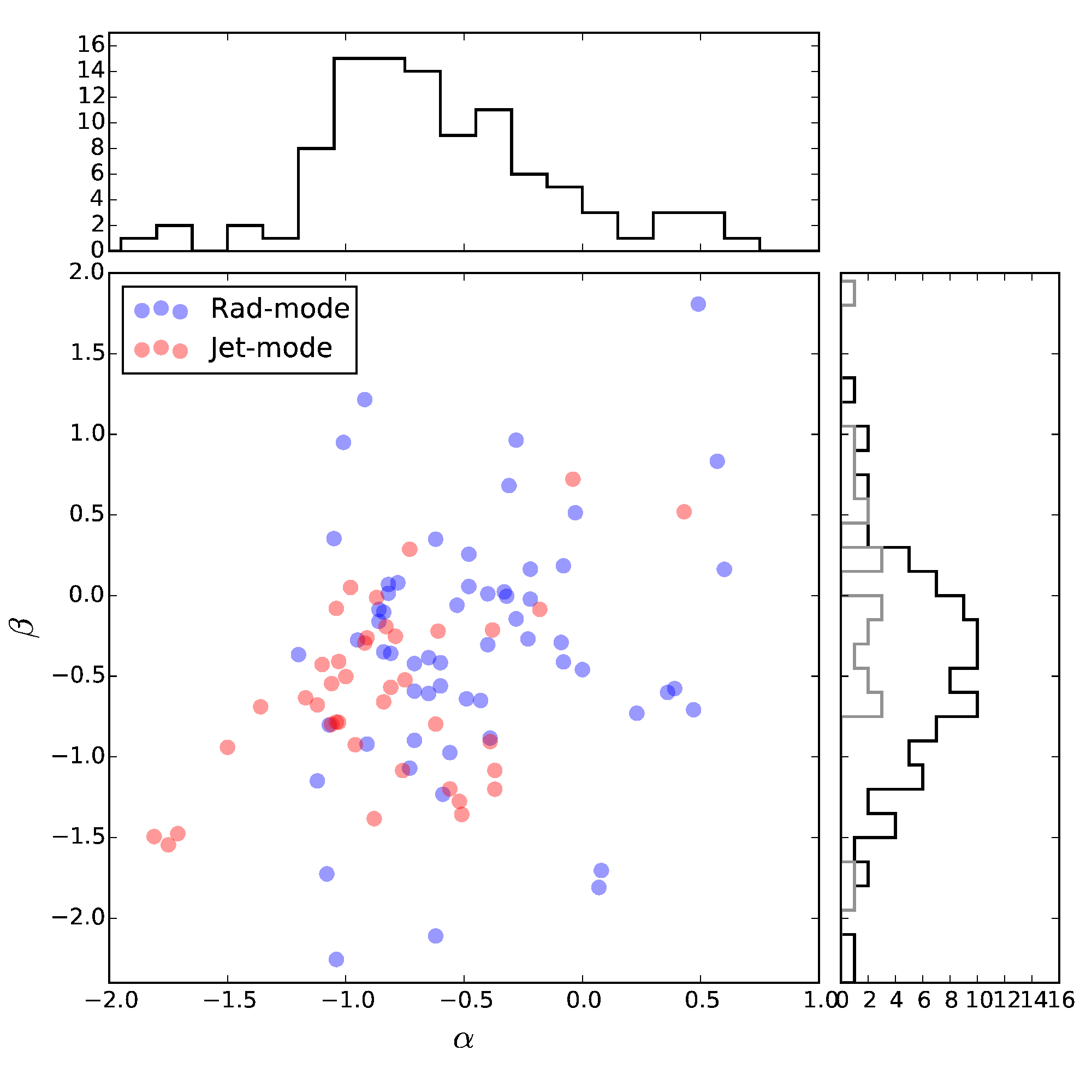} 
    \caption{ {\small Plot of the power-law index, $\beta$, of $p(\lambda)\propto\lambda^\beta$ versus the total intensity spectral index, $\alpha$, 
    with radiative-mode sources in blue and jet-mode sources in red. Histograms of $\alpha$ and $\beta$ are shown on the top and on the right, respectively. 
    The $\beta$ histogram also shows the flat-spectrum sources (grey line), with all sources shown by the black line. } }
    \label{spixbeta}
\end{figure}

Here we compare our results with FGC14 by also fitting a power-law to the 16~cm band polarization data, in an attempt to identify the main causes of the different $p(\lambda)$ behaviours. 
Figure~\ref{spixbeta} shows $\alpha$ versus $\beta$ for all sources in our sample (for direct comparison with FGC14, fig.~7), while also color-coding the points to identify the radiative-mode and jet-mode sources. Overall, we have fewer flat spectrum sources in our sample compared to FGC14 (22\% vs.~$\sim$35\%). This is not unexpected given that the FGC14 catalog has a significant contribution from data at 20 GHz, where flat spectrum sources become more prominent (for a fixed flux density limit) due to both their spectral index and approximately constant $p(\lambda)$. However, we still find that 23\% of all our sources have $\beta > 0$, in excellent agreement with the FGC14 results. Similar to FGC14, we also find that a greater fraction of flat spectrum sources have $\beta>0$ than steep spectrum sources (39\% versus 18\%, respectively). 

\begin{figure} 
\centering
    \includegraphics[angle=0, clip=true, trim=0cm 0cm 0cm 0cm, width=.5\textwidth]{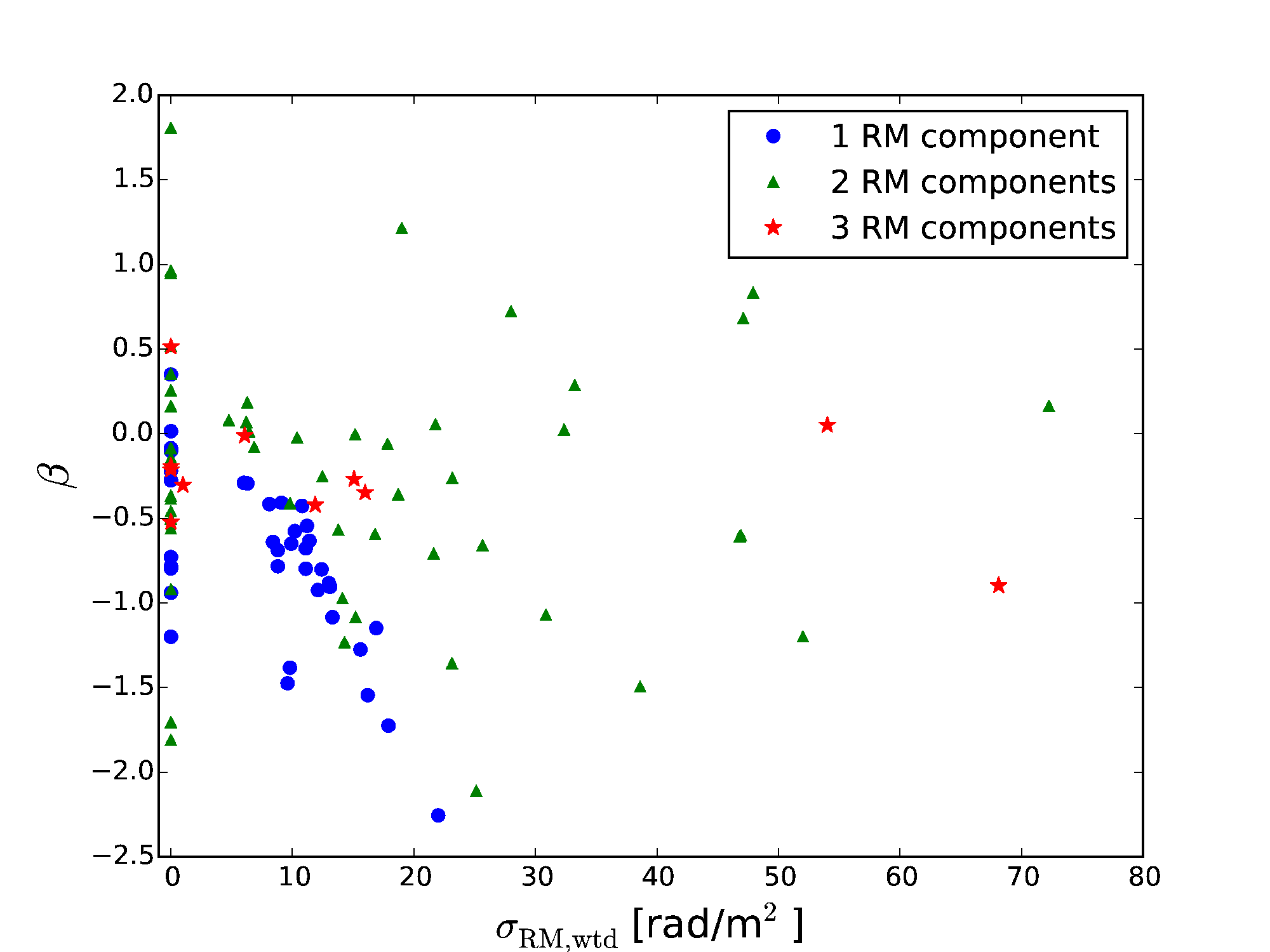} 
    \caption{ {\small Plot of $\beta$ versus the polarization-weighted RM dispersion ($\sigmaRMwtd$). Sources best fit by one RM component 
    are plotted as blue circles, two RM component sources with green triangles and three RM components with red stars. } }
    \label{sigmaRMbeta}
\end{figure}  

In Figure~\ref{sigmaRMbeta}, we plot $\beta$ versus $\sigmaRMwtd$ and color-code the points by the number of RM components to better identify the causes of different $p(\lambda)$ behaviours. For sources with one RM component, $\beta$ is anti-correlated with $\sigmaRMwtd$ (cc: $-0.6$, p-value: $10^{-5}$) showing that Faraday depolarization is the dominant effect in this case. The sources that deviate from this anti-correlation have two or three RM components, clearly showing that the cause of $\beta>0$ is multiple RM components in a source (as discussed as the most likely scenario in FGC14).

\begin{figure} 
\centering
    \includegraphics[angle=0, clip=true, trim=0cm 0.2cm 0cm 1cm, width=.5\textwidth]{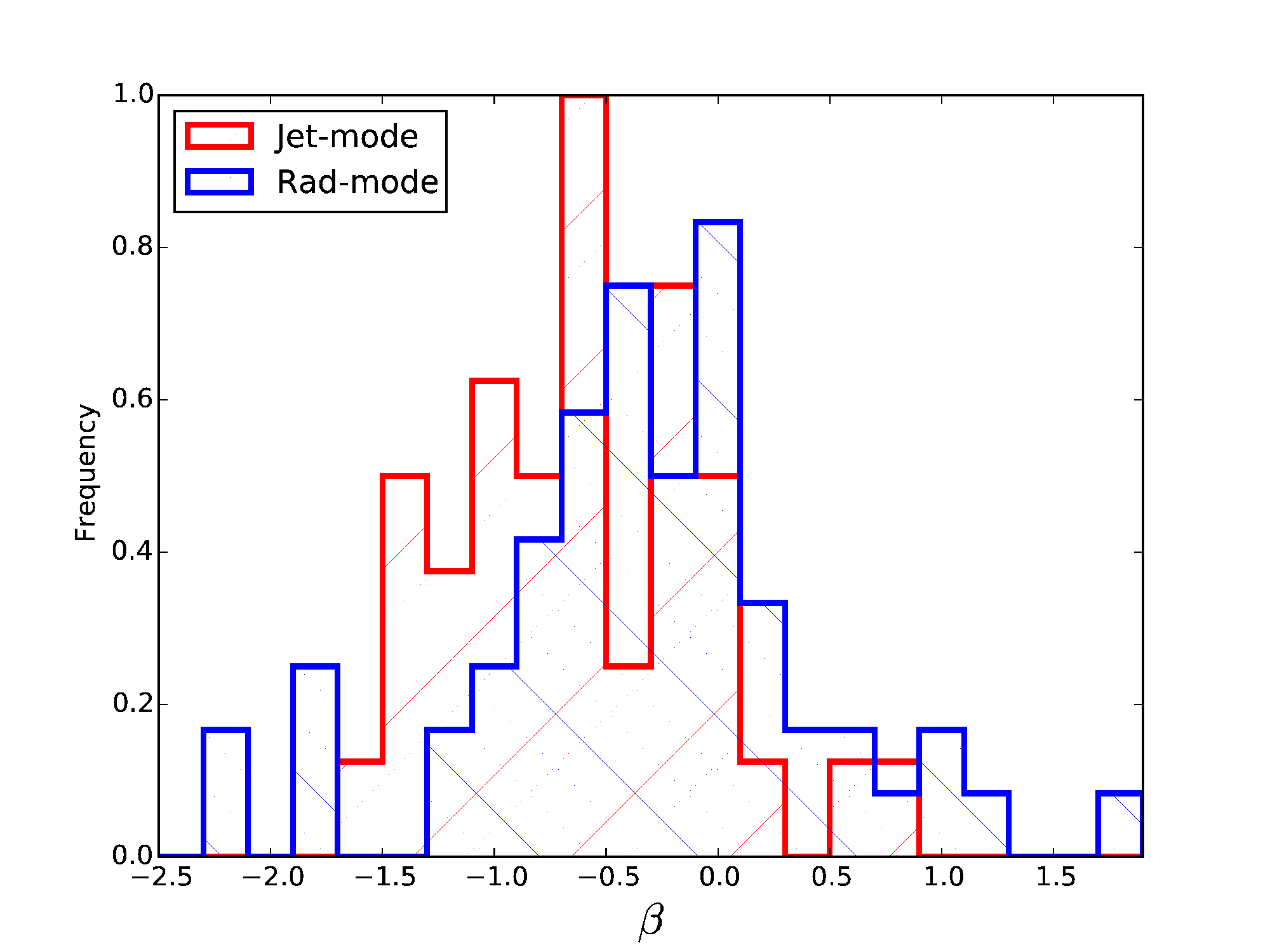} 
    \caption{ {\small Histogram of the power-law index, $\beta$, of $p(\lambda)\propto\lambda^\beta$, split into radiative-mode (blue) and 
    jet-mode sources (red). } }
    \label{histbeta}
\end{figure}

FGC14 also suggested that there was a bimodal distribution in $\beta$ for steep spectrum sources (i.e.~two populations of depolarizing sources). In Figure~\ref{histbeta}, we show the distribution of $\beta$ for radiative-mode and jet-mode sources. By eye, the peaks of the histograms for the two types of sources appear to be shifted from each other, with a median value of $\beta_{\rm med} = -0.36$ ($\beta_{\rm med} = -0.68$) for steep-spectrum radiative-mode (jet-mode) sources. However, a KS test indicates that this difference is not statistically significant (p-value: 0.03). Larger samples are required to determine if there really are two populations of depolarizing steep-spectrum sources, and that these correspond to the radiative-mode and jet-mode division. 

\subsection{Redshift evolution} \label{sec_z}
\begin{figure} 
\centering
    \includegraphics[angle=0, clip=true, trim=0cm 0.1cm 0cm 1cm, width=.5\textwidth]{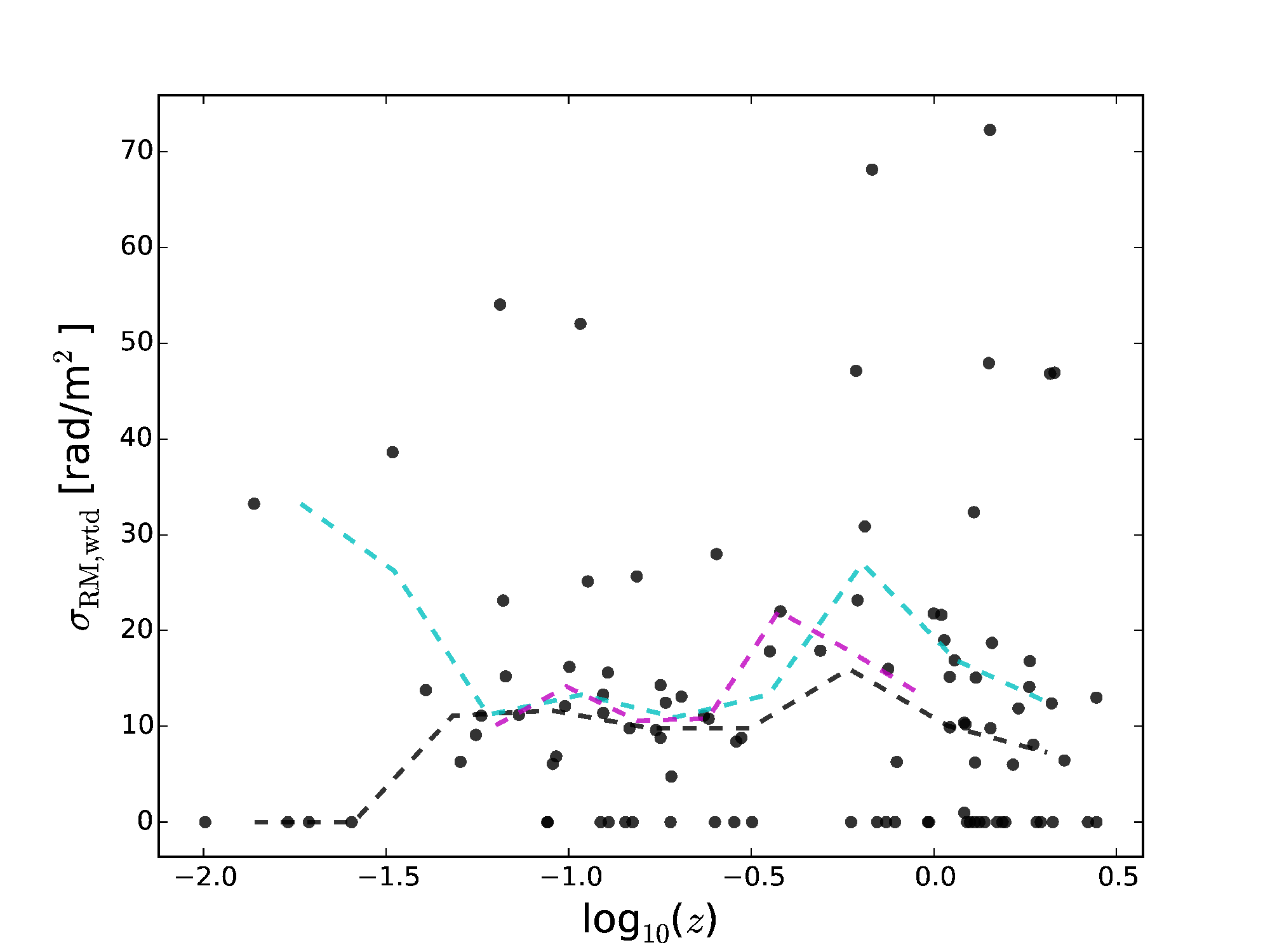} 
    \caption{ {\small Plot of the polarization-weighted RM dispersion ($\sigmaRMwtd$) versus the redshift ($z$) for all sources. 
    Running medians shown for all sources (black dashed line), for steep spectrum sources only (cyan dashed line), and for steep-spectrum 
    sources with only one RM component (magenta dashed line). } }
    \label{sigmaRMz}
\end{figure}

Considering the wide redshift range of our sample ($0.01 < z < 2.8$), it is worth investigating whether or not we can measure any redshift evolution in the magneto-ionic properties of our sample. Figure~\ref{sigmaRMz} shows $\sigmaRMwtd$ versus $log_{10}(z)$ for the full sample. 
We use a Spearman rank correlation test to determine the correlation coefficient (cc) and the significance of the correlation (in terms of the p-value). 
The black-dashed line shows the running median for all sources (cc: $-0.05$, p-value: 0.59). The cyan-dashed line shows the running median for only steep-spectrum sources ($\alpha < -0.3$) and excludes Faraday thin sources (cc: $-0.00$, p-value: 0.97). 
Finally, the magenta-dashed line considers only sources with one RM component, with a steep-spectrum and again excluding the Faraday thin sources (cc: 0.17, p-value: 0.43). 
In all cases, the Spearman rank correlation test finds no evidence for a correlation of $\sigmaRMwtd$ versus redshift. 

\begin{figure} 
\centering
    \includegraphics[angle=0, clip=true, trim=0cm 0.1cm 0cm 1cm, width=.5\textwidth]{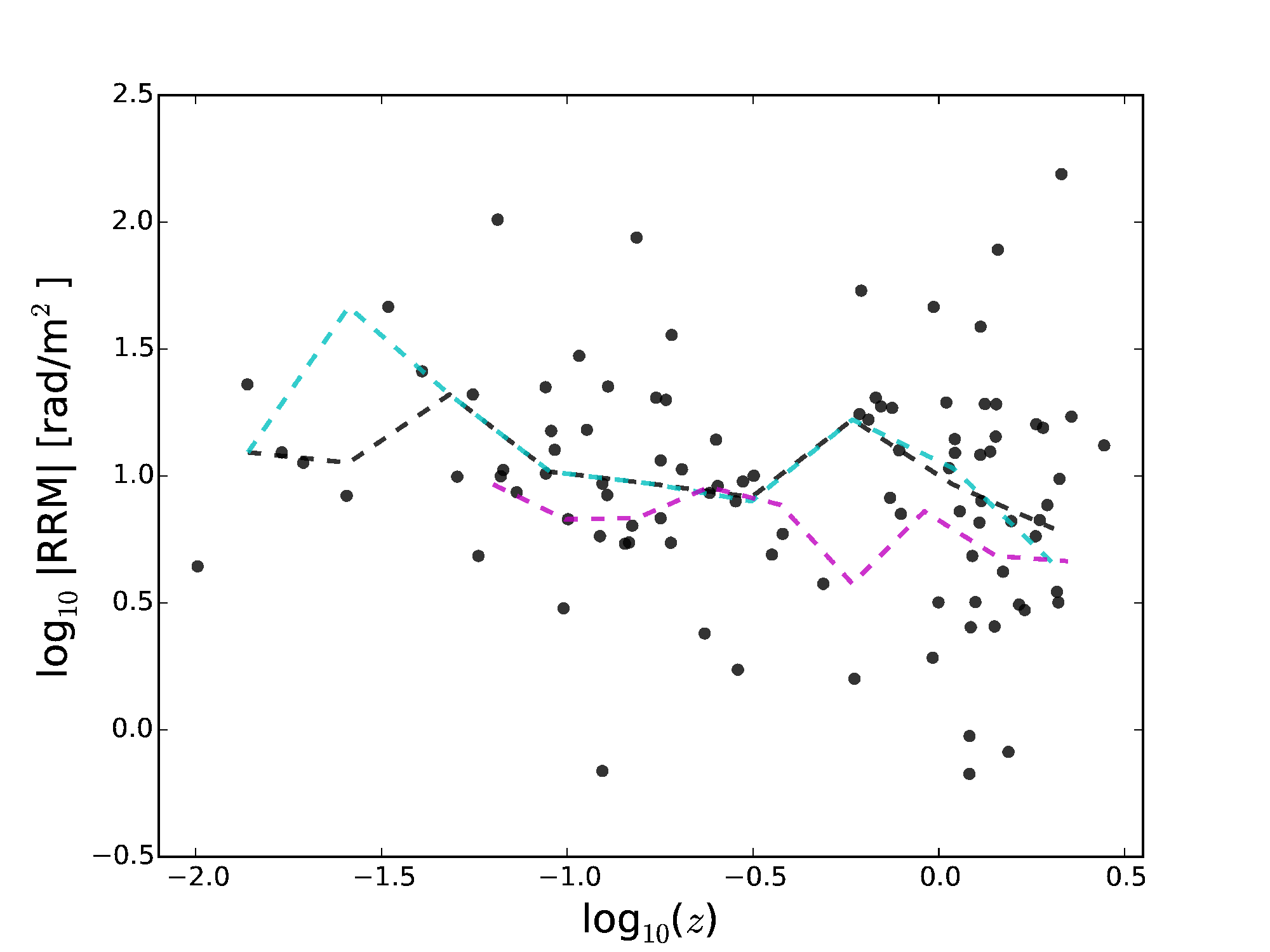} 
    \caption{ {\small Plot of the residual rotation measure (RRM) versus the redshift ($z$) for all sources. 
    Running medians shown for all sources (black dashed line), for steep spectrum sources only (cyan dashed line), and for steep-spectrum 
    sources with only one RM component (magenta dashed line). } }
    \label{RRMz}
\end{figure}

In Figure~\ref{RRMz}, we show the absolute value of the RRM versus redshift. The dashed lines represent all sources (black), only steep spectrum sources (cyan) and only steep spectrum sources with one RM component (magenta). For all sources, there is an slight indication of an anti-correlation of RRM with redshift (cc: $-0.2$, p-value: 0.04). However, there is a more significant anti-correlation between GRM and redshift (cc: $-0.2$, p-value: 0.01), and no correlation between RM and redshift (cc: $-0.06$, p-value 0.5), indicating that the GRM signature has remained in the RRM. For the steep spectrum sources and one RM component sources, the results are similar.
We also checked for a correlation between the difference in RM for the two RM component sources (${\rm dRM} = |{\rm RM}_1 - {\rm RM}_2|$) and redshift, finding no evidence for a correlation (cc: 0.01, p-value: 0.95). 

\subsection{Summary}
Here we summarise the main results from our study. 
\begin{enumerate}
\item We have implemented a general QU-fitting algorithm which provides accurate model fits to the broadband polarization (1 to 3 GHz) behaviour of the 100 sources in our sample. The signal-to-noise ratio (S/N) in polarization ranges from 10 to 70, with a median S/N of 35.5. 
 \item The majority of sources are best fit by a two RM component model (52\%), with the fraction increasing at higher S/N ($\sim$62\% with ${\rm S/N} > 35.5$). A total of 11\% of sources require a three RM component model ($\sim$20\% for ${\rm S/N} > 35.5$), while 37\% are best fit by one RM component ($\sim$18\% for ${\rm S/N} > 35.5$).
 \item Almost all steep spectrum sources are spatially resolved ($\sim$94\%), with a median linear size of $l_{\rm med}=102$~kpc. Of the sources that remain unresolved (at $\sim$3''), 80\% are flat spectrum sources. 
 \item There is no simple relationship between the total intensity morphology of a source and the number of RM components. 
 \item We find no evidence for a correlation between the linear size of a source and its magneto-ionic properties ($\sigmaRMwtd$, RRM) or the mean intrinsic degree of polarization, $\langle p_0 \rangle$. 
 \item 24\% of our sample have best-fit `Faraday thin' models (i.e.~$\sigmaRMwtd=0$ and $\DeltaRMwtd=0$). Excluding the Faraday thin sources, the median values of $\sigmaRMwtd$ and $\gradRMwtd$ are 14.1~rad~m$^{-2}$ and 60.6~rad~m$^{-2}$, with median errors of 1.8~rad~m$^{-2}$, and 3.3~rad~m$^{-2}$, respectively. The median value of the RRM is 9.7~rad~m$^{-2}$, with a median error of 7.5~rad~m$^{-2}$.
 \item Only 9\% of sources have best-fit models that require the $\gradRM$ parameter for Faraday depolarization. The majority (6/9) of these sources are flat spectrum sources, and the median value of $\gradRMwtd$ is 50~rad~m$^{-2}$. The three steep spectrum sources have a lower median $\gradRMwtd$ of 30~rad~m$^{-2}$. 
 \item The median values of $\langle p_0 \rangle$ are 6.6\% for steep spectrum sources and 2.6\% for flat spectrum sources.  
 \item We find an anti-correlation between the degree of polarization at 1.4~GHz ($p_{\rm 1.4\,GHz}$) and Faraday depolarization ($\sigmaRMwtd$), with a Spearman rank correlation-coefficient of $-0.44$ (p-value of 0.0002). 
 \item Of the steep spectrum sources, 37 are classified as jet-mode AGN and 41 as radiative-mode AGN. The jet-mode AGN have significantly higher values of $\langle p_0 \rangle$ than radiative-mode AGN (KS test p-value: $10^{-4}$). No significant difference is observed in magneto-ionic properties ($\sigmaRMwtd$, RRM) between jet-mode and radiative-mode AGN. Therefore, $\langle p_0 \rangle$ is the dominant variable in the observed difference in $p_{\rm 1.4\,GHz}$ between radiative-mode and jet-mode AGN (OS15). 
 \item The intrinsic polarization angle ($\psi_0$) for jet-mode AGN shows a clear preference for being aligned with the jet direction ($\sim$94\% of jet-mode sources with $\psi_0$ within 20$^\circ$ of jet direction), while there is no clear preference for radiative-mode AGN. 
 \item The full range of power-law behaviour of sparsely sampled, broadband polarization SEDs (Farnes et al.~2014a) is well understood from our sample in terms of Faraday depolarization and the presence of multiple RM components. We also find marginal evidence for two populations of depolarizing sources, that may correspond to the radiative-mode/jet-mode division. 
 \item We find no evidence for a redshift evolution in the magneto-ionic properties ($\sigmaRMwtd$, RRM) of our sample ($0.01 < z < 2.8$).
\end{enumerate}
It should be kept in mind that these results are for a relatively small sample of highly-polarized AGN. In future, significantly larger samples ($10^3$ to $10^6$ polarized sources) will allow much more robust statistics on the broadband polarization and Faraday rotation properties of extragalactic radio sources.

\section{Discussion}
\subsection{Extracting physically meaningful parameters from broadband polarization data}
In the coming era of large-area radio surveys with modest angular resolution $O(10")$, the ability to accurately characterise the broadband spectro-polarimetric behaviour of the integrated emission from extragalactic radio sources is crucial for a wide range of scientific goals related to radio galaxy physics and cosmic magnetism.\footnote{https://www.skatelescope.org/wp-content/uploads/2011/03/SKA-Astophysics-Vol1.pdf} 
In this paper, we have introduced a physically meaningful model (Section~\ref{model}) 
that captures a wide range of possible polarization and Faraday rotation behaviour. The success of our model-fitting approach shows that we are able to reliably characterise the broadband polarization behaviour of a wide range of source types, enabling us to disentangle the intrinsic magnetic field properties ($p_0$, $\psi_0$) of individual sources from the magneto-ionic material causing the Faraday rotation and depolarization (RM, $\sigmaRM$, $\DeltaRM$). 

The results of this work are valid for low to medium signal-to-noise ratio in polarization ($10<{\rm S/N}<70$). We fit up to three RM components to each source, and to account for the dependence of the number of RM components on S/N, we have adopted polarization-weighted parameters (RM$_{\rm wtd}$, $\sigmaRMwtd$, $\gradRMwtd$) for comparison between sources. 
It is possible that higher signal-to-noise ratio detections may require more detailed model-fitting approaches (such as those that would describe skewed or more complex Faraday dispersion functions). Anderson et al.~(2016) used ``super-Gaussian'' functions to describe Faraday dispersion functions (FDF) ranging from a Gaussian shape to effectively a top-hat shape. This approach proved very successful for accurately describing a wide range of complex polarization behaviour in data from 1 to 10 GHz. In principle, our approach using Eqn.~\ref{modeleqn} can describe the same range of behaviours, with the exponential-$\sigmaRM$ part representing a Gaussian FDF and the sinc-$\DeltaRM$ part representing a top-hat FDF, with combinations of these parameters describing the behaviour in between. However, in practice, none of our sources require both $\sigmaRM$ and $\DeltaRM$ parameters in a single RM component. This may indicate a limitation of the model-fitting procedure in that it cannot easily differentiate between the exponential-$\sigmaRM$ behaviour and the sinc-$\DeltaRM$ behaviour in $p(\lambda^2)$ over a limited range in $\lambda^2$ (or possibly that the BIC model-selection criteria penalises additional parameters too heavily). Therefore, further work is required to determine which of these methods (or an alternative) provides the optimum approach. 

\subsection{Origin of the difference in $p_{1.4\,{\rm GHz}}$ between radiative-mode and jet-mode AGN}
OS15 found a significant difference in the integrated degree of polarization at 1.4~GHz ($p_{1.4\,{\rm GHz}}$) between a large sample of radiative-mode and jet-mode AGN, with the jet-mode sources extending up to values of $p_{1.4\,{\rm GHz}} \lesssim 30\%$, while the radiative-mode AGN were restricted to $p_{1.4\,{\rm GHz}} \lesssim 15\%$. OS15 discussed the origin of this observed difference in terms of the intrinsic magnetic field properties and the large scale magnetised environment of the two classes of sources. Without being able to directly distinguish between depolarization and intrinsic magnetic field disorder due to the lack of broadband polarization data, OS15 were limited to indirect measures in order to differentiate between the two. Due to the observed effects of the radio morphology and the local galaxy-density environment on $p_{1.4\,{\rm GHz}}$, in combination with knowledge from the literature on the range of environments of radiative-mode and jet-mode AGN, OS15 suggested that the observed difference in $p_{1.4\,{\rm GHz}}$ was most likely due to the local environments of the radio sources, in terms of both the ambient gas density and the magnetoionic properties of this gas. 

Our analysis shows that jet-mode sources with high integrated polarization at 1.4 GHz have low RM dispersion, are typically dominated by a single RM component, and have a steep spectral index. We find no significant difference in the magneto-ionic properties of jet-mode AGN and radiative-mode AGN. Indeed, it is the intrinsically disordered magnetic field structures of the radiative-mode AGN that leads to the lower integrated polarization seen at 1.4 GHz (see Section~\ref{radjetresults}). 
Therefore, Faraday depolarization is not the dominant effect causing the observed difference in $p_{1.4\,{\rm GHz}}$ between radiative-mode and jet-mode AGN. The dominant effect of the intrinsic magnetic field disorder can be due to intrinsically random magnetic fields in a straight jet and/or uniform fields in a bent jet structure. It remains unclear which of these effects is key, and what role the density of the environment plays in influencing the morphology and overall magnetic field structure of the jet. 
This result is similar to the conclusion of Lamee et al.~(2016) in which they found that the intrinsic magnetic field disorder, and not depolarization, was the dominant effect producing the observed integrated degrees of polarization at 2.3~GHz and 1.4~GHz. 

As the majority of jet-mode sources have lower radio luminosity than the radiative-mode sources, these differences can also be seen as differences in the polarization properties of high and low radio-luminosity sources (e.g.~Banfield et al.~2011). 
Further investigation is required, but one possibility is that the low luminosity (jet-mode) sources that can achieve high integrated degrees of polarization are predominantly FRI radio galaxies with jets that do not have any significant bends in the jet structure out to large scales. Their jets may also be orientated in the plane of the sky such that the Faraday dispersion function of the two radio jet/lobes is very similar (requiring just one RM component). Furthermore, the `double$+$core' morphology jet-mode sources have the highest values of $\langle p_0 \rangle$ (Section~\ref{radjetresults}). These sources are mostly like FRI sources in which the inner jet dominates the integrated polarized emission (e.g.~Figures~\ref{images1}g, \ref{images1}i, \ref{images3}m). Although there are exceptions to this (see Figure~\ref{images5}i). 

Since the inner jet is expected to have a highly ordered magnetic field structure (e.g.~Laing \& Bridle 2014) this may help explain the high values of $\langle p_0 \rangle$. Also, Kharb et al.~(2005) found that the cores of FRI sources had high degrees of polarization on parsec-scales, in contrast to the weak or undetectable polarization in FRII cores. 
FRI source morphologies become less common at higher luminosities with FRII sources tending to dominate (e.g.~Best et al.~2009), and those high-luminosity sources that have FRI morphology are typically found in rich galaxy group and cluster environments (e.g.~Gendre et al.~2013), where the jet can be strongly disturbed and the Faraday depolarization is much higher. Indeed, the sources with low values of $\langle p_0 \rangle$ and `double+core' morphology tend to be more like FRII sources (e.g.~Figure~\ref{images1}k, \ref{images2}g, \ref{images2}q). 

In general, the angular resolution is too low at 16~cm and the sensitivity and image quality is too poor at 4cm to definitively classify all sources into FRI and FRII classes. 
High fidelity, broadband polarization imaging at higher angular resolution for a larger sample of sources (e.g.~as will be provided by the VLASS) is required to determine the true nature of the difference in $\langle p_0 \rangle$ between radiative-mode and jet-mode sources. 

\subsection{Depolarization models}
\subsubsection{External Faraday dispersion}
The most common type of model used in the literature to describe the depolarization of radio galaxy emission is known at `External Faraday Dispersion' or the `Burn-law' model (Burn 1966, eq.~23). In this model, the degree of polarization is modified due to a Gaussian distribution of Faraday rotating cells external to the emission region. The most likely origin of this Faraday rotation material is local to the source, for example in the magnetised intracluster medium (e.g.~Laing et al.~2008)\footnote{The `Burn-law' model is considered unlikely to hold at longer wavelengths than observed here (Tribble 1991).}. 
Our Eqn.~\ref{modeleqn} is equivalent to this model when $\DeltaRM=0$ and $\sigmaRM\ne0$. In the majority of cases (67\% of all sources, 72\% of steep spectrum sources) the observed depolarization in our sample is well described by external Faraday dispersion, even though in some cases the integrated $p(\lambda^2)$ can display much more complicated behaviour than the simple exponential decrease that is expected for a well resolved emission region. In our model-fitting, this complicated behaviour is described in terms of multiple RM components (i.e.~distinct emission regions of the source with different polarization and/or Faraday dispersion properties). Therefore, one would expect that the integrated emission from a double-lobed radio galaxy should be described by two RM components. This is true for the majority of sources we can identify as sources with distinct double-lobed structure ($\sim$$71\%$, Table~\ref{cmpntsmorph}), with high signal-to-noise ratio in polarization. 

The `double' sources best described by two RM components possibly have their lobes inclined in the plane of the sky such that the polarized emission from the more distant lobe will pass through more magneto-ionic material and will therefore have higher RM dispersion (Laing 1988, Garrington et al.~1988). Alternatively, large asymmetries in the source environment may influence the jet and lobe propagation such that their morphological and intrinsic polarization properties may be significantly different (e.g.~van Breugel et al.~1985).

Of the eleven `double' sources with two RM components, five are Faraday thin. These may be FRII sources whose compact hotspots dominate the polarized emission and have little Faraday depolarization. Of the remaining six sources, five display behaviour more typical of the `Laing-Garrington' effect with large asymmetries in Faraday depolarization between the two RM components, with only one having very different values of $p_0$. 
The `double' sources that are best described by one RM component (Table~\ref{cmpntsmorph}) may have their lobes orientated such that their polarized emission propagates through approximately equal amounts of magneto-ionic material, or one lobe may be completely depolarized across our observing band. 

\subsubsection{Rotation measure gradients}
Only a minority (9\%) of sources show depolarization that is best described by the sinc-$\DeltaRM$ function (with $\sigmaRM=0$), and only 3\% are steep-spectrum sources. In the case of a well resolved emission region, such depolarization behaviour is usually associated with internal Faraday rotation (Burn 1966), however for unresolved sources it is also possible that a smooth gradient in RM across the emission region is causing this behaviour (Sokoloff et al.~1998). 

It is interesting that most sources that show this behaviour are flat spectrum sources (i.e.~blazars). It is worth noting that the applicability of our model for optically thick or synchrotron self-absorbed regions is theoretically questionable since it does not account for the intrinsic degree of polarization varying as a function of frequency due to the unity optical depth surface moving further upstream in the jet at higher frequencies. We suspect that such an effect is not a large concern over our frequency range (1 to 3 GHz) but could certainly be important over much larger ranges of frequency (e.g.~O'Sullivan \& Gabuzda.~2009b). 

Smooth gradients in Faraday rotation both perpendicular to and parallel to blazar jets on parsec scales have been found by many studies (e.g.~Asada et al.~2002, O'Sullivan \& Gabuzda 2009a, Hovatta et al.~2012). Transverse RM gradients are often associated with the toroidal or helical magnetic field structure of the jet, while RM gradients along the jet are natural to expect for an expanding jet in a stratified environment. While we cannot distinguish between these possible scenarios, it does suggest that broadband spectro-polarimetry of the integrated emission from blazars can potentially select interesting targets for follow-up studies with VLBI (see Anderson et al.~2016 for further discussion on this topic). 

Two of the steep-spectrum `double' sources described by the $\DeltaRM$ depolarization have two RM components, while the other has three RM components. In each case, the depolarization of one component is much larger than the other(s), suggesting large differences in the magneto-ionic environment of both lobes. 
Observations showing large-scale RM gradients across radio galaxy lobes are not very common, however there have been some recent examples (e.g.~Guidetti et al.~2011, Guidetti et al.~2012, Gabuzda et al.~2015). Indeed, Guidetti et al.~(2012) proposed a model of a radio galaxy interacting with its environment in a way which could produce these ordered RM structures and also had large depolarization asymmetries between the two lobes. 
These studies strongly suggest that these smooth RM structures are directly related to the radio galaxy and are important for investigating the physics of radio galaxies and how they interact with their environment. Follow-up studies with better, arcsecond-scale polarization imaging would provide important tests of the reliability of our polarization modelling technique to identify these type of sources in future large-area radio surveys. 

\subsection{Magnetic field geometry}
In Section~\ref{bfieldgeometry} we found that the intrinsic polarization angle was preferentially aligned with the jet direction in the jet-mode AGN. This result is obviously biased towards sources with straight jets because these are the source in which we could reliably measure the jet direction. Sources with complicated or bent morphologies were excluded from the analysis.  

The typical kpc-scale magnetic field geometry found in straight FRI radio sources is initially longitudinal but becomes mainly toroidal in the more extended regions of the source (e.g.~Laing \& Bridle 2014). While there is not a one-to-one correspondence between jet-mode AGN and FRIs, the jet-mode AGN dominate the low-luminosity end of the 1.4~GHz radio luminosity function ($<10^{25}$~W~Hz$^{-1}$) while the traditional FRI/FRII luminosity divide also occurs at $\sim10^{25}$~W~Hz$^{-1}$. Therefore, it is likely that the sources we detect with their intrinsic magnetic field orientation perpendicular to the jet direction are FRI sources in which their extended jet/lobe regions dominate the integrated polarized flux. 
This discovery also provides more confidence that our model-fitting technique can reliably separate the intrinsic magnetic field properties from the magneto-ionic properties of radio sources and is extracting physically meaningful parameters. 

There was also a clear difference in the results between jet-mode and radiative-mode AGN, since the radiative-mode AGN had no clear preference for the orientation of the intrinsic polarization angle with respect to the jet direction. However, the radiative-mode AGN were more commonly best-fit by two RM components compared to the jet-mode AGN which have a larger fraction of one RM component sources. This means that in general the precision with which the intrinsic polarization angles are determined in radiative-mode sources is lower than in jet-mode sources. There is some slight evidence for peaks at $45^\circ$ and $90^\circ$ for the radiative-mode sources (Figs.~\ref{histjetpa1},~\ref{histjetpa23}) but a larger sample with better signal-to-noise ratio and/or better angular resolution is required to determine if these peaks are real. 

Similar studies have been conducted before (e.g.~Gardner \& Whiteoak 1966, Clarke et al.~1980) finding peaks near both 0$^\circ$ and 90$^\circ$. However, these studies did not account for the presence of multiple RM components in their integrated polarization measurements. 
High angular resolution studies of FRII radio galaxies find that the magnetic fields tend to be longitudinal, however there was no preferred orientation at the hotspots (Saikia \& Salter 1988). This is important because the highly-polarized hotspot emission is likely to dominate the integrated emission in FRII radio sources. Therefore, if the high-luminosity radiative-mode AGN are dominated by FRII morphologies, then it is unsurprising we do not detect any clear relationship between the intrinsic polarization angle and the jet direction.  

\subsection{Cosmic evolution of magnetic fields}
Mapping out the evolution of cosmic magnetic fields is a key science goal for the SKA and its pre-cursors, as well as being the topic of many studies to date (c.f.~Johnston-Hollitt et al.~2015, Gaensler et al.~2015, Taylor et al.~2015). Recently, Farnes et al.~(2014b, 2016) provided one of the most statistically robust assessments of the evolution of magnetic fields in galaxies, indicating that a galactic `dynamo' can rapidly amplify the strength and coherence of magnetic fields by $z\sim1$, to similar levels as seen in galaxies in the local Universe. However, the sources studied in this paper are typically more sensitive to magnetic fields on larger scales, such as the intragroup/cluster medium and the extended halo of the radio source host galaxy. 

The redshift of the Faraday screen is expected to decrease the rest-frame RM and RM dispersion to the observed values by a factor of $(1+z)^2$ (e.g.~Hammond et al.~2012). Therefore, for a sample of sources with identical magneto-ionic properties but placed at different $z$, we expect to observe an anti-correlation between RRM (and $\sigmaRM$) with $z$. However, for the RRM there are several possible contributors along the line of sight, such as the magneto-ionic material local to the source, in the intergalactic medium, and from intervening galaxies. Additionally, the subtraction of the GRM model also introduces significant uncertainties (Oppermann et al.~2015). 
The location of the material influencing $\sigmaRM$ is much clearer. Many studies have concluded that, outside of lines of sight through the Galactic plane, the dominant contribution is local to the source, for example in the intracluster/group and host galaxy medium in which the radio galaxy is embedded (Leahy 1987, Garrington \& Conway 1991, Laing et al.~2008, Farnes et al.~2014a, Lamee et al.~2016). 

The fact that we observe no anti-correlation of RRM or $\sigmaRM$ with $z$ (Section~\ref{sec_z}) suggests that either the magneto-ionic material surrounding radio galaxies was denser at earlier times or that the scatter in our measurements is too large to detect such an evolution given our small sample size. Indeed,  if the electron density in the intergalactic medium was higher in the early Universe, as seems reasonable to assume (e.g.~Rees \& Reinhardt~1972, Kronberg et al.~1972, Pshirkov et al.~2015), then $B_{||}$ is naively expected to decrease with $z$. Hammond et al.~(2012) analysed the largest sample of RRM($z$) to date of 3,651 sources and found no evolution in the observed RRM with $z$ (over a range of $0<z<5.3$). This is consistent with the results from our much smaller sample, but which has much higher precision RRM measurements. 

Lamee et al.~(2016) detected weak evidence of an anti-correlation of depolarization with $z$ for a sample of 49 steep-spectrum, strongly depolarizing sources. Even when we consider only those sources with one RM component, with a steep-spectrum and with $\sigmaRM \ne 0$ (24 sources), we still do not detect any statistically significant relation between $\sigmaRM$ and $z$. This discrepancy may be due to the different sample selections (highly polarized sources selected at 1.4~GHz here versus $\sim$unbiased selection at 2.3~GHz in Lamee et al.~2016) or due to the different ways of quantifying the depolarization ($\sigmaRM$ versus $D=p_{1.4\,{\rm GHz}} / p_{2.3\,{\rm GHz}}$). Although we also do no find a relation with $z$ if we estimate the depolarization $D$ in an identical manner. 
Clearly, much larger samples are required, in comparison with realistic models, to determine the true nature of the evolution of magneto-ionic material with redshift. 

\section{Conclusions}
This paper presents the results of a broadband, spectro-polarimetry analysis of 100 radio AGN with the Australia Telescope Compact Array, using data from 1 to 10 GHz. From the Hammond et al.~(2012) catalog, the sources were selected to be highly polarized at 1.4~GHz, to have a known redshift, and to be at least 20$^\circ$ in latitude from the Galactic plane. 
We have implemented a general purpose QU-fitting procedure that accurately describes the wide range of broadband polarization behaviour observed from 1 to 3~GHz in our sample of radio AGN. The higher frequency data from 4 to 10 GHz were used to determine the morphology of the sources at high angular resolution. Most importantly, this study shows how we can reliably determine the intrinsic magnetic field properties ($p_0$, $\psi_0$) and the magneto-ionic properties (RM, $\sigmaRM$, $\DeltaRM$) from the integrated emission of extragalactic radio sources, using our polarization model-fitting approach.

We find that 37\%/52\%/11\% of our sample requires one/two/three RM components to describe the observed broadband polarization behaviour (for polarization signal-to-noise ratios ranging from 10 to 70). However, the fraction of two and three RM components increases with signal-to-noise ratio. A total of 24\% of sources are classified as Faraday thin, in the sense that the best-fit model did not require Faraday depolarization ($\sigmaRM=0$, $\DeltaRM=0$).  
Most steep spectrum sources are resolved (94\%), with a median linear size of 102~kpc, while the sources that remain unresolved at $\sim$3'' angular resolution are mainly flat spectrum sources (80\%). 

In general, our analysis shows that sources with high integrated polarization at 1.4 GHz ($p_{1.4\,{\rm GHz}}$) have low Faraday depolarization, are typically dominated by a single RM component, have a steep spectral index, and a high intrinsic degree of polarization. 
There is no simple relationship between the total intensity morphology of a source and the number of RM components. 
More generally, we show that previous results on the broadband, sparsely-sampled polarization behaviour of radio sources can be well understood in terms of multiple RM components combined with the effect of Faraday depolarization. 

We find no evidence for a correlation between the linear size of the radio sources and their magneto-ionic or intrinsic polarization properties. 
There is no evidence for an observed redshift evolution of the magneto-ionic properties of our sample ($0<z<2.8$). However, the strength of these conclusions are limited by the small sample size. 

By identifying the accretion mode in the host galaxy of the radio sources, we classified our sample into radiative-mode and jet-mode AGN.
This allows us to investigate the origin of the observed difference in $p_{1.4\,{\rm GHz}}$ between radiative-mode and jet-mode AGN, as found in O'Sullivan et al.~(2015). 
While we find an anti-correlation between $p_{1.4\,{\rm GHz}}$ and the Faraday depolarization for all sources, we find no significant difference between the magneto-ionic environments in jet-mode and radiative-mode AGN. However, there is a statistically significant difference in the mean intrinsic degree of polarization, $\langle p_0 \rangle$, between the two types, with the jet-mode sources having more intrinsically ordered magnetic field structures than the radiative-mode sources.
We also find a preferred perpendicular orientation of the intrinsic magnetic field structure of jet-mode AGN with respect to the jet direction, while no clear preference is found for the radiative-mode sources. 
Double-lobed sources with a bright core/inner jet region have the highest integrated degrees of polarization in our sample, but interestingly it is only the jet-mode AGN which show this. The physical origin of this behaviour is unclear, but may be related to the inner jet regions of FRI radio galaxies where the magnetic field is expected to have a high degree of order. 

Overall, this study paves the way for future broadband and spectro-polarimetric studies with large-area surveys (e.g.~VLASS, ASKAP-POSSUM) that will measure the polarization and Faraday rotation properties of hundreds of thousands of radio-loud AGN. This will enable much greater statistical power for determining the magnetised properties of radio AGN and their environments, and in using these sources as accurate statistical probes of foreground magneto-ionic material. 

\section{Acknowledgements}
S.P.O acknowledges the support of the Australian Research Council through grant FS100100033 and from UNAM through the PAPIIT project IA103416. 
The Dunlap Institute is funded through an endowment established by the David Dunlap family and the University of Toronto. B.M.G. acknowledges the support of the Natural Sciences and Engineering Research Council of Canada (NSERC) through grant RGPIN-2015-05948, and of the Canada Research Chairs program. 
The Australia Telescope Compact Array is part of the Australia Telescope National Facility which is funded by the Commonwealth of Australia for operation as a National Facility managed by CSIRO. Parts of this research were conducted by the Australian Research Council Centre of Excellence for All-sky Astrophysics (CAASTRO), through project number CE110001020.
This research has made use of: the NASA/IPAC Extragalactic Database (NED) which is operated by the Jet Propulsion Laboratory, California Institute of Technology, under contract with NASA; NASA's Astrophysics Data System Abstract Service; the 2dF/6dF Redshift Surveys, which were compiled from observations made with the 2-degree Field and 6-degree Field on the Anglo-Australian Telescope; TOPCAT, an interactive graphical viewer and editor for tabular data (Taylor 2005); Ned Wright's online Javascript cosmology calculator (Wright 2006). 

\bibliography{bflux_bib}

\begin{thebibliography}{}
\makeatletter
\relax
\def\mn@urlcharsother{\let\do\@makeother \do\$\do\&\do\#\do\^\do\_\do\%\do\~}
\def\mn@doi{\begingroup\mn@urlcharsother \@ifnextchar [ {\mn@doi@}
  {\mn@doi@[]}}
\def\mn@doi@[#1]#2{\def\@tempa{#1}\ifx\@tempa\@empty \href
  {http://dx.doi.org/#2} {doi:#2}\else \href {http://dx.doi.org/#2} {#1}\fi
  \endgroup}
\def\mn@eprint#1#2{\mn@eprint@#1:#2::\@nil}
\def\mn@eprint@arXiv#1{\href {http://arxiv.org/abs/#1} {{\tt arXiv:#1}}}
\def\mn@eprint@dblp#1{\href {http://dblp.uni-trier.de/rec/bibtex/#1.xml}
  {dblp:#1}}
\def\mn@eprint@#1:#2:#3:#4\@nil{\def\@tempa {#1}\def\@tempb {#2}\def\@tempc
  {#3}\ifx \@tempc \@empty \let \@tempc \@tempb \let \@tempb \@tempa \fi \ifx
  \@tempb \@empty \def\@tempb {arXiv}\fi \@ifundefined
  {mn@eprint@\@tempb}{\@tempb:\@tempc}{\expandafter \expandafter \csname
  mn@eprint@\@tempb\endcsname \expandafter{\@tempc}}}

\bibitem[\protect\citeauthoryear{{Anderson}, {Gaensler}  \& {Feain}}{{Anderson}
  et~al.}{2016}]{anderson2016}
{Anderson} C.~S.,  {Gaensler} B.~M.,   {Feain} I.~J.,  2016, \mn@doi [\apj]
  {10.3847/0004-637X/825/1/59}, \href
  {http://adsabs.harvard.edu/abs/2016ApJ...825...59A} {825, 59}

\bibitem[\protect\citeauthoryear{Asada, Inoue, Uchida, Kameno, Fujisawa, Iguchi
   \& Mutoh}{Asada et~al.}{2002}]{asada2002}
Asada K.,  Inoue M.,  Uchida Y.,  Kameno S.,  Fujisawa K.,  Iguchi S.,   Mutoh
  M.,  2002, PASJ, 54, L39

\bibitem[\protect\citeauthoryear{{Banfield}, {George}, {Taylor}, {Stil},
  {Kothes}  \& {Scott}}{{Banfield} et~al.}{2011}]{banfield2011}
{Banfield} J.~K.,  {George} S.~J.,  {Taylor} A.~R.,  {Stil} J.~M.,  {Kothes}
  R.,   {Scott} D.,  2011, \mn@doi [ApJ] {10.1088/0004-637X/733/1/69}, \href
  {http://adsabs.harvard.edu/abs/2011ApJ...733...69B} {733, 69}

\bibitem[\protect\citeauthoryear{{Banfield}, {Schnitzeler}, {George}, {Norris},
  {Jarrett}, {Taylor}  \& {Stil}}{{Banfield} et~al.}{2014}]{banfield2014}
{Banfield} J.~K.,  {Schnitzeler} D.~H.~F.~M.,  {George} S.~J.,  {Norris} R.~P.,
   {Jarrett} T.~H.,  {Taylor} A.~R.,   {Stil} J.~M.,  2014, \mn@doi [MNRAS]
  {10.1093/mnras/stu1411}, \href
  {http://adsabs.harvard.edu/abs/2014MNRAS.444..700B} {444, 700}

\bibitem[\protect\citeauthoryear{{Best}}{{Best}}{2009}]{best2009}
{Best} P.~N.,  2009, \mn@doi [Astronomische Nachrichten]
  {10.1002/asna.200811152}, \href
  {http://adsabs.harvard.edu/abs/2009AN....330..184B} {330, 184}

\bibitem[\protect\citeauthoryear{{Best} \& {Heckman}}{{Best} \&
  {Heckman}}{2012}]{bestheckman2012}
{Best} P.~N.,  {Heckman} T.~M.,  2012, \mn@doi [MNRAS]
  {10.1111/j.1365-2966.2012.20414.x}, \href
  {http://adsabs.harvard.edu/abs/2012MNRAS.421.1569B} {421, 1569}

\bibitem[\protect\citeauthoryear{{Bonafede} et~al.,}{{Bonafede}
  et~al.}{2015}]{bonafede2015}
{Bonafede} A.,  et~al., 2015, Advancing Astrophysics with the Square Kilometre
  Array (AASKA14), \href {http://adsabs.harvard.edu/abs/2015aska.confE..95B}
  {p.~95}

\bibitem[\protect\citeauthoryear{{Brentjens} \& {de Bruyn}}{{Brentjens} \& {de
  Bruyn}}{2005}]{bdb2005}
{Brentjens} M.~A.,  {de Bruyn} A.~G.,  2005, \mn@doi [A\&A]
  {10.1051/0004-6361:20052990}, \href
  {http://adsabs.harvard.edu/abs/2005A%26A...441.1217B} {441, 1217}

\bibitem[\protect\citeauthoryear{{Burn}}{{Burn}}{1966}]{burn1966}
{Burn} B.~J.,  1966, MNRAS, 133, 67

\bibitem[\protect\citeauthoryear{{Clarke}, {Kronberg}  \&
  {Simard-Normandin}}{{Clarke} et~al.}{1980}]{clarke1980}
{Clarke} J.~N.,  {Kronberg} P.~P.,   {Simard-Normandin} M.,  1980, \mn@doi
  [\mnras] {10.1093/mnras/190.2.205}, \href
  {http://adsabs.harvard.edu/abs/1980MNRAS.190..205C} {190, 205}

\bibitem[\protect\citeauthoryear{{Colless} et~al.,}{{Colless}
  et~al.}{2001}]{colless2001}
{Colless} M.,  et~al., 2001, \mn@doi [\mnras]
  {10.1046/j.1365-8711.2001.04902.x}, \href
  {http://adsabs.harvard.edu/abs/2001MNRAS.328.1039C} {328, 1039}

\bibitem[\protect\citeauthoryear{{Condon}, {Cotton}, {Greisen}, {Yin},
  {Perley}, {Taylor}  \& {Broderick}}{{Condon} et~al.}{1998}]{condon1998}
{Condon} J.~J.,  {Cotton} W.~D.,  {Greisen} E.~W.,  {Yin} Q.~F.,  {Perley}
  R.~A.,  {Taylor} G.~B.,   {Broderick} J.~J.,  1998, \mn@doi [AJ]
  {10.1086/300337}, \href {http://adsabs.harvard.edu/abs/1998AJ....115.1693C}
  {115, 1693}

\bibitem[\protect\citeauthoryear{{Conway}, {Haves}, {Kronberg}, {Stannard},
  {Vallee}  \& {Wardle}}{{Conway} et~al.}{1974}]{conway1974}
{Conway} R.~G.,  {Haves} P.,  {Kronberg} P.~P.,  {Stannard} D.,  {Vallee}
  J.~P.,   {Wardle} J.~F.~C.,  1974, MNRAS, \href
  {http://adsabs.harvard.edu/abs/1974MNRAS.168..137C} {168, 137}

\bibitem[\protect\citeauthoryear{{Farnes}, {Gaensler}  \& {Carretti}}{{Farnes}
  et~al.}{2014a}]{farnes2014a}
{Farnes} J.~S.,  {Gaensler} B.~M.,   {Carretti} E.,  2014a, \mn@doi [ApJS]
  {10.1088/0067-0049/212/1/15}, \href
  {http://adsabs.harvard.edu/abs/2014ApJS..212...15F} {212, 15}

\bibitem[\protect\citeauthoryear{{Farnes}, {O'Sullivan}, {Corrigan}  \&
  {Gaensler}}{{Farnes} et~al.}{2014b}]{farnes2014b}
{Farnes} J.~S.,  {O'Sullivan} S.~P.,  {Corrigan} M.~E.,   {Gaensler} B.~M.,
  2014b, \mn@doi [\apj] {10.1088/0004-637X/795/1/63}, \href
  {http://adsabs.harvard.edu/abs/2014ApJ...795...63F} {795, 63}

\bibitem[\protect\citeauthoryear{{Farnes}, {Rudnick}, {Gaensler}, {Haverkorn},
  {O'Sullivan}  \& {Curran}}{{Farnes} et~al.}{2016}]{farnes2016}
{Farnes} J.~S.,  {Rudnick} L.,  {Gaensler} B.~M.,  {Haverkorn} M.,
  {O'Sullivan} S.~P.,   {Curran} S.~J.,  2016, preprint, \href
  {http://adsabs.harvard.edu/abs/2016arXiv160901623F} {} (\mn@eprint {arXiv}
  {1609.01623})

\bibitem[\protect\citeauthoryear{{Farnsworth}, {Rudnick}  \&
  {Brown}}{{Farnsworth} et~al.}{2011}]{farnsworth2011}
{Farnsworth} D.,  {Rudnick} L.,   {Brown} S.,  2011, \mn@doi [AJ]
  {10.1088/0004-6256/141/6/191}, \href
  {http://adsabs.harvard.edu/abs/2011AJ....141..191F} {141, 191}

\bibitem[\protect\citeauthoryear{{Gabuzda}, {Knuettel}  \&
  {Bonafede}}{{Gabuzda} et~al.}{2015}]{gabuzda2015}
{Gabuzda} D.~C.,  {Knuettel} S.,   {Bonafede} A.,  2015, \mn@doi [\aap]
  {10.1051/0004-6361/201527185}, \href
  {http://adsabs.harvard.edu/abs/2015A%26A...583A..96G} {583, A96}

\bibitem[\protect\citeauthoryear{{Gaensler} et~al.,}{{Gaensler}
  et~al.}{2015}]{gaensler2015}
{Gaensler} B.,  et~al., 2015, Advancing Astrophysics with the Square Kilometre
  Array (AASKA14), \href {http://adsabs.harvard.edu/abs/2015aska.confE.103G}
  {p.~103}

\bibitem[\protect\citeauthoryear{{Gardner} \& {Whiteoak}}{{Gardner} \&
  {Whiteoak}}{1966}]{gardnerwhiteoak1966}
{Gardner} F.~F.,  {Whiteoak} J.~B.,  1966, \mn@doi [\araa]
  {10.1146/annurev.aa.04.090166.001333}, \href
  {http://adsabs.harvard.edu/abs/1966ARA%26A...4..245G} {4, 245}

\bibitem[\protect\citeauthoryear{{Garrington} \& {Conway}}{{Garrington} \&
  {Conway}}{1991}]{garringtonconway1991}
{Garrington} S.~T.,  {Conway} R.~G.,  1991, MNRAS, \href
  {http://adsabs.harvard.edu/abs/1991MNRAS.250..198G} {250, 198}

\bibitem[\protect\citeauthoryear{{Garrington}, {Leahy}, {Conway}  \&
  {Laing}}{{Garrington} et~al.}{1988}]{garrington1988}
{Garrington} S.~T.,  {Leahy} J.~P.,  {Conway} R.~G.,   {Laing} R.~A.,  1988,
  \mn@doi [Nature] {10.1038/331147a0}, \href
  {http://adsabs.harvard.edu/abs/1988Natur.331..147G} {331, 147}

\bibitem[\protect\citeauthoryear{{Gendre}, {Best}, {Wall}  \& {Ker}}{{Gendre}
  et~al.}{2013}]{gendre2013}
{Gendre} M.~A.,  {Best} P.~N.,  {Wall} J.~V.,   {Ker} L.~M.,  2013, \mn@doi
  [MNRAS] {10.1093/mnras/stt116}, \href
  {http://adsabs.harvard.edu/abs/2013MNRAS.430.3086G} {430, 3086}

\bibitem[\protect\citeauthoryear{{George}, {Stil}  \& {Keller}}{{George}
  et~al.}{2012}]{george2012}
{George} S.~J.,  {Stil} J.~M.,   {Keller} B.~W.,  2012, \mn@doi [\pasa]
  {10.1071/AS11027}, \href {http://adsabs.harvard.edu/abs/2012PASA...29..214G}
  {29, 214}

\bibitem[\protect\citeauthoryear{{Gugliucci}, {Taylor}, {Peck}  \&
  {Giroletti}}{{Gugliucci} et~al.}{2007}]{gugliucci2007}
{Gugliucci} N.~E.,  {Taylor} G.~B.,  {Peck} A.~B.,   {Giroletti} M.,  2007,
  \mn@doi [\apj] {10.1086/515560}, \href
  {http://adsabs.harvard.edu/abs/2007ApJ...661...78G} {661, 78}

\bibitem[\protect\citeauthoryear{{Guidetti}, {Laing}, {Bridle}, {Parma}  \&
  {Gregorini}}{{Guidetti} et~al.}{2011}]{guidetti2011}
{Guidetti} D.,  {Laing} R.~A.,  {Bridle} A.~H.,  {Parma} P.,   {Gregorini} L.,
  2011, \mn@doi [MNRAS] {10.1111/j.1365-2966.2011.18321.x}, \href
  {http://adsabs.harvard.edu/abs/2011MNRAS.413.2525G} {413, 2525}

\bibitem[\protect\citeauthoryear{{Guidetti}, {Laing}, {Croston}, {Bridle}  \&
  {Parma}}{{Guidetti} et~al.}{2012}]{guidetti2012}
{Guidetti} D.,  {Laing} R.~A.,  {Croston} J.~H.,  {Bridle} A.~H.,   {Parma} P.,
   2012, \mn@doi [MNRAS] {10.1111/j.1365-2966.2012.20961.x}, \href
  {http://adsabs.harvard.edu/abs/2012MNRAS.423.1335G} {423, 1335}

\bibitem[\protect\citeauthoryear{{Hammond}, {Robishaw}  \&
  {Gaensler}}{{Hammond} et~al.}{2012}]{hammond2012}
{Hammond} A.~M.,  {Robishaw} T.,   {Gaensler} B.~M.,  2012, ArXiv e-prints:
  1209.1438v3, \href {http://adsabs.harvard.edu/abs/2012arXiv1209.1438H} {}

\bibitem[\protect\citeauthoryear{{Hardcastle}, {Evans}  \&
  {Croston}}{{Hardcastle} et~al.}{2007}]{hardcastle2007}
{Hardcastle} M.~J.,  {Evans} D.~A.,   {Croston} J.~H.,  2007, \mn@doi [MNRAS]
  {10.1111/j.1365-2966.2007.11572.x}, \href
  {http://adsabs.harvard.edu/abs/2007MNRAS.376.1849H} {376, 1849}

\bibitem[\protect\citeauthoryear{{Haverkorn} et~al.,}{{Haverkorn}
  et~al.}{2015}]{haverkorn2015}
{Haverkorn} M.,  et~al., 2015, Advancing Astrophysics with the Square Kilometre
  Array (AASKA14), \href {http://adsabs.harvard.edu/abs/2015aska.confE..96H}
  {p.~96}

\bibitem[\protect\citeauthoryear{{Heckman} \& {Best}}{{Heckman} \&
  {Best}}{2014}]{heckmanbest2014}
{Heckman} T.~M.,  {Best} P.~N.,  2014, \mn@doi [ARA\&A]
  {10.1146/annurev-astro-081913-035722}, \href
  {http://adsabs.harvard.edu/abs/2014ARA%26A..52..589H} {52, 589}

\bibitem[\protect\citeauthoryear{{Hovatta}, {Lister}, {Aller}, {Aller},
  {Homan}, {Kovalev}, {Pushkarev}  \& {Savolainen}}{{Hovatta}
  et~al.}{2012}]{hovatta2012}
{Hovatta} T.,  {Lister} M.~L.,  {Aller} M.~F.,  {Aller} H.~D.,  {Homan} D.~C.,
  {Kovalev} Y.~Y.,  {Pushkarev} A.~B.,   {Savolainen} T.,  2012, \mn@doi [\aj]
  {10.1088/0004-6256/144/4/105}, \href
  {http://adsabs.harvard.edu/abs/2012AJ....144..105H} {144, 105}

\bibitem[\protect\citeauthoryear{{Ideguchi}, {Takahashi}, {Akahori}, {Kumazaki}
   \& {Ryu}}{{Ideguchi} et~al.}{2014}]{ideguchi2014}
{Ideguchi} S.,  {Takahashi} K.,  {Akahori} T.,  {Kumazaki} K.,   {Ryu} D.,
  2014, \mn@doi [\pasj] {10.1093/pasj/pst007}, \href
  {http://adsabs.harvard.edu/abs/2014PASJ...66....5I} {66, 5}

\bibitem[\protect\citeauthoryear{{Ishwara-Chandra}, {Saikia}, {Kapahi}  \&
  {McCarthy}}{{Ishwara-Chandra} et~al.}{1998}]{ishwara-chandra1998}
{Ishwara-Chandra} C.~H.,  {Saikia} D.~J.,  {Kapahi} V.~K.,   {McCarthy} P.~J.,
  1998, \mn@doi [\mnras] {10.1046/j.1365-8711.1998.01906.x}, \href
  {http://adsabs.harvard.edu/abs/1998MNRAS.300..269I} {300, 269}

\bibitem[\protect\citeauthoryear{{Johnston-Hollitt} et~al.,}{{Johnston-Hollitt}
  et~al.}{2015}]{johnston-hollitt2015}
{Johnston-Hollitt} M.,  et~al., 2015, Advancing Astrophysics with the Square
  Kilometre Array (AASKA14), \href
  {http://adsabs.harvard.edu/abs/2015aska.confE..92J} {p.~92}

\bibitem[\protect\citeauthoryear{{Jones} et~al.,}{{Jones}
  et~al.}{2009}]{jones2009}
{Jones} D.~H.,  et~al., 2009, \mn@doi [\mnras]
  {10.1111/j.1365-2966.2009.15338.x}, \href
  {http://adsabs.harvard.edu/abs/2009MNRAS.399..683J} {399, 683}

\bibitem[\protect\citeauthoryear{{Kharb}, {Shastri}  \& {Gabuzda}}{{Kharb}
  et~al.}{2005}]{kharb2005}
{Kharb} P.,  {Shastri} P.,   {Gabuzda} D.~C.,  2005, \mn@doi [\apjl]
  {10.1086/497984}, \href {http://adsabs.harvard.edu/abs/2005ApJ...632L..69K}
  {632, L69}

\bibitem[\protect\citeauthoryear{{Kronberg}, {Conway}  \& {Gilbert}}{{Kronberg}
  et~al.}{1972}]{kronberg1972}
{Kronberg} P.~P.,  {Conway} R.~G.,   {Gilbert} J.~A.,  1972, \mn@doi [\mnras]
  {10.1093/mnras/156.3.275}, \href
  {http://adsabs.harvard.edu/abs/1972MNRAS.156..275K} {156, 275}

\bibitem[\protect\citeauthoryear{{Kumazaki}, {Akahori}, {Ideguchi}, {Kurayama}
  \& {Takahashi}}{{Kumazaki} et~al.}{2014}]{kumazaki2014}
{Kumazaki} K.,  {Akahori} T.,  {Ideguchi} S.,  {Kurayama} T.,   {Takahashi} K.,
   2014, \mn@doi [\pasj] {10.1093/pasj/psu030}, \href
  {http://adsabs.harvard.edu/abs/2014PASJ...66...61K} {66, 61}

\bibitem[\protect\citeauthoryear{{Laing}}{{Laing}}{1988}]{laing1988}
{Laing} R.~A.,  1988, \mn@doi [Nature] {10.1038/331149a0}, \href
  {http://adsabs.harvard.edu/abs/1988Natur.331..149L} {331, 149}

\bibitem[\protect\citeauthoryear{{Laing} \& {Bridle}}{{Laing} \&
  {Bridle}}{2014}]{laingbridle2014}
{Laing} R.~A.,  {Bridle} A.~H.,  2014, \mn@doi [\mnras]
  {10.1093/mnras/stt2138}, \href
  {http://adsabs.harvard.edu/abs/2014MNRAS.437.3405L} {437, 3405}

\bibitem[\protect\citeauthoryear{{Laing}, {Bridle}, {Parma}  \&
  {Murgia}}{{Laing} et~al.}{2008}]{laing2008}
{Laing} R.~A.,  {Bridle} A.~H.,  {Parma} P.,   {Murgia} M.,  2008, \mn@doi
  [MNRAS] {10.1111/j.1365-2966.2008.13895.x}, \href
  {http://adsabs.harvard.edu/abs/2008MNRAS.391..521L} {391, 521}

\bibitem[\protect\citeauthoryear{{Lamee}, {Rudnick}, {Farnes}, {Carretti},
  {Gaensler}, {Haverkorn}  \& {Poppi}}{{Lamee} et~al.}{2016}]{lamee2016}
{Lamee} M.,  {Rudnick} L.,  {Farnes} J.~S.,  {Carretti} E.,  {Gaensler} B.~M.,
  {Haverkorn} M.,   {Poppi} S.,  2016, \mn@doi [\apj]
  {10.3847/0004-637X/829/1/5}, \href
  {http://adsabs.harvard.edu/abs/2016ApJ...829....5L} {829, 5}

\bibitem[\protect\citeauthoryear{{Leahy}}{{Leahy}}{1987}]{leahy1987}
{Leahy} J.~P.,  1987, \mn@doi [\mnras] {10.1093/mnras/226.2.433}, \href
  {http://adsabs.harvard.edu/abs/1987MNRAS.226..433L} {226, 433}

\bibitem[\protect\citeauthoryear{O'Sullivan \& Gabuzda}{O'Sullivan \&
  Gabuzda}{2009a}]{osullivangabuzda2009a}
O'Sullivan S.~P.,  Gabuzda D.~C.,  2009a, \mn@doi [MNRAS]
  {10.1111/mnr.2009.393.issue-2}, 393, 429

\bibitem[\protect\citeauthoryear{{O'Sullivan} \& {Gabuzda}}{{O'Sullivan} \&
  {Gabuzda}}{2009b}]{osullivangabuzda2009b}
{O'Sullivan} S.~P.,  {Gabuzda} D.~C.,  2009b, \mn@doi [MNRAS]
  {10.1111/j.1365-2966.2009.15428.x}, \href
  {http://adsabs.harvard.edu/abs/2009MNRAS.400...26O} {400, 26}

\bibitem[\protect\citeauthoryear{{O'Sullivan} et~al.,}{{O'Sullivan}
  et~al.}{2012}]{osullivan2012}
{O'Sullivan} S.~P.,  et~al., 2012, MNRAS, \href
  {http://adsabs.harvard.edu/abs/2012arXiv1201.3161O} {421, 3300}

\bibitem[\protect\citeauthoryear{{O'Sullivan}, {Gaensler}, {Lara-L{\'o}pez},
  {van Velzen}, {Banfield}  \& {Farnes}}{{O'Sullivan}
  et~al.}{2015}]{osullivan2015}
{O'Sullivan} S.~P.,  {Gaensler} B.~M.,  {Lara-L{\'o}pez} M.~A.,  {van Velzen}
  S.,  {Banfield} J.~K.,   {Farnes} J.~S.,  2015, \mn@doi [\apj]
  {10.1088/0004-637X/806/1/83}, \href
  {http://adsabs.harvard.edu/abs/2015ApJ...806...83O} {806, 83}

\bibitem[\protect\citeauthoryear{{Oppermann} et~al.,}{{Oppermann}
  et~al.}{2015}]{oppermann2015}
{Oppermann} N.,  et~al., 2015, \mn@doi [A\&A] {10.1051/0004-6361/201423995},
  \href {http://adsabs.harvard.edu/abs/2015A%26A...575A.118O} {575, A118}

\bibitem[\protect\citeauthoryear{{Pasetto}, {Kraus}, {Mack}, {Bruni}  \&
  {Carrasco-Gonz{\'a}lez}}{{Pasetto} et~al.}{2016}]{pasetto2016}
{Pasetto} A.,  {Kraus} A.,  {Mack} K.-H.,  {Bruni} G.,
  {Carrasco-Gonz{\'a}lez} C.,  2016, \mn@doi [\aap]
  {10.1051/0004-6361/201526963}, \href
  {http://adsabs.harvard.edu/abs/2016A%26A...586A.117P} {586, A117}

\bibitem[\protect\citeauthoryear{{Planck Collaboration. } et~al.,}{{Planck
  Collaboration. } et~al.}{2014}]{planck2014}
{Planck Collaboration. } et~al., 2014, \mn@doi [A\&A]
  {10.1051/0004-6361/201321591}, \href
  {http://adsabs.harvard.edu/abs/2014A%26A...571A..16P} {571, A16}

\bibitem[\protect\citeauthoryear{{Porth}, {Fendt}, {Meliani}  \&
  {Vaidya}}{{Porth} et~al.}{2011}]{porth2011}
{Porth} O.,  {Fendt} C.,  {Meliani} Z.,   {Vaidya} B.,  2011, \mn@doi [ApJ]
  {10.1088/0004-637X/737/1/42}, \href
  {http://adsabs.harvard.edu/abs/2011ApJ...737...42P} {737, 42}

\bibitem[\protect\citeauthoryear{{Pracy} et~al.,}{{Pracy}
  et~al.}{2016}]{pracy2016}
{Pracy} M.~B.,  et~al., 2016, \mn@doi [\mnras] {10.1093/mnras/stw910}, \href
  {http://adsabs.harvard.edu/abs/2016MNRAS.460....2P} {460, 2}

\bibitem[\protect\citeauthoryear{Raftery}{Raftery}{1995}]{raftery1995}
Raftery A.~E.,  1995, Sociological Methodology, 25, 111

\bibitem[\protect\citeauthoryear{{Rees} \& {Reinhardt}}{{Rees} \&
  {Reinhardt}}{1972}]{reesreinhardt1972}
{Rees} M.~J.,  {Reinhardt} M.,  1972, \aap, \href
  {http://adsabs.harvard.edu/abs/1972A%26A....19..189R} {19, 189}

\bibitem[\protect\citeauthoryear{{Saikia} \& {Salter}}{{Saikia} \&
  {Salter}}{1988}]{saikiasalter1988}
{Saikia} D.~J.,  {Salter} C.~J.,  1988, \mn@doi [ARA\&A]
  {10.1146/annurev.aa.26.090188.000521}, \href
  {http://adsabs.harvard.edu/abs/1988ARA%26A..26...93S} {26, 93}

\bibitem[\protect\citeauthoryear{{Schnitzeler}, {Banfield}  \&
  {Lee}}{{Schnitzeler} et~al.}{2015}]{schnitzeler2015}
{Schnitzeler} D.~H.~F.~M.,  {Banfield} J.~K.,   {Lee} K.~J.,  2015, \mn@doi
  [\mnras] {10.1093/mnras/stv708}, \href
  {http://adsabs.harvard.edu/abs/2015MNRAS.450.3579S} {450, 3579}

\bibitem[\protect\citeauthoryear{{Sokoloff}, {Bykov}, {Shukurov},
  {Berkhuijsen}, {Beck}  \& {Poezd}}{{Sokoloff} et~al.}{1998}]{sokoloff1998}
{Sokoloff} D.~D.,  {Bykov} A.~A.,  {Shukurov} A.,  {Berkhuijsen} E.~M.,  {Beck}
  R.,   {Poezd} A.~D.,  1998, \mn@doi [MNRAS]
  {10.1046/j.1365-8711.1998.01782.x}, \href
  {http://adsabs.harvard.edu/abs/1998MNRAS.299..189S} {299, 189}

\bibitem[\protect\citeauthoryear{{Stil} \& {Keller}}{{Stil} \&
  {Keller}}{2015}]{stilkeller2015}
{Stil} J.,  {Keller} B.,  2015, Advancing Astrophysics with the Square
  Kilometre Array (AASKA14), \href
  {http://adsabs.harvard.edu/abs/2015aska.confE.112S} {p.~112}

\bibitem[\protect\citeauthoryear{{Sun} et~al.,}{{Sun} et~al.}{2015}]{sun2015}
{Sun} X.~H.,  et~al., 2015, \mn@doi [\aj] {10.1088/0004-6256/149/2/60}, \href
  {http://adsabs.harvard.edu/abs/2015AJ....149...60S} {149, 60}

\bibitem[\protect\citeauthoryear{{Taylor}}{{Taylor}}{2005}]{taylor2005}
{Taylor} M.~B.,  2005, in {Shopbell} P.,  {Britton} M.,   {Ebert} R.,  eds,
  Astronomical Society of the Pacific Conference Series Vol. 347, Astronomical
  Data Analysis Software and Systems XIV. p.~29

\bibitem[\protect\citeauthoryear{{Taylor}, {Stil}  \& {Sunstrum}}{{Taylor}
  et~al.}{2009}]{taylor2009}
{Taylor} A.~R.,  {Stil} J.~M.,   {Sunstrum} C.,  2009, \mn@doi [ApJ]
  {10.1088/0004-637X/702/2/1230}, \href
  {http://adsabs.harvard.edu/abs/2009ApJ...702.1230T} {702, 1230}

\bibitem[\protect\citeauthoryear{{Taylor} et~al.,}{{Taylor}
  et~al.}{2015}]{taylor2015}
{Taylor} R.,  et~al., 2015, Advancing Astrophysics with the Square Kilometre
  Array (AASKA14), \href {http://adsabs.harvard.edu/abs/2015aska.confE.113T}
  {p.~113}

\bibitem[\protect\citeauthoryear{{Tribble}}{{Tribble}}{1991}]{tribble1991}
{Tribble} P.~C.,  1991, MNRAS, \href
  {http://adsabs.harvard.edu/abs/1991MNRAS.250..726T} {250, 726}

\bibitem[\protect\citeauthoryear{{Vacca} et~al.,}{{Vacca}
  et~al.}{2015}]{vacca2015}
{Vacca} V.,  et~al., 2015, Advancing Astrophysics with the Square Kilometre
  Array (AASKA14), \href {http://adsabs.harvard.edu/abs/2015aska.confE.114V}
  {p.~114}

\bibitem[\protect\citeauthoryear{{Wilson} et~al.,}{{Wilson}
  et~al.}{2011}]{wilson2011}
{Wilson} W.~E.,  et~al., 2011, MNRAS, \href
  {http://adsabs.harvard.edu/abs/2011arXiv1105.3532W} {416, 832}

\bibitem[\protect\citeauthoryear{Wright}{Wright}{2006}]{wright2006}
Wright E.~L.,  2006, PASP, 118, 1711

\bibitem[\protect\citeauthoryear{{van Breugel}, {Miley}, {Heckman}, {Butcher}
  \& {Bridle}}{{van Breugel} et~al.}{1985}]{vanbreugel1985}
{van Breugel} W.,  {Miley} G.,  {Heckman} T.,  {Butcher} H.,   {Bridle} A.,
  1985, \mn@doi [\apj] {10.1086/163007}, \href
  {http://adsabs.harvard.edu/abs/1985ApJ...290..496V} {290, 496}

\makeatother
\end{thebibliography}
\bibliographystyle{mnras}


\appendix
\section{Spectropolarimetric data}
\label{sec:quplots}
In Figure~\ref{quplots1}, we plot the $q(\lambda^2)$ and $u(\lambda^2)$ data and show the best-fit models (Table~\ref{modelfits}) overlaid on the data for each source. 
We have made this data available online as supplementary material, and it can also be found on the VizieR website (http://vizier.u-strasbg.fr/viz-bin/VizieR). 
Table~\ref{exdata} provides an example of the format of the data.

\begin{figure*}
\centering
\includegraphics[clip=true, trim=0cm 0cm 0cm 0cm, width=16.7cm]{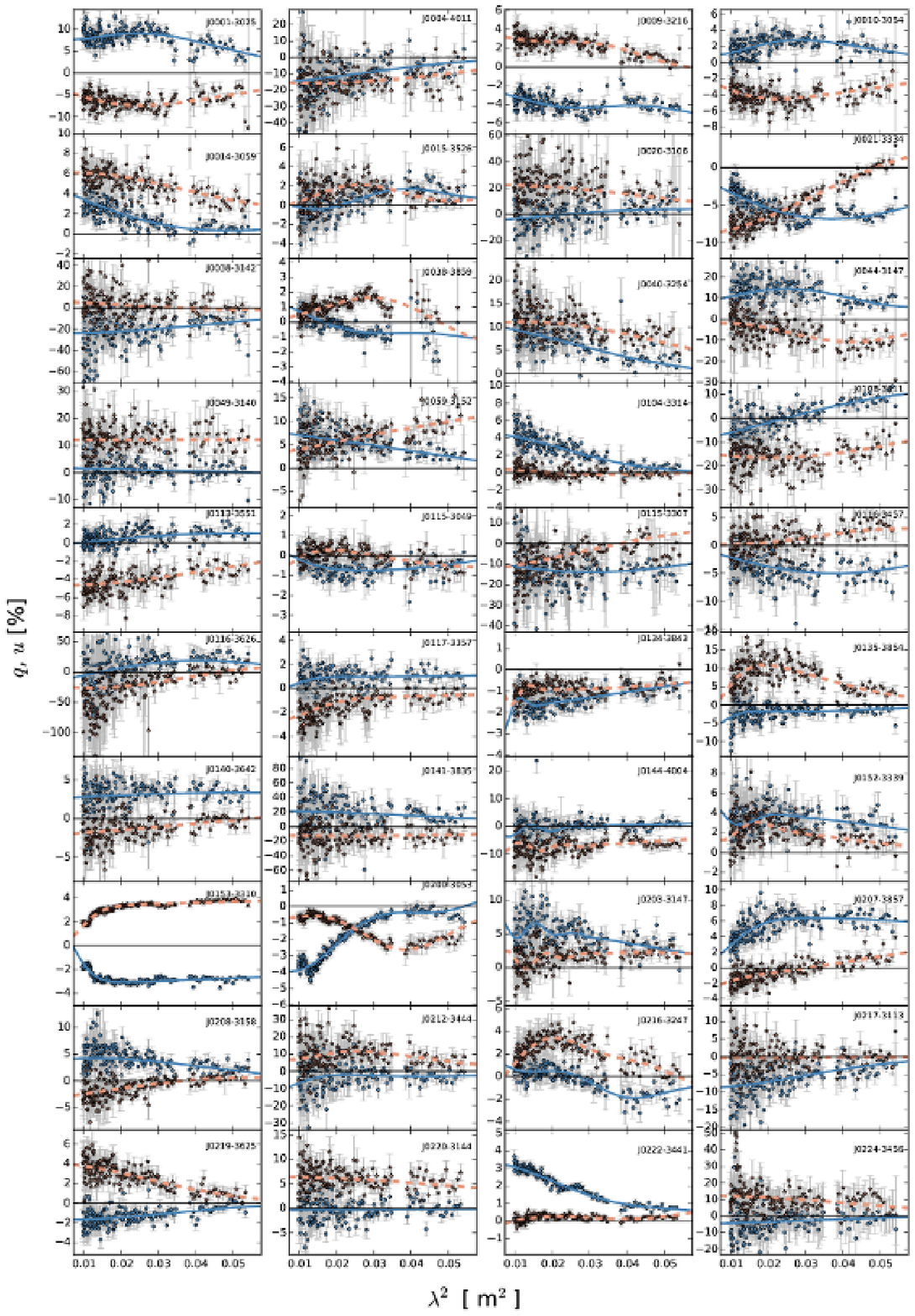}
\caption{Plots of $q(\lambda^2)$ (steel-blue, solid) and $u(\lambda^2)$ (dark-salmon, dashed) for all sources, overlaid with the best-fit model. }
\label{quplots1}
\end{figure*}

\begin{figure*}
\centering
\includegraphics[clip=true, trim=0cm 0cm 0cm 0cm, width=16.7cm]{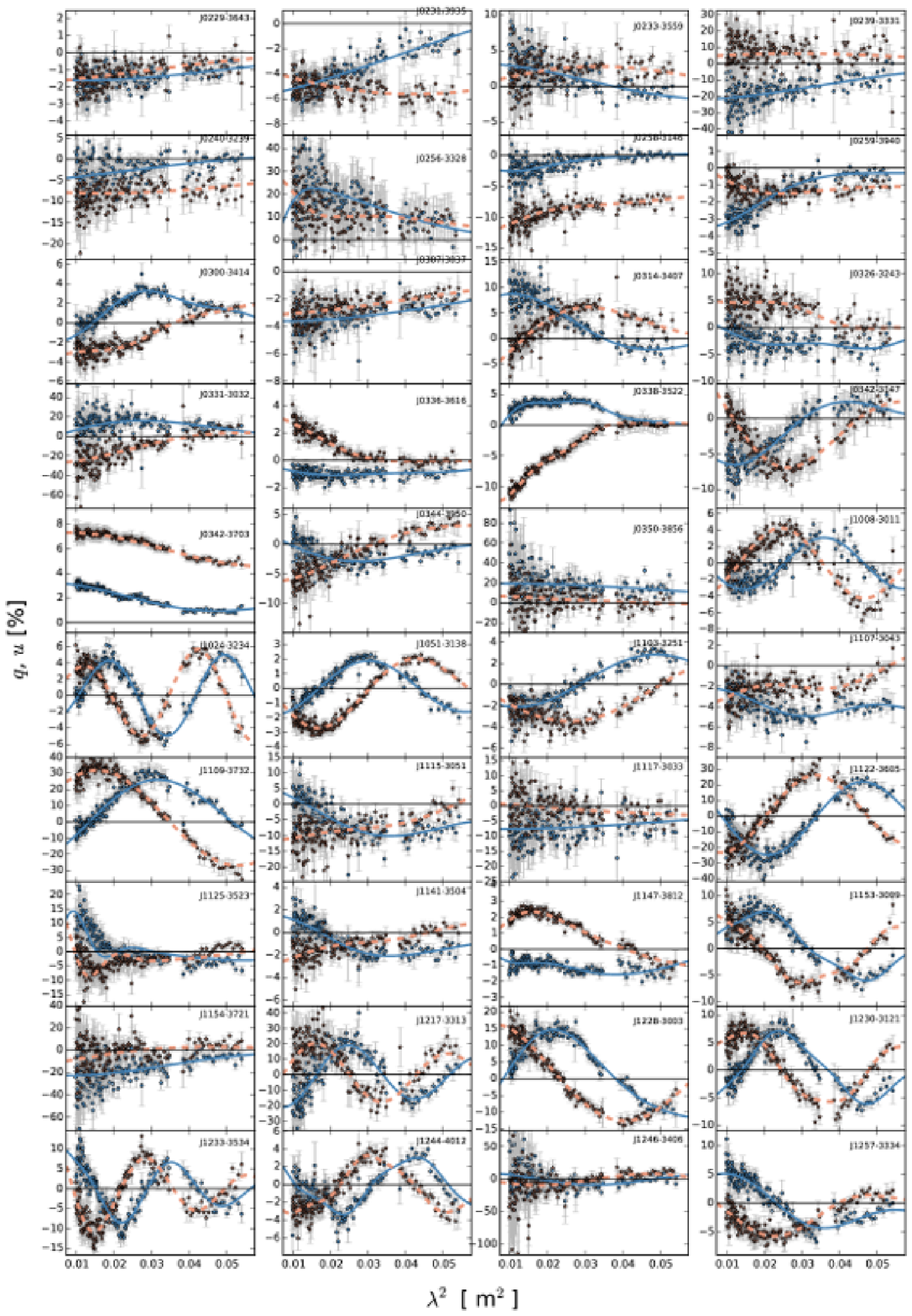}
\contcaption{Plots of $q(\lambda^2)$ (steel-blue, solid) and $u(\lambda^2)$ (dark-salmon, dashed) for all sources, overlaid with the best-fit model. }
\label{quplots2}
\end{figure*}

\begin{figure*}
\centering
\includegraphics[clip=true, trim=0cm 0cm 0cm 0cm, width=16.7cm]{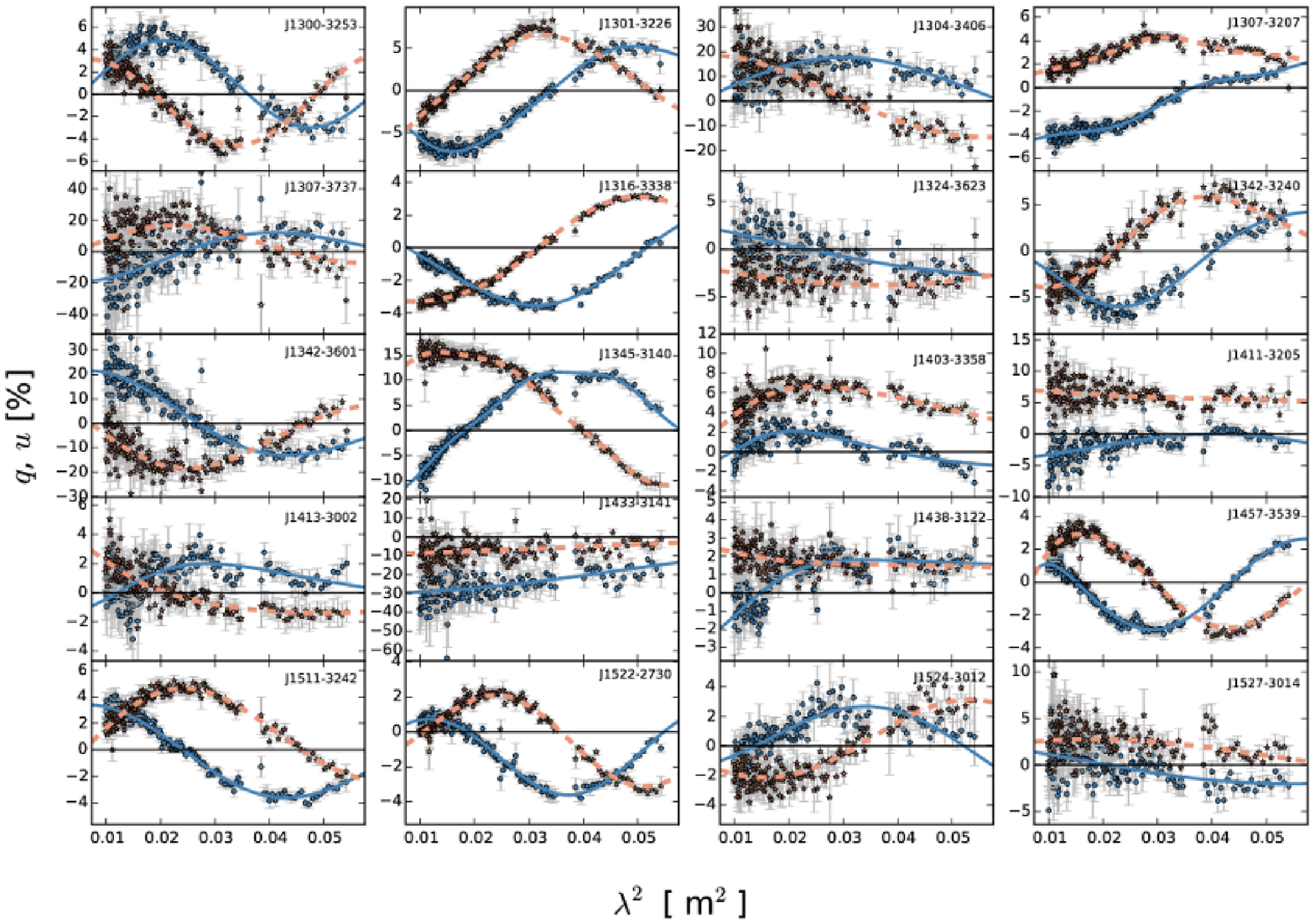}
\contcaption{Plots of $q(\lambda^2)$ (steel-blue, solid) and $u(\lambda^2)$ (dark-salmon, dashed) for all sources, overlaid with the best-fit model. }
\label{quplots3}
\end{figure*}

\begin{table*}
	\centering
	\caption{This table is an excerpt of the $q(\lambda^2)$ and $u(\lambda^2)$ data for each source shown in Figure~\ref{quplots1}, and that is available in full online.}
	\label{exdata}
	\begin{tabular}{lccccc} 
		\hline
		Name & $\lambda^2$ & $q$ & $u$ & $q_{\rm error}$ & $u_{\rm error}$ \\
		(J2000) & (m$^2$) &   &  &  &  \\
		\hline
J0001$-$3025  &   0.0541252  &   0.0549955  &   $-0.1237774$  &   0.1296630  &   0.1644325 \\
J0001$-$3025  &   0.0532954  &   0.0288238  &   $-0.1098046$  &   0.0246340  &   0.0358045 \\
J0001$-$3025  &   0.0524846  &   0.0371329  &   $-0.0327055$  &   0.0187818  &   0.0190200 \\
J0001$-$3025  &   0.0516920  &   0.0403654  &   $-0.0282388$  &   0.0112305  &  0.0119184 \\
J0001$-$3025  &   0.0509174  &   0.0302855  &   $-0.0326857$  &   0.0122068  &   0.0139447 \\
J0001$-$3025  &   0.0501599  &   0.0597910  &   $-0.0721186$  &   0.0088106  &   0.0085630 \\
J0001$-$3025  &   0.0494193  &   0.0756755  &   $-0.0537177$  &   0.0092737  &   0.0106487 \\
J0001$-$3025  &   0.0486950  &   0.0570419  &   $-0.0522955$  &   0.0087066  &   0.0100261 \\
J0001$-$3025  &   0.0479864  &   0.0671745  &   $-0.0585362$  &   0.0102211  &   0.0116907 \\
J0001$-$3025  &   0.0472932  &   0.0574044  &   $-0.0618616$  &   0.0082491  &   0.0085874 \\
.  &   .  &   .  &   .  &   .  &   . \\
.  &   .  &   .  &   .  &   .  &   . \\
.  &   .  &   .  &   .  &   .  &   . \\
		\hline
	\end{tabular}
\end{table*}


\section{Images of all sources}
\label{sec:images}
\begin{figure*} 
\centering
    \includegraphics[angle=0, clip=true, trim=0cm 0cm 0cm 0cm, width=1.\textwidth]{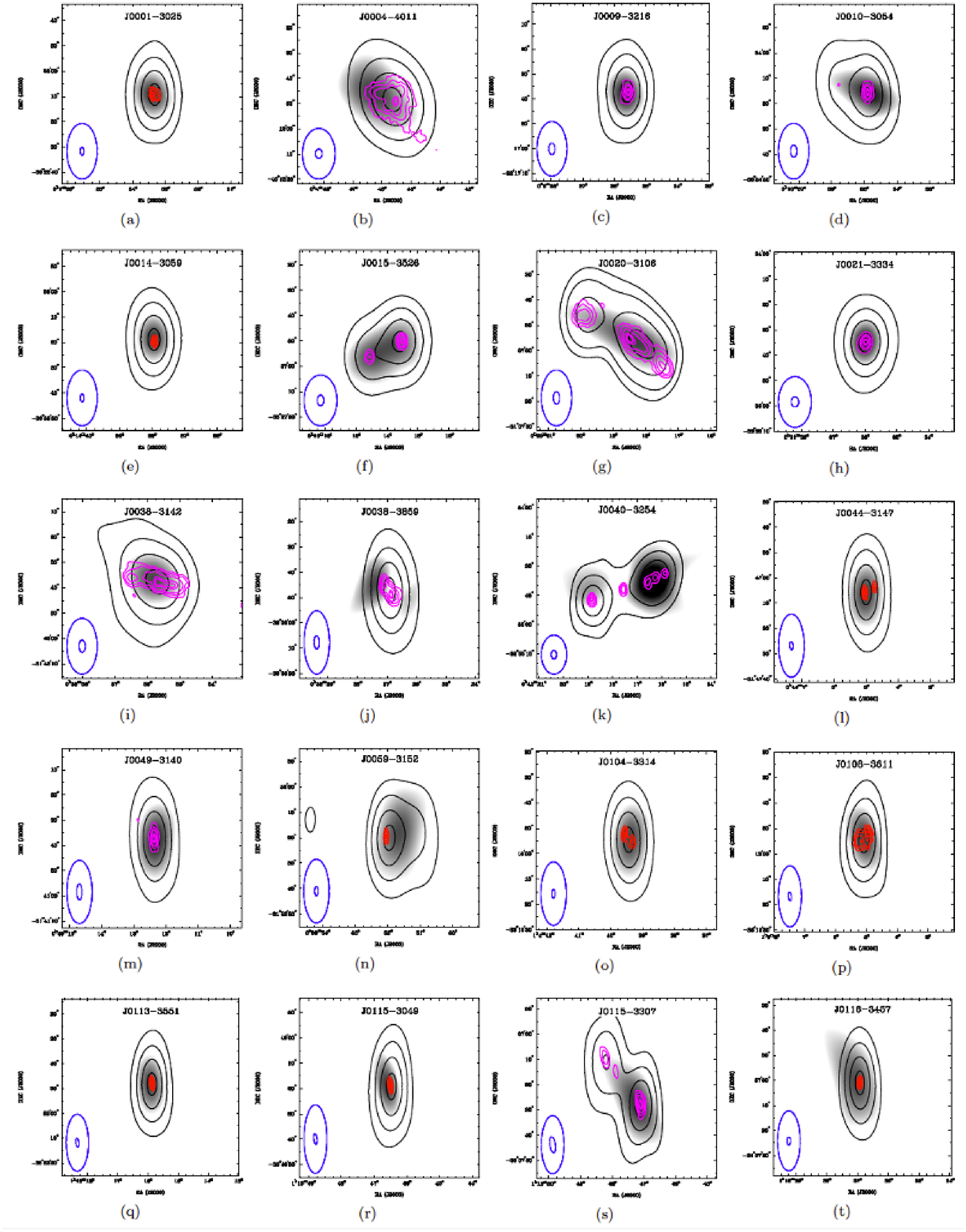}
  \caption{ Images of the total intensity (black contours) and polarization morphology (greyscale) for each source from 
  the 16~cm data. High resolution 4~cm images shown in magenta contours (4.5 to 6.5 GHz) and red contours (8 to 10 GHz). 
  Beam-sizes of low and high resolution images shown in bottom-left corner of each sub-image. }\label{images1}
\end{figure*}

\begin{figure*} 
\centering
    \includegraphics[angle=0, clip=true, trim=0cm 0cm 0cm 0cm, width=1.\textwidth]{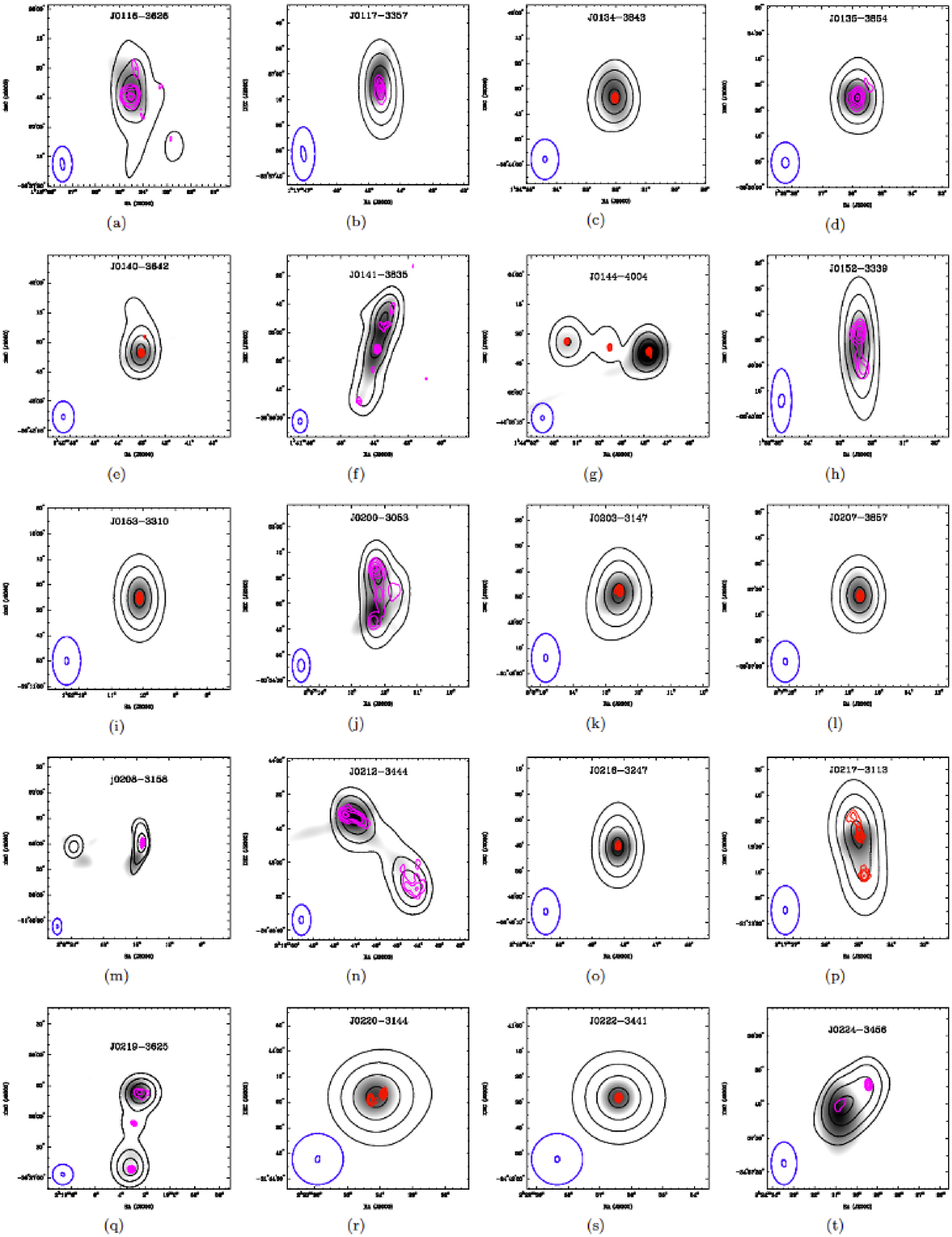}
  \caption{ See Figure~\ref{images1} caption. }\label{images2}
\end{figure*}

\begin{figure*} 
\centering
    \includegraphics[angle=0, clip=true, trim=0cm 0cm 0cm 0cm, width=1.\textwidth]{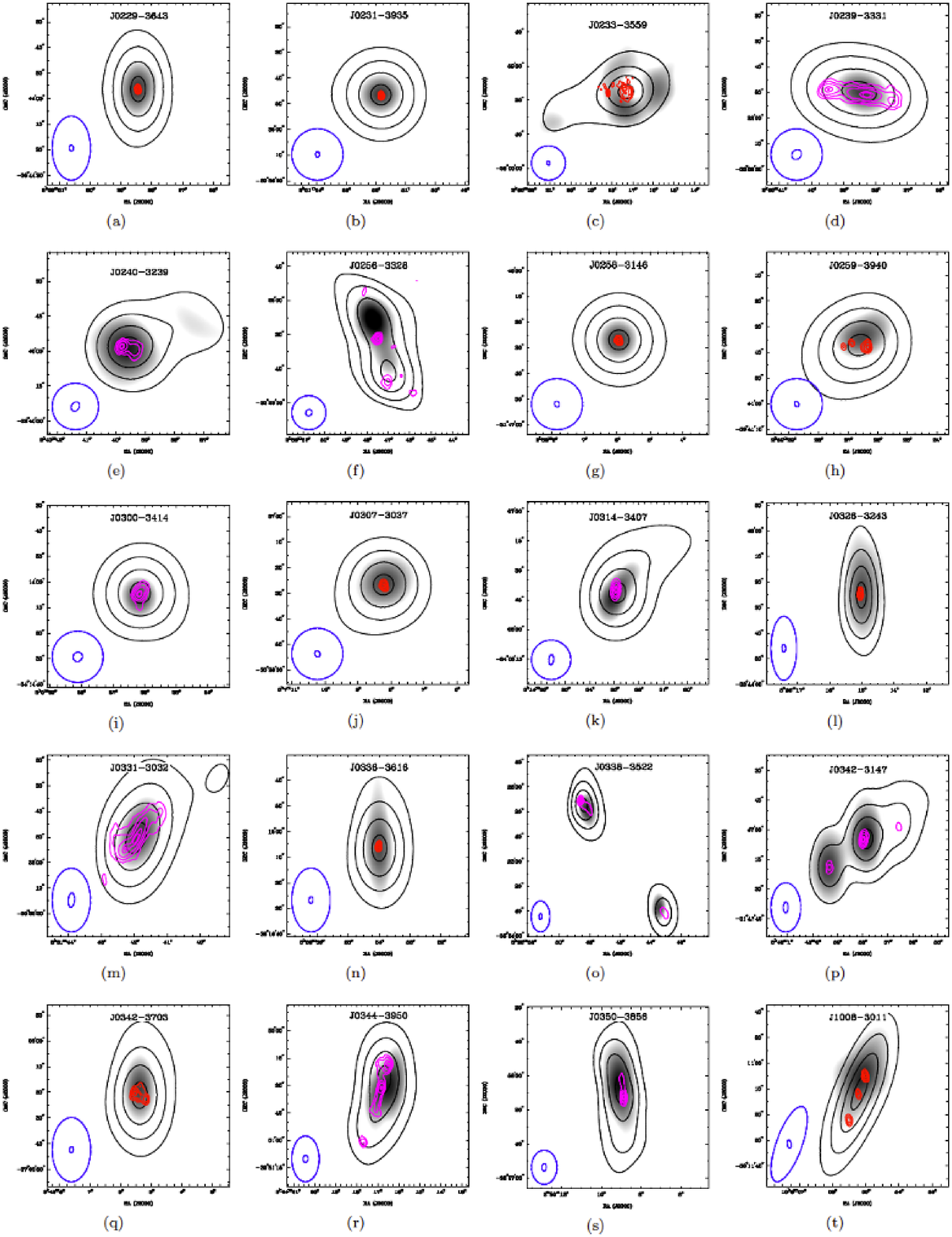}
  \caption{ See Figure~\ref{images1} caption. }\label{images3}
\end{figure*}

\begin{figure*} 
\centering
    \includegraphics[angle=0, clip=true, trim=0cm 0cm 0cm 0cm, width=1.\textwidth]{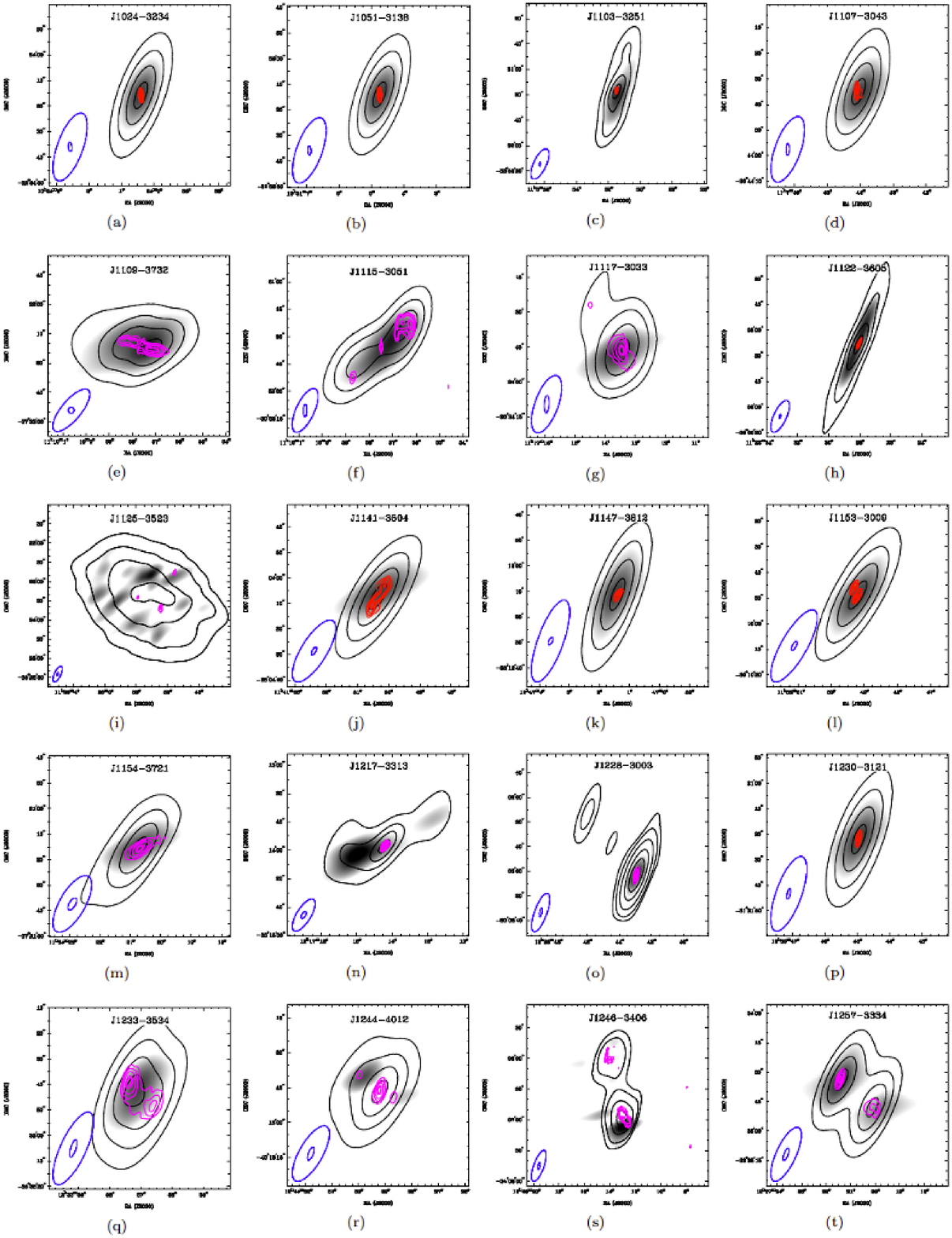}
  \caption{ See Figure~\ref{images1} caption. }\label{images4}
\end{figure*}

\begin{figure*} 
\centering
    \includegraphics[angle=0, clip=true, trim=0cm 0cm 0cm 0cm, width=1.\textwidth]{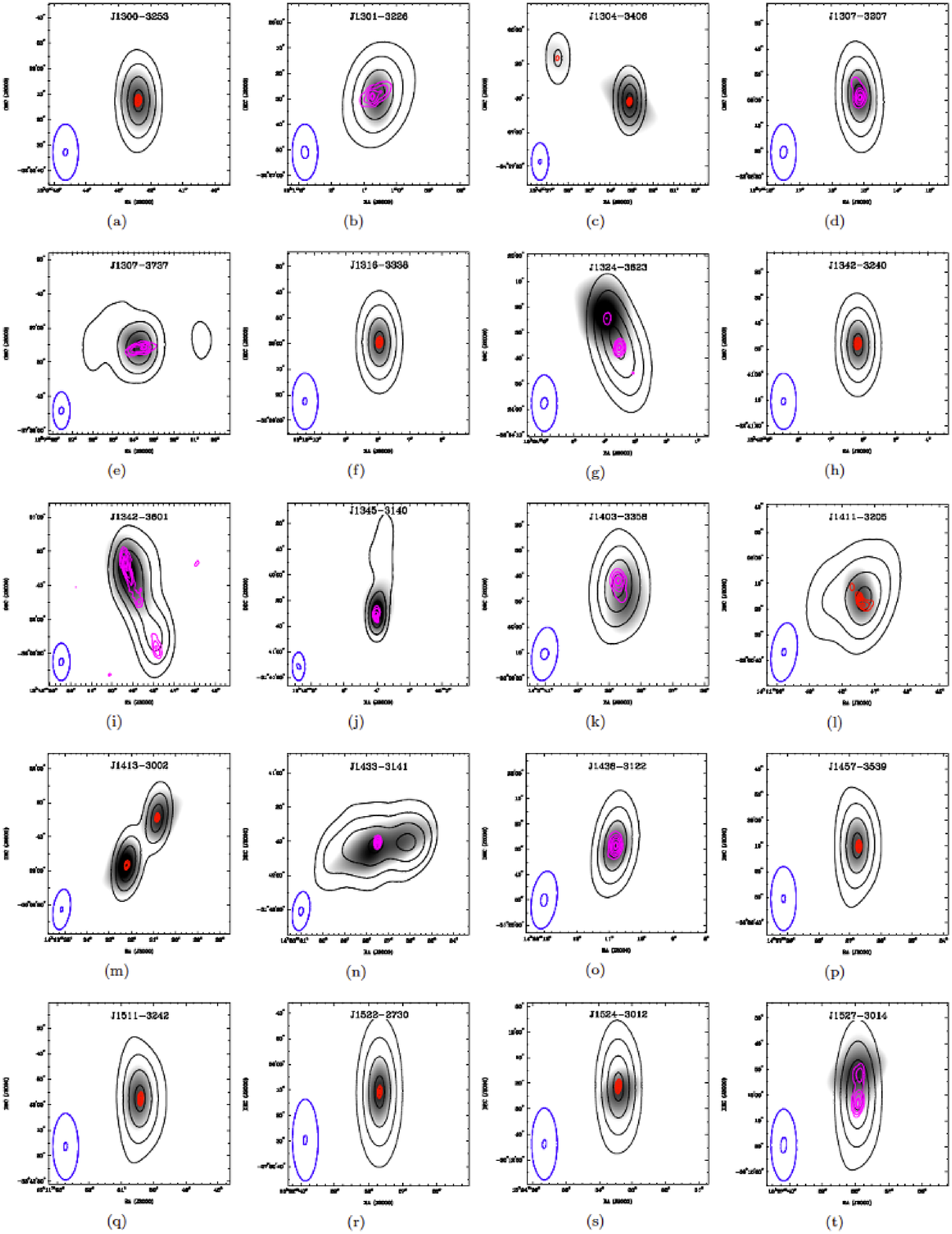}
  \caption{ See Figure~\ref{images1} caption. }\label{images5}
\end{figure*}

In Figures~\ref{images1}, \ref{images2}, \ref{images3}, \ref{images4}, \ref{images5}, we include the full-bandwidth total intensity images for all sources from 1 to 3 GHz, 4.5 to 6.5 GHz and 8 to 10 GHz, as well as the band-averaged polarized intensity distribution from 1 to 3 GHz, obtained by using RM synthesis.

\bsp
\label{lastpage}

\end{document}